\documentclass[a4paper,10pt]{book}
\usepackage{epsf}
\usepackage[dvips]{graphicx}

\begin{document}
\title{Biological Evolution and Statistical Physics}
\author{Barbara Drossel\\Department of Condensed Matter Physics\\
Tel Aviv University\\Ramat Aviv, Tel Aviv 69978, Israel}
\maketitle

{\bf Abstract}

This review is an introduction to theoretical models and mathematical
calculations for biological evolution, aimed at physicists. The
methods in the field are naturally very similar to those used in
statistical physics, although the majority of publications appeared in
biology journals. The review has three parts, which can be read
independently. The first part deals with evolution in fitness
landscapes and includes Fisher's theorem, adaptive walks, quasispecies
models, effects of finite population sizes, and neutral evolution. The
second part studies models of coevolution, including evolutionary game
theory, kin selection, group selection, sexual selection, speciation,
and coevolution of hosts and parasites. The third part discusses
models for networks of interacting species and their extinction
avalanches. Throughout the review, attention is paid to giving the
necessary biological information, and to pointing out the assumptions
underlying the models, and their limits of validity.

\tableofcontents

\chapter{Introduction}
\label{intro}

{\small Then God said: ``Let the land produce vegetation: seed-bearing
plants and trees on the land that bear fruit with seeds in it,
according to their various kinds.'' And it was so. The land produced
vegetation: plants bearing seed according to their kinds and trees
bearing fruit with seeds in it according to their kinds. And God saw
that it was good. And there was evening, and there was morning - the
third day. 

(...) 

And God said: ``Let the water teem with living creatures, and let
birds fly above the earth across the expanse of the sky.'' So God
created the great creatures of the sea and every living and moving
thing with which the water teems, according to their kinds, and every
winged bird according to its kind. And God saw that it was good.
(...) And there was evening, and there was morning - the fifth day.

And God said: ``Let the land produce living creatures according to
their kinds: livestock, creatures that move on the ground, and wild
animals, each according to its kind.'' And it was so. God made the
wild animals according to their kinds, the livestock according to
their kinds, and all creatures that move along the ground according to
their kinds. And God saw that it was good.  Then God said, ``Let us
make man in our image, in our likeness, and let them rule over the
fish of the sea and the birds of the air, over the livestock, over all
the wild animals, and over all the creatures that move along the
ground.'' So God created man in his own image, in the image of God he
created him; male and female he created them. (...) God saw all that
he had made, and it was very good. And there was evening, and there
was morning - the sixth day.

\hfill (From Genesis 1)}

\medskip

This ancient biblical account states that God is the ultimate cause of
everything, including life, calling it into existence through the
power of his word, and giving it value and meaning by affirming its
goodness. The focus of this text is on these metaphysical questions
rather than on details of how it came about, and its language contains
many poetical elements. Already hundreds of years ago, famous Jewish
scholars and Christian church fathers held the view that the ``days''
represent different facets of creation and are not meant to be literal
units of duration nor to describe the temporal sequence of events.

The question concerning the sequence of events and the natural
mechanisms and laws at work belongs to the realm of science, which has
revealed much of the history of life on earth during the past two
centuries. Although evolutionary ideas are old, they received
widespread acceptance only after Darwin suggested a plausible
mechanism for evolution and backed it by a detailed analysis of
observations in nature and among breeding stocks. This mechanism is
natural selection which increases the relative survival chances of
those naturally occurring variants that are best adapted, or
fittest. Thus, he reasoned, through a long chain of small steps, each
of which selected the fittest variants, the rich variety of today's
life forms could have evolved from one or a few simple original
forms. Darwin admitted that the causing agents of variation are not
known. Furthermore, he argued that the fossil record, the geographic
distribution of species, the similarities in embryonic development and
in structural elements even if their function differs, and the
grouping of species into a hierarchy of taxonomic orders point to the
common descent of all life
\cite{darwin}.

The first step towards the science of genetics was made by Darwin's
contemporary Mendel who discovered that parental traits do not blend
in their offspring, but are preserved as distinct units under the laws
of inheritance. With the discovery of chromosomes and genes, the
material basis for inheritance and the storage of genetic information
was revealed, and mutations were identified as a source of genetic
variation. The ``synthetic theory'', also called ``Neodarwinism'',
unified the findings of genetics with Darwin's principle of natural
selection into a theory of evolution where spontaneous, random
mutations in the germ cells produce the variation on which natural
selection acts. It is said that acquired characteristics cannot be
inherited nor does the environment influence the direction of the
mutations. Natural selection is considered to be the only force
capable of driving the evolutionary process.

The last decades have seen rapid progress in molecular biology. The
deciphering of the genetic code and the sequencing of many genes
confirmed the common origin of all life forms, and provided new means
of reconstructing phylogenic trees. It was also found that the
molecular processes involved in the reproduction, expression, and
rearrangement of genetic material are of breathtaking complexity, and
subject to sophisticated control and feedback mechanisms. Experiments
with bacteria, accompanied by gene sequencing, continue to reveal
their phenotypic and genotypic responses to selection, and the
connection between the two. Studies of {\it Drosophila} uncovered the
genetics of development in multicellular organisms.  These and other 
findings create the need to refine and to modify the theory. The
debate about which mechanisms are important for large-scale evolution
is lively and diverse.

The theoretical approaches lag far behind the experimental
findings. While existing theoretical models and mathematical
calculations cover a certain range of phenomena, verbal arguments and
plausible stories prevail in many other areas, creating the need for
more theoretical efforts.  It is the purpose of this review to give an
introduction to the existing theoretical literature on biological
evolution for physicists. Since evolution deals with genes in a gene
pool, with individuals in a population, or with species in an
ecosystem, it can in fact be considered as a topic for statistical
physics.  Although the majority of theoretical models and calculations
appeared in biology journals so far, the methods they employ are quite
similar to those used in statistical physics, and physicists have been
involved in evolutionary research for a long time. Ensemble averages,
diffusion equations, energy landscapes, cellular automata, and
computer simulations of networks of interacting units, belong to the
tool box of evolutionary theory. Parallels between statistical physics
problems and population genetics models are for instance pointed out
in \cite{higgs95}.

The outline of this reviews is as follows: The next chapter gives
an overview of the theory of evolution by random mutation,
selection, and drift in fixed fitness landscapes. Then, in chapter 3,
the effect of other individuals on fitness is taken into account, and
models for coevolution are discussed. Chapter 4 reviews models of
networks of many interacting species, with the focus on extinction
avalanches. Finally, the concluding chapter gives a brief summary and
points out challenges to be taken up in the future.

The main aim of this review is not to give a complete overview of the
literature (which is simply impossible), but rather to give an
introduction to the ideas, concepts, models, and methods used in the
field. Representative examples and instructive models were preferred
over a complete list of references on a given subject. Special
attention was paid to pointing out the underlying assumptions of the
models, and the limits of their validity. The reader will find a
sufficient amount of biological background information and references
to biological data, so that (s)he has a good starting point for
doing her/his own research.

Many interesting and relevant topics could not be covered.  These
include computer experiments and artificial life simulations on
evolution, autocatalytic and genetic networks, and the reconstruction
of phylogenies from molecular data.

\chapter{Evolution on fitness landscapes}

\section{Introduction}

This chapter is concerned with models for evolution of populations on
fixed fitness landscapes. The fitness of an individual is a measure of
the degree of its adaptation to its environment, and is defined in
mathematical terms as the expected number of
offspring that reach adulthood. A fitness landscape is a mapping of
genotype space onto fitness, i.e., onto real numbers. The underlying
assumption of this mapping is that the fitness of an individual can be
expressed as a function of its genotype alone. This assumption is
generally not correct. First, changes in climate and other external
conditions (including those during embryonic development) affect the
fitness. Part of these could be taken into account using dynamic
fitness landscapes that change in time. (For a review on dynamic
fitness landscapes, see \cite{wilke99}.) Second, the fitness of an
individual may depend on the composition of the population: Mate
preferences, age structure, and specialization within a population can
affect the survival and reproductive success of a given individual.
Third, other species strongly affect the fitness of individuals of a
given species, like competitors, predators, preys, symbionts, or
parasites. Some of these will be taken into account in the next
chapter which discusses coevolution.  Nevertheless, a static fitness
landscape that depends only on the genotype can be a good
approximation on short time scales, if all the mentioned factors
remain essentially constant, or under laboratory conditions which can
be held constant over many generations.  In fitness landscape models,
it is usually assumed that the total population size is fixed and
determined by the carrying capacity of the environment, e.g., the
availability of food. For this reason, absolute fitness values are not
important, but only fitness ratios between different genotypes.

Even if the fitness landscape can be considered fixed, finding such a
landscape and modelling the change of the genetic composition of a
population in it during the course of time is a complex
task: The mapping from genotype to fitness is extremely complicated
due to nonlinear interactions between genes and complex regulatory
mechanisms controlling gene activity, and the features of the
correct fitness landscape are therefore essentially unknown.
Furthermore, mutation and reproduction must be implemented in some
way. In most models, mutations are assumed to be copying errors in
form of point mutations. However, it is known that the genome can
undergo far more sophisticated rearrangements than mere random point
mutations, and that these changes may also be influenced by the external
environment.  Reproduction is for many species sexual and involves
recombination between homologous chromosomes during meiosis, and
mating is often non random. Usually only models with simple
fitness landscapes make assumptions that are more complicated than
asexual reproduction or random mating. 

This chapter is structured in the following way: In the next two
sections, we will give a very short overview over the biology of the
genotype-phenotype mapping, and of mutational processes. The knowledge
of these biological facts is necessary in order to appreciate the
simplifications contained in the models, and to define meaningful
directions of future theoretical research.  In the subsequent
sections, we will summarize various models and concepts for evolution
on fitness landscapes and discuss their relevance to real-life
situations. Among them are Fisher's theorem of natural selection, NK
landscapes and the adaptive walk model, Eigen's quasispecies model and
the error threshold, and models with finite population sizes and
Mueller's ratchet. Then we will focus on the question of neutral
evolution and flat or partially flat fitness landscapes. RNA evolution
will be mentioned as well as some new models that contain neutral
directions in the fitness landscape.  Finally, the effect of
recombination and other phenomena that were not discussed in the
earlier parts are considered.

\section{From genotype to phenotype}

The information given in this and the next section can be found in
textbooks on genetics, and the reader who wants to work in the field
of evolution is strongly recommended to acquire a solid knowledge of
the workings of the genome by studying a recent edition of one of the
textbooks.

The genetic information is stored in the four-letter alphabet of the
DNA. These four letters are the deoxyribonucleic acids adenine and
guanine (the two {\it purines}), and thymine and cytosine (the two
{\it pyramidines}). They are arranged in a stable double helix
structure consisting of two parallel strands of DNA wound around each
other, with hydrogen bonds connecting partners like rungs of a ladder.
Adenine is (almost) always paired with thymine, and cytosine with
guanine, ensuring the faithful replication of a genome by separating
the two strands and attaching the matching partners to each strand. In
prokaryotes, such as bacteria, the main DNA is arranged in a single
chromosome, which has the shape of a ring. Smaller rings or strands
that contain more special genetic information that is not vital in all
situations are usually also present and are called plasmids. In
eucaryotes, the genetic information is distributed among several
chromosomes, which are confined in the cell's nucleus.  The typical
size of a bacterial genome is of the order of $10^6$ base pairs. The
human genome has $3.4\times 10^9$ base pairs.

Part of the DNA codes for proteins. A section of DNA coding for a
protein (or something else) is called a {\it gene}. The chromosomal
location of a gene is called a {\it locus}, and alternative forms of
the gene at a given locus are called {\it alleles}. Proteins are
polymers consisting of amino acids. Since there are 20 different
primary amino acids, three letters of the DNA alphabet are needed in
order to code for one amino acid. The 64 possible three-letter
combinations lead to some redundancy, which is usually found in the
third of the three letters. Such a three-letter combination is called
a {\it codon}. Three of the codons do not code for amino acids, but
are {\it stop codons} which are the signals for ending
translation. The generation of a protein is a two-step process: First,
the relevant section of DNA is transcribed into a RNA copy.  (RNA is
very similar to DNA, but with thymine being replaced by uracil, and
with ribose instead of deoxyribose as its backbone sugar moiety.)  In
eucaryotes, the RNA is further processed by cutting out intervening
sequences, called {\it introns}, that do not code for the
protein. This cutting may be done in different ways to produce
different proteins. The resulting product is called the {\it messenger
RNA}, or mRNA. In procaryotes, one mRNA strand usually codes for
several proteins that function together (e.g., because they are
involved in the digestion of the same sugar). Next, the mRNA is
translated into the protein(s) by associating the appropriate amino
acid with each codon and by weaving the amino acids together.  This
process is mediated by the {\it transfer RNA} (tRNA) which transports
amino acids to the corresponding codon, and by {\it ribosomes} that
weave the amino acids together. The function of a protein is not
determined as much by its sequence as by the three-dimensional
structure into which it folds. The connection between the amino-acid
sequence of a protein and its three-dimensional structure is
essentially an unsolved problem.  Proteins play a variety of important
roles, for instance as structural material ( e.g., for cell walls), as
motor molecules in muscles, as enzymes regulating important cell
functions (including those that produce the proteins themselves!), as
transport vehicles (like hemoglobine transporting oxygen), as carriers
of information inside an organism (like hormones), or as antibodies in
the immune system.  Because many proteins are needed only in certain
situations or - for multicellular organisms - in certain cells, gene
expression is regulated by a variety of mechanisms and control cycles,
and often depends also on stimuli from the external environment.

Those regions of DNA that do not code for proteins, fulfil a variety
of different functions. Some genes code for RNA that fulfils certain
catalytic tasks in the cell, like the above-mentioned tRNA, and the
ribosomal RNA, which is part of the ribosomes. The DNA in the
neighborhood of a gene contains regions for the attachment of the
enzyme that catalyzes transcription, and of repressor or activator
proteins involved in gene regulation. Other sequences of DNA, which
are highly repetitive, play a role in recombination processes that
rearrange the genome. Still other parts of the DNA are believed to
play a role in determining the structural properties of the
chromosomes.  (However, many people tend to consider these non-coding
parts simply as ``junk DNA''.) There are also many gene duplications
in the genome.  Some gene copies may differ slightly from each other
and be active at different stages in the developmental process, while
others appear to be completely inactive. Many remnants of genes still
exist which were once active in some ancestor, but are no longer used,
and are often damaged by mutations.

In summary, the phenotype is the result of an extremely
complex interplay of the different parts of the genome, and of its
 environment. In fact, the phenotype comprises not only the
physical makeup of an individual, but also its behavioural pattern.
Here the influence of the environment becomes especially clear.
Particularly in higher organisms, a certain part of their behaviour
is not programmed in the genes, but acquired by learning.

\section{Heritable genetic changes}
\label{mutations}

A change in the DNA sequence that is passed on to daughter cells is
called a {\it mutation}. One source of mutations are copying errors
during DNA replication, such as the insertion of the wrong nucleic
acid - a {\it point mutation}, or the deletion or insertion of a
nucleic acid, called {\it frameshift mutations}. While a point
mutation in a coding region at worst changes one amino acid, a
frameshift mutation changes the entire sequence of amino acids from
the point of the insertion or deletion, because the codons are now
different.  Deletions of several nucleic acids may also occur. As the
vast majority of such copying errors are deleterious, sophisticated
proof-reading and error-correcting mechanisms try to avoid them.
X-rays, radioactive radiation, and chemicals, may also induce
mutations. They are usually deleterious and are corrected as far as
possible by repair mechanisms.  Uncorrected deleterious mutations are
quickly removed from the gene pool by negative natural selection, if
the fitness is decreased sufficiently.  Some random mutations may have
no effect on fitness, and very rarely are they beneficial. It seems to
be a widespread belief that these very rare beneficial random
mutations are the main source for the genetic variation on which
positive natural selection acts, thus driving the process of
evolution.

However, there exist a variety of other mechanisms by which the genome
is changed in a less random way, and which must also play a role in
evolutionary processes. First, it is known that mutations occur more
often at some sites (``mutational hot spots'') than at others, and
that the overall rate of mutations can also be subject to change and
selection. Second, transposable genetic elements, i.e., DNA segments
that can move from one place in the genome to another, cause changes
in the level of gene expression, in the timing of the developmental
pathway, and even rearrangements of genes or chromosomes. One type of
these elements are {\it insertion sequences} (IS) that usually contain
translational and transcriptional stop signals, and sometimes start
signals. Therefore, they may interfere with the regulation and
expression of adjacent genes.  Site-specific recombination can take
place between two identical IS elements. This leads to deletion,
inversion, or duplication of sequences flanked by two identical IS
elements. The second type of transposable elements are {\it
transposons}, which are larger than insertion sequences and carry
other genes besides those necessary for transposition. They can
therefore move or copy genes from one place in the genome to
another. It is held by some scientists that transposable elements play
a pivotal role in evolution because they allow for coordinated changes
in the genome subject to biologically meaningful feedback
\cite{shapiro}. Third, it is believed that duplication and subsequent
differentiation of genes has played an important role in the creation
of new genes. Such duplication may occur through the mediation of
transposable elements, or by so-called tandem duplications. Fourth,
genetic material can be transferred horizontally between bacteria and
even across species. This is done either by incorporating free DNA from
the environment, or through viruses, or by physical contact between
bacteria via a process called conjugation.

A recent report on our current knowledge of these and other mechanism
can be found in the proceedings of the conference ``Molecular
strategies in biological evolution'' \cite{msibe}.

\section{General properties of theoretical models}

Given the immense complexity of the genotype-fitness mapping,
theoretical models have to make a variety of simplifying assumptions.
Most models in biological literature focus on the effect of one or a
few genetic loci on the fitness of individuals in a population,
assuming that each of the considered loci can be occupied by a limited
number of different alleles that have different effects on the
fitness, and that the rest of the genome is part of the invariant
environment.  Based on this approximation, the first attempts to
obtain analytical results for changes in the
gene pool of a population under the influence of inheritance,
selection and mutation were made in the late 1920s and early 1930s by
R.A.  Fisher, J.B.S. Haldane, and S. Wright, who founded the field of
population genetics. Initially ridiculed by some as ``beanbag
genetics'' their method of randomly drawing the genes of the daughter
population from the pool of parent genes, with weights proportional to
the fitness, proved to be very successful at calculating the evolution
of allele frequencies from one generation to the next, or the chances
of a new mutation to spread through a population, even taking into
account various patterns of mating, dominance effects, nonlinear
effects between different genes, etc.  Wright was also the first to
envisage the concept of a genotype space and a mapping from this
genotype space to fitness. Population genetics has since then
developed into a mature field with a sophisticated mathematical
apparatus, and with wide-ranging applications. For an introduction to
this field, the reader is recommended the textbook by Crow and Kimura
\cite{crow}.

With the advent of molecular genetics, models were introduced that map
a DNA or RNA sequence to fitness. The fitness of a RNA or of a protein
encoded by DNA gauges its efficiency at performing a certain
catalytic task. If one assumes that the efficiency affects the fitness
of the entire individual, selection will act upon the underlying
sequence. In models for prebiotic evolution, such as the quasispecies
model introduced by Eigen, the fitness of a macromolecule is simply
given by its reproduction rate.

Still another approach tries to model evolution of a population in its
full fitness landscape. Because the structure of the full fitness
landscape is unknown and complex beyond any modelling capabilities, toy
landscapes are introduced that may hopefully reflect some features of
the real landscape. These landscapes are often rugged, with many peaks
and valleys. The genotype is usually represented by a binary
string. Each bit may either stand for a genetic locus with two alleles
(in the classical picture), or for a purine or pyramidine (in the
molecular picture). The choice of a binary representation should not
affect the conclusions drawn from these models.

In the following, we will present models and results for all three
mentioned approaches (population genetics, molecular genetics, and toy
landscapes). All discussed models have in common that the environment
and the genome length are fixed, that fitness is independent of the
composition of the population, and that mutation rates are constant in
time and independent of the place at which they occur. Only in the
last section will we loosen some of these constraints.

We will start with Fisher's fundamental theorem of natural selection,
which considers the effect of selection on an infinite population
without mutations. Next, we introduce rare mutations that spread
through the entire population if they are advantageous. The properties
of the resulting ``adaptive walks'' will be discussed for a variety of
fitness landscapes. The next logical step is to consider
genetically mixed populations, but to still neglect fluctuations in
the composition of the population. This will lead us to Eigen's
quasispecies model. Then, the effect of finite population sizes will
be discussed, followed by the topic of neutral evolution.
Finally, some possible extensions are mentioned.

The reader is also recommended to consult the review by Baake and
Gabriel \cite{baake}, which covers many of these topics and has an
extensive list of references. The lecture notes by Peliti
\cite{pelnotes} are a good pedagogical introduction into the the topic
of fitness landscapes.

\section{Fisher's Theorem of Natural Selection}
\label{fundamentaltheorem}

The fundamental theorem of natural selection states that in the
absence of mutations in an infinitely large population the rate of
change in mean fitness is equal to the additive genetic variance in
fitness within the population. An immediate consequence is that the
mean fitness never decreases, and that it remains constant only when
all individuals of the population have the same fitness.

The additive genetic variance in fitness is the variance of the
additive contribution of each gene or chromosome (depending on which
are the fundamental units that can be shuffled through recombination
and sex) to the deviation of the fitness of a genotype from the
population mean. These additive contributions are obtained by
performing a least-square fit to the actual fitness deviations from
the mean, using only additive gene (or chromosome) contributions.

This theorem is very general.  In the following, we will first
consider a sexually reproducing population and proof the theorem for a
specific case, where the fitness is determined by a single gene which
occurs in two variants (``alleles'') in a randomly mating population.
Generalization to more alleles is easily possible, however, the
generalization to more genes with possibly nonlinear interactions is
more complicated and can be found in \cite{crow}. The original
derivation by Fisher \cite{fisher} is very difficult to understand.
Then, we will give a general proof for an asexual population.

Let $p_1$ be the frequency of allele 1 and $p_2$ the frequency of
allele 2 in the population, and let $W_{ij}$ be the fitness of an
individual carrying allele $i$ and allele $j$. (As mentioned above,
the fitness of an individual is proportional to the expected number of
offspring that reach adulthood.) The mean fitness of an individual
carrying allele 1 is $W_1=p_1 W_{11} + p_2 W_{12}$, and the mean
fitness of the population is
\begin{equation}\bar W = p_1^2 W_{11} + 2 p_1p_2 W_{12} + p_2^2 W_{22}.
\label{wbar}
\end{equation}
We assume that the entire population is updated in parallel and
replaced with the daughter population after one generation.  In the
next generation, the frequency of allele 1 will be
$$p_1'=W_1p_1/\bar W=p_1+\Delta p$$ with $\Delta p = p_1(W_1-\bar
W)/\bar W$, and that of allele 2 $$p_2'=p_2-\Delta p.$$ From
Eq.~(\ref{wbar}) one obtains an expression for the change in mean
fitness,
$$\Delta \bar W = 2 \Delta p[(W_1-\bar W)-(W_2-\bar W)],$$
which gives finally
$$\frac{\Delta \bar W}{ \bar W } = \frac{2[p_1(W_1-\bar
W)^2+p_2(W_2-\bar W)^2]}{\bar W^2}.$$ The numerator is the additive
genetic variance in fitness. (This can be checked by using the
definition of the additive genetic variance given at the beginning of
this subsection and minimizing the quantity $p_1^2(2a-(W_{11}-\bar
W))^2 + 2p_1p_2(a+b-(W_{12}-\bar W))^2 + p_2^2(2b-(W_{22}-\bar W))^2$
with respect to the additive terms $a$ and $b$, yielding $a=W_1-\bar
W$ and $b=W_2-\bar W$). The factor 2 is due to the fact that
individuals are diploid, i.e., have two sets of genes. Fisher's
original derivation works with a model that is continuous in time. We
can obtain his result from ours if we assume that the gene frequencies
and therefore the fitness values $W_1$ and $W_2$ change little from
one generation to the next, and that the fitness values are identical
to the expected numbers of offspring (instead of merely being
proportional to them). The mean fitness $\bar W$ is then close to 1;
subtracting 1 from the fitness values gives the reproduction rates
(which are the fitness values in continuous-time models), and we obtain
Fisher's equation for the rate of change in mean fitness,
$$\frac{d\bar W}{dt} = 2[p_1(W_1-\bar W)^2+p_2(W_2-\bar W)^2].$$

For an asexually reproducing population without recombination, the
general derivation of the theorem is straightforward: Let $p_i$ be the
frequency of genotype $i$ in the population, and $W_i$ its
fitness. The change in $p_i$ from one generation to the next is then
$$\Delta p_i=p_i(W_i-\bar W)/\bar W,$$
leading to a change in mean fitness
$$\frac{\Delta \bar W}{\bar W} = \frac{\sum_i W_i \Delta p_i}{\bar W}
= \frac{\sum_i p_i(W_i^2-\bar W^2)}{\bar W^2},$$ which is 
proportional to the genetic variance in fitness. If the fitness
changes from one generation to the next are small, this becomes again
an equation which states that the rate of change in fitness is
identical to the genetic variance in fitness.

Consequently, if such a population is placed in a fitness landscape,
it will climb uphill, until it reaches the highest peak accessible
without ever moving downhill, and all individuals of the population
will finally have the optimal phenotype. Natural selection chooses
among the existing genetic options and increases the weights of those
genomes that have higher fitness. This does not necessarily imply that
all individuals will also have the same genotype, because different
genotypes may have the same fitness.

Clearly, the theorem is based on a series of assumptions that are
generally not satisfied. First, taking mutations and finite population
sizes into account, some populations will move downhill, as we will
see below. Second, it is usually incorrect to assume that the fitness
of an individual depends only on its genotype and not on the other
individuals, as we have mentioned above.  Nevertheless, it is
appropriate to neglect mutations and the effect of other individuals
on fitness in situations of strong directional selection: For
instance, under breeding conditions those individuals that are closer
to the desired phenotype are chosen and given a chance to reproduce,
leading to an increase in frequency of the desired (``fitter'')
phenotypes (e.g., long fur, or high milk yield). A famous example of
natural directional selection is that of the peppered moth in Great
Britain, which changed its preferred colour from light, peppered, to
dark, melanic, due to the darkening of tree trunks by environmental
pollution. The good camouflage of these moths reduces their predation
by birds. The frequency of the initially rare melanic version
increased so much that it became the dominant phenotype and
constituted more than 90 percent of the population in polluted areas
in the mid-twentieth century.

\section{Adaptive walks and rugged fitness landscapes}

The first logical step away from the pure selection model discussed in
the previous subsection is a model that includes rare mutations.  The
concept of adaptive walks was introduced by S.A. Kauffman
\cite{kauff87,kauffman}. The underlying assumption is that all members
of a population have the same genotype most of the time. If a mutation
occurs that increases the fitness, it quickly spreads to all members;
mutations that reduce fitness are quickly eliminated. As a
consequence, populations always move uphill in the fitness landscape,
until they reach a local maximum, where they stay forever. The main
application of the adaptive-walk concept is in computer simulations or
analytical calculations for exploring and describing the properties
and the statistical structure of fitness landscapes.  But this model
is also a plausible limiting case of the evolution of a real
population in which the probability of finding a fitter variant of the
gene that is under selection pressure is very low per generation,
compared with the fitness difference between the new and the old
allele \cite{gil84}.

\subsection{The probability of an advantageous mutation to become fixed}

Before considering adaptive walks in a fitness landscape, let us first
calculate the probability that a mutation that confers a small fitness
advantage spreads through the entire population (which is assumed to
be very large and fixed in size). We assume that the fitness of an
individual that carries the advantageous mutation is larger by a
factor $(1+s)$ than the fitness of an individual that does not carry
it, and that $s$ is small.  We denote by $P_s(m)$ the probability that
the mutation will spread through the population if it is initially
carried by $m$ individuals.  Now, there are several ways to calculate
the composition of the population's next generation, corresponding to
different biological situations. Let us use a ``sequential update''
according to the following rule: We choose one individual at random to
be killed.  Then we choose another individual with a probability
proportional to its fitness to produce an offspring, i.e., an
identical copy. A short calculation shows that the population is
$(1+s)$ times more likely to make a transition from $m$ to $(m+1)$
individuals carrying the advantageous allele, than to $(m-1)$
individuals carrying it. It follows
$$P_s(m)=\frac{1-s/2}{2}P_s(m-1) + \frac{1+s/2}{2}P_s(m+1).$$
Because $s$ is small, the $P_s(m)$ change slowly with $m$, and we can perform a Taylor expansion, leading to
$$s\frac{dP_s(m)}{dm}=- \frac{d^2P_s(m)}{dm^2}.$$
The solution is 
\begin{equation}
P_s(m) = 1-e^{-sm},\label{fix}
\end{equation}
where we have made use of the limits $P_s(0)=0$ and
$P_s(\infty)=1$. If initially there is one individual that carries the
advantageous mutation, this mutation will spread with a probability
$s$ through the entire population. If the advantageous mutation occurs
with probability $U$ per unit time and per individual, the mean
waiting time for a given advantageous mutation sweeping through the
population is of the order $(MUs)^{-1}$, with $M$ being the population
size. This result depends on the details of the rule for updating the
population. Many authors perform a parallel update of the entire
population by choosing the individuals of the next generation
according to a multinomial sampling rule, leading to a probability of
$2s$ for a slightly advantageous mutation to become fixed. (See for
instance \cite{crow}.) In section \ref{finite}, we will give a more
formal derivation which comprises the different cases.

\subsection{Representation of the genotype, and the Fujiyama landscape}

In the following, we want to study the properties of landscapes with
varying degrees of ruggedness and describe the behaviour of adaptive
walks in these landscapes.  In order to be able to define a fitness
landscape, we must first give a representation of the genotype.  The
usual representation is a ``spin chain'' $\sigma =
(\sigma_1,...,\sigma_N)$, with $\sigma_i = \pm 1$, and the fitness by
a function $W(\sigma)$ that will be specified further below.
Furthermore, we assume that at each ``time step'' (usually a step in a
computer simulation) a point mutation occurs in one of the individuals
that flips one of the ``spins''. If the resulting genotype has a
higher fitness than the original one, the entire population adopts
this new genotype, otherwise the genotype of the population remains
unchanged. It is useful to define the distance $d_H$ between two
genotypes $\sigma$ and $\sigma'$ as the number of ``spins'' in which
they differ,
$$d_H = \frac{1}{2} \sum_i (1-\sigma_i \sigma_i').$$
This quantity is usually called the ``Hamming distance'', and 
is identical to the number of mutations separating the two genotypes.
The overlap between two genotypes is given by
$$q=\frac{1}{N}\sum_i \sigma_i \sigma_i' = 1-\frac{2}{N} d_H.$$

In a ``Fujiyama landscape'', which has a single peak that can be reached
from everywhere by always going uphill, an adaptive walk that starts
with a random genome always reaches the place of highest fitness. A
realization of such a landscape is given by an additive fitness function
$$W(\sigma) = \sum_i J_i \sigma_i$$ with arbitrary nonzero constants
$J_i$.  Each spin has an optimum value which is independent of the
other spins. In an adaptive-walk simulation, one spin is chosen per
unit time and flipped if this increases the fitness. Consequently, the
overlap $q$ with the optimum sequence changes with time according to
$$\dot q = \frac{1}{N} (1-q),$$ leading to $$q(t) = 1 -
(1-q(0))e^{-t/N}.$$ The distance to the peak decays exponentially with
time.

\subsection{Spin-glass landscapes}

Fitness landscapes will generally be more complicated than a Fujiyama
landscape and have several peaks due to conflicting constraints.  Two
different types of random fitness landscapes are found in the
literature. The first one was introduced by Derrida \cite{der81},
originally in the context of spin glasses.  It has the form
$$W(\sigma) = \sum_{i_1,i_2,...,i_{p}} J_{i_1i_2...i_{p}}
\sigma_{i_1} \sigma_{i_2} \cdots \sigma_{i_{p}},$$
with random
couplings $J$ that follow a Gaussian distribution with zero mean and variance $J_0^2$.
The given fitness function is identical to the Hamiltonian of the
$p$-spin model in spin-glass theory. For instance, for
$p=2$ we have the Sherrington-Kirkpatrick model of spin glasses, which is known to have a very complex energy landscape.
Derrida \cite{der81} proved that the fitness correlations in such a
landscape decay as
$$\frac{\langle W(\sigma)W(\sigma')\rangle}{\langle W(\sigma)^2\rangle} \simeq 
q^{p} \, ,
$$
if $N$ is much larger than $p$.  This can be shown in the following
way: $W(\sigma)$ is the sum of $N \choose p$ random terms that are
Gaussian distributed with zero mean and variance $J_0^2$. Of the terms
contributing to $W(\sigma')$, most are identical to those contributing
to $W(\sigma)$, and some have the opposite sign. If $p \ll N$, most of
the terms with opposite sign differ in the sign of one spin in the
spin product, and terms that differ in the sign of 3 or any higher odd
number of spins can be neglected. The number of such terms with
difference in the sign of one spin is $d_H {N-d_H \choose p-1}$. The
first factor is the number of possibilities to choose a spin that
has a different sign in $\sigma$ and $\sigma'$; the second factor is
the number of possibilities to choose the $p-1$ remaining spins among
the $N-d_H$ spins that agree in sign. Let $W_1$ be the sum of all
those terms that are identical for the two spin configurations, and
$W_2$ the sum of all those terms whose signs become reversed in
$W(\sigma')$. Then we have
\begin{eqnarray}
\langle W(\sigma)W(\sigma')\rangle &=& \langle (W_1+W_2)(W_1-W_2)\rangle
= \langle W_1^2\rangle - \langle W_2^2\rangle \nonumber\\
&=& \langle W(\sigma)^2\rangle - 2\langle W_2^2\rangle \nonumber\\
&\simeq& \langle W(\sigma)^2\rangle \left(1-2d_H \frac{{{N-d_H} \choose {p-1}}}{{N \choose p}} \right)\nonumber\\
&\simeq&\langle W(\sigma)^2\rangle \left(1-2d_H \frac{p}{N} \right)\\
&\simeq& \langle W(\sigma)^2\rangle q^{p}.\nonumber \label{corr}
\end{eqnarray}
As $p$ increases, the fitness correlations decay faster,
and the fitness landscape becomes more rugged. Beyond some finite
distance in genotype space, fitness values are essentially uncorrelated.
Some properties
of such an uncorrelated rugged landscape will be given below in the
context of the NK model.

\subsection{NK landscapes}
\label{NK}

Kauffman introduced a different but related family of random
landscapes where the fitness contribution of each of the $N$ spins
depends on the value of $K$ other spins \cite{kauffman}. For each spin
$i$, these $K$ spins are chosen at random, or according to some other
rule. Let their indices be $i_1,...,i_K$. Kauffman assumes that
the fitness of each of the $2^{K+1}$ possible configurations of spin
$i$ and its $K$ ``neighbours'' is an independent random number chosen
from a uniform distribution in the interval [0,1]. This means
$$W(\sigma) = \frac{1}{N}\sum_i
J(\sigma_i,\sigma_{i_1},\sigma_{i_2},...,\sigma_{i_K}).$$
An explicit
expression for $J$ contains a sum over $2^K$ different spin products.
For instance, for $K=1$, the fitness can be written as
\begin{eqnarray*}
W(\sigma) &=& \frac{1}{4N}\sum_i
\left[J_i^{(1)}(\sigma_i+1)(\sigma_{i_1}+1) +
  J_i^{(2)}(\sigma_i+1)(\sigma_{i_1}-1)\right.  \\
 &+&  \left. J_i^{(3)}(\sigma_i-1)(\sigma_{i_1}+1) +
  J_i^{(4)}(\sigma_i-1)(\sigma_{i_1}-1) \right] .
\end{eqnarray*}
In contrast to the
spin-glass model, the sum is taken for each $i$ only over one set of
partners, which means that there is less frustration in the system.
Also, while in the $p$-spin glass only products of $p$ spins occur, in
the NK model spin products of any number of up to $K+1$ spins occur.
The case $K=1$ corresponds to a spin glass where each spin has on an
average two random neighbours, and a random field. This model is different
from the Sherrington-Kirkpatrick mean-field spin glass mentioned
above.

In the case $K=0$, we have a Fujiyama landscape with a single maximum.
As $K$ increases, more and more conflicting constraints are imposed on
the system. Computer simulations of adaptive walks result in the
following \cite{kauffman}: For small $K$, the constraints can be
satisfied to a large extent, and the highest peaks in the fitness
landscape are close to each other. For $K=2$, these highest peaks have
the largest basins of attraction. With increasing $K$, these basins
become smaller, the height of the largest peaks decreases, and the
distance between them increases.  These numerical findings are in
agreement with an analytical study \cite{wei91}, which finds among
other results that the correlations in fitness decay faster for larger
$K$, with a law equivalent to Eq.~(\ref{corr}).

In the extreme case $K=N-1$, each genotype has a fitness which is an
independent random variable. This limit is equivalent to the
random-energy model which was introduced and analytically solved by
Derrida \cite{der81}.  Kauffman \cite{kauffman} gives a variety of
analytical results for this fitness landscape that can be easily
derived (see also \cite{fl92} for a detailed analytical analysis):

\begin{itemize}
\item The probability that a given configuration is a local maximum is $1/(N+1)$ (because each configuration has $N$ neighboring configurations).
  
\item The mean number of steps for an adaptive walk to reach a maximum
  is proportional to $\log_2 N$ (because at each flip the number of
  configurations with higher fitness is halved), and the mean time
  needed to reach this maximum is proportional to $N$ (because after
  this time $N+1$ configurations have been probed).
  
\item The larger $N$ is, the more likely it is that a local maximum is
  close to a typical value. This follows from the central limit
  theorem: For large $N$, the probability distribution of $W(\sigma)$
  is a Gaussian distribution with mean 1/2 and variance 1/(12N), and a
  typical maximum differs from the mean by the order of
  $\sqrt{\ln(N+1)/6N}$.  (The last expression can be obtained by
  calculating the maximum of $n$ independent Gaussian variables that
  are chosen from a distribution with mean $a$ and variance
  $\sigma^2$. The maximum is also a Gaussian variable, with mean
  $a+\sigma\sqrt{2\ln n}$ and variance $\sigma^2/2\ln n$.)  A similar
  conclusion can be drawn if $K$ is smaller than $N-1$, and if $K/N$
  is kept fixed with increasing $N$.

\end{itemize}

Kauffman calls this last effect a ``complexity catastrophe'': With
increasing complexity peaks are more likely to be close to typical
values. The more conflicting constraints on the fitness exist, the
poorer are the attainable compromises. He concludes that natural
selection and random mutation can do no more than pulling the genotype
slightly away from ``typical'' configurations. And the ``typical
properties'' themselves are according to him determined by laws of
self organization rather than by natural selection. Only if there are
not too many conflicting constraints (i.e., if $K$ remains small) can
the complexity catastrophe be averted, because the landscape is
sufficiently smooth to retain high optima as $N$ increases.

Another problem, apart from the complexity catastrophe, posed by
adaptive walks on rugged fitness landscapes is that populations get
stuck at low local maxima and cannot reach the higher maxima even when
they exist. There are several hypotheses of how populations
manage to reach peaks of high adaptation. The least accepted one is
that of ``hopeful monsters'', i.e., of larger genetic changes which
carry the genotype to a random point in the fitness landscape that is
farther away. It seems very unlikely that such a big blind jump leads
to a higher fitness. Another way to reach higher maxima would be to
allow the adaptive walk to also move downhill sometimes.  The
population could thus escape from one peak and find another one which
might be higher.  We will see in the section after the next that this
naturally happens if we take a finite population size into account.
However, it is quite uncertain that real fitness landscapes are as
rugged as in the random-energy model.  As the analysis of Kauffman has
shown, less rugged landscapes allow for high degrees of
adaptation. Furthermore, it might well be that true fitness landscapes
have many neutral directions which would allow the population to drift
through genome space without loosing much fitness, and thus to come
near the higher peaks. The topic of neutral evolution and of fitness
landscapes with neutral directions will be discussed later.
It should also be noted that the representation of the genome by a
binary string of qualitatively equivalent bits, and of mutations as
single-spin flips is far from the hierarchical structure of real
genomes and the nonlocal rearrangements occurring in them, as sketched
at the beginning of this chapter.  Finally, as we have discussed
further above, the picture of a fixed fitness landscape that depends
only on the genotype is inadequate in particular for evolution on
longer time scales. Wilke and Martinetz have shown that an adaptive
walk can percolate through a time-dependent NK landscape
\cite{wil99}. The last two points may hide the true solution to the
puzzle of how species reach states of high adaptation.

\section{The quasispecies model and the error threshold}

In contrast to the adaptive walk model, the quasispecies model assumes
that the mutation rate is finite, and it therefore takes into account
the genetic variability within a population. However, it neglects
fluctuations in the composition of the population. This approach is
appropriate if the population size is so large that most occurring
genotypes are represented by many individuals. For models that
consider only the contribution of one or a few genes to the fitness,
this approximation is valid. It might also be valid for clonal
populations if they have a recent common ancestor. But even clonal
species accumulate a considerable genetic variety after a longer time.
This was tested experimentally for E.Coli bacteria who were analysed
during a 10,000 generation experiment \cite{pap99}. For sexual
species, usually no two individuals carry the same genotype, and the
model is therefore inadequate if the entire genome is considered. 

The quasispecies model was introduced by M. Eigen \cite{eigen} in the
context of a model for prebiotic evolution. In this model, the
individuals are replicating macromolecules in a chemical tank. A
constant flow is maintained through the tank, supplying the building
blocks and removing reaction products. The fitness of a molecule
depends on its monomer sequence and is the expected number of copies
made from it during its stay in the tank, divided by the time it
spends in the tank. (Division by the stay time is necessary because
some macromolecules may exist longer than others. With this
definition, the fitness is identical to the growth rate. Previously,
the discussion was based on a biological population in which all
genotypes have the same generation time.)  Point mutations occur at a
constant small probability $\mu$ per site and per replication.  This
model represents a mathematical confirmation that even inanimate
replicating molecules are capable of adaptation, as is also
demonstrated by in-vitro experiments, the oldest one of which is
described in
\cite{spie71}. 

In the following, we give an overview of the quasispecies model in
various fitness landscapes. We consider only haploid models
with asexual reproduction, as is appropriate for macromolecules. The
diploid case (where each individual has two sets of chromosomes, one
inherited from the father, and one from the mother if reproduction is
sexual) is for instance discussed in \cite{higgs94,wie95}; it
shows an even richer scenario than the haploid case. A review over the
quasispecies model is given by Eigen, McCaskill, and Schuster
\cite{eigenreview}, and more recent literature is discussed by Baake
and Gabriel \cite{baake}.

We represent each molecule again by a sequence of
$\pm 1$ spins, and we denote the fraction of molecules with sequence
$\sigma$ at time $t$ by $n_t(\sigma)$. We then obtain the quasispecies
equation for the temporal evolution of these fractions: 
\begin{equation}
n_{t+1}(\sigma) = \frac{\sum_{\sigma'}r_{\sigma' \sigma}W(
\sigma') n_t(\sigma')}{\bar W_t}\, ,
\label{quasispecies}
\end{equation}
where the transition probabilities are given by $r_{\sigma' \sigma} =
\mu^{d_H} (1-\mu)^{N-d_H}$, including the case $\sigma=\sigma'$.
They are normalized according to $\sum_\sigma r_{\sigma' \sigma}=1$. 
$\bar W_t = \sum_\sigma n_t(\sigma) W(\sigma)$ is the mean
reproduction rate.  Often the continuous-time version of this equation
is used. In the discrete version given here, the time interval must be
chosen shorter than the time between the production of a macromolecule
and of its first daughter molecule.  In a modified quasispecies model
mutation rates are defined per unit of time instead of per
generation. This model was mapped on a quantum spin model and solved
for various cases in \cite{baake97,wagner}.

For $\mu=0$, the species with the highest reproduction rate wins, and
Fisher's fundamental theorem is recovered.  For non-vanishing mutation
rates, the effects of combined selection and mutation can be very
different under different conditions.  The most interesting feature of
the quasispecies model is the existence of an {\it error threshold} in
many fitness landscapes.  For sufficiently small mutation rates, one
can expect the population to be centered around a peak (where it is
called a {\it quasispecies}), while it is spread over genome space if
the mutation rate exceeds a critical value.

\subsection{Quasispecies in a multiplicative Fujiyama landscape}

Let us first consider a multiplicative Fujiyama
landscape and show that no error threshold exists in this case. We
assume a fitness function $W(\sigma) = \exp(A+\sum_i J_i \sigma_i)$,
and we search for a stationary distribution $n_{t+1}(\sigma) =
n_t(\sigma) \equiv n^*(\sigma)$. From Eq.~(\ref{quasispecies}) we can
see that the dynamics of the $n(\sigma)$ do not change if all
fitness values are multiplied by the same factor. Without loss of
generality we assume therefore that $A$ is such that $\bar W^*=1$, and
the stationarity condition becomes
$$n^*(\sigma) =
\sum_{\sigma'}r_{\sigma'\sigma}W(\sigma')n^*(\sigma').$$
Now, $r$ and
$W$ both factorize into their contributions from the single spin
components, and therefore $n^*(\sigma)$ factorizes into a
product of the single spin frequencies $n^*(\sigma_i)$. The stationary
distribution of the value of spin $i$ does not depend on the value for
spin $j$ and is given by
$$n^*(\sigma_i)=\sum_{\sigma_i'=\pm 1}
r_{\sigma_i'\sigma_i} e^{A_i+J_i\sigma_i'}n^*(\sigma_i').$$
This is a set of two equations for the two quantities $A_i$ and $n^*(\sigma_i=1)$, and it has a unique solution, 
$$e^{-A_i}=(1-\mu)\cosh{J_i}+\sqrt{(1-\mu)^2\cosh^2{J_i}-(1-2\mu)}$$ 
and 
$$n^*(\sigma_i=1)=\frac{e^{-A_i}-e^{-J_i}}{e^{J_i}- e^{-J_i}}.
$$
This result can be found for instance in \cite{woodcock}.
Often, the infinite genome limit $N\to \infty$ with fixed mean
number of mutations per molecule $\mu N$ is considered. Because $\mu$
is small in this case, the results simplify to
$e^{-A_i}\simeq(1-\mu)e^{J_i}$ and 
\begin{equation}
n^*(\sigma_i=1) \simeq 1-\mu e^{J_i}/(e^{J_i}-e^{-J_i}).
\label{fujiyama}
\end{equation}
 (Without loss of generalization, we have assumed that $J_i>0$.) Thus,
the probability that a spin is flipped out of its favourable position
is proportional to $\mu$, and the mean number of mutations away from
the optimal sequence is proportional to $\mu N$. Most molecules do not
sit at the top of the peak, but at a distance of the order $\mu N$
away from it. This is the place where the uphill force due to
selection, and the downhill force due to mutation balance each
other. At this distance from the peak the fitness of a molecule
relative to the maximum fitness $W_0$ is given for small $\mu$ by
$W(\sigma) = W_0 e^{-\mu N}$.  With increasing mutation rate, the
population slides further down the slope.  If the peak has only a
finite height over a surrounding flat landscape, one can expect that
the population reaches the bottom of the mountain and moves away from
it when $\mu N$ becomes of the order $\ln(W_0/W_1)$, where $W_1$ is the
average fitness in the ``plain'' surrounding the mountain. This is
demonstrated explicitly in \cite{wiehe} for a model in which the
fitness depends only on the distance to the wild type.

\subsection{The sharp peak landscape}

This escape from a peak can be demonstrated most easily by choosing a
fitness landscape where one genotype, the wild type $\sigma_0$, has a
fitness $W_0$, while all mutants have a lower fitness $W_1$
\cite{eigenreview}. For the frequency of the wild type we obtain then
the stationarity condition
$$n^*(\sigma_0) = \frac{r_{\sigma_0\sigma_0} W_0
    n^*(\sigma_0)+\sum_{\sigma'\neq \sigma_0}r_{\sigma'\sigma_0}W_1
    n^*(\sigma')}{W_0n^*(\sigma_0)+W_1(1-n^*(\sigma_0))}.
  $$
  When the length $N$ of the molecules is large enough and
  $n^*(\sigma_0)$ is finite, the second term in the numerator can be
  neglected, because back mutations to the wild type are extremely
  unlikely. In order to get a meaningful limit $N\to \infty$, the
  mutation rate per site must scale as $1/N$, such that the product
  $\mu N$ remains fixed. The stationary frequency of the wild type can
  then easily be calculated and is
  $$n^*(\sigma_0)=\frac{e^{-\mu N} -W_1/W_0}{1-W_1/W_0}.$$
  A finite
  fraction of all molecules have the wild-type sequence, and the rest
  of the population has one or a few mutations, with the mean number
  of mutations being of the order $\mu N$. The population is localized
  around the wild-type sequence, and is called a {\it quasispecies}.
  This quasispecies solution exists only if $(1-\mu)^N \simeq
  e^{-\mu N} > W_1/W_0$. The critical mutation rate $(\mu N)^*$, for
  which the inequality becomes an equality, is the {\it error
    threshold}.  For higher mutation rates, our starting assumption
  that $n^*(\sigma_0)$ is nonzero, breaks down, and the population
  becomes delocalized and wanders through sequence space. The
  quasispecies equation is no longer valid in this case, because a
  given sequence occurs with only a small probability, and
  fluctuations in the composition of the population are high. This
  situation of neutral evolution is discussed in a later section.

  A modified version, which contains not only a narrow high peak, but
  also a broader, slightly lower peak, has been studied in
  \cite{swe88}. The authors find that the population is localized at
  the high peak for small mutation rates, and that it switches to the
  lower peak at higher mutation rates, because the mean fitness of the
  population can be increased this way. For even higher mutation
  rates, the population becomes completely delocalized. A variety of
  further results for the quasispecies model in the sharp-peak
  landscape are given in \cite{eigenreview}, and a recent study was
  performed by Galluccio et al~\cite{gal96,gal97} by mapping the
  system on a directed polymer in a random medium and applying the
  transfer matrix method.
  
\subsection{Truncated landscapes}

  The error threshold or ``delocalization'' of a population from a
  peak has been compared to a phase transition in equilibrium
  statistical physics. Mutation plays a role similar to temperature by
  pulling the configurations away from the energetically preferred
  ones. A phase transition is well defined only in the thermodynamic
  limit, which corresponds to the limit $N \to \infty$ for the
  quasispecies model.  Just as for equilibrium statistical physics,
  not all types of fitness landscapes lead to a phase transition. We
  have already discussed the example of the multiplicative Fujiyama
  landscape. Another example is a ``truncation landscape''
  \cite{wiehe}, where all individuals with more than $k$ mutations
  away from the wild type have zero fitness.  Clearly, the population
  is confined in genotype space to a volume element of radius $k$
  around the wild type, and no delocalization can occur. 
  
\subsection{Uncorrelated random landscapes}

  There are also examples where no localized state occurs. For
  instance, in a random landscape where the fitness of each molecule
  is a random number chosen from a uniform distribution in the
  interval $[W_{min},W_{max}]$, a non-vanishing fraction of sequences
  has a fitness within a relative distance of the order $1-e^{-\mu N}$
  of $W_{max}$. Since these sequences are spread over the entire
  sequence space, no localized state exist.  (Furthermore,
  since the solution changes smoothly as function of $\mu$, there
  is no phase transition between non-localized states.)
  
  This situation changes for random fitness landscapes where the ratio
  between largest and smallest fitness increases with increasing $N$.
  Let us consider a random fitness landscape where the fitness has the
  form $$W(\sigma) = e^{k F(\sigma)},$$
  with $F$ being random
  variables from a Gaussian distribution 
  $$P(F(\sigma)=E) dE = e^{-E^2/N}/\sqrt{\pi N} dE.$$
  This situation
  was studied in depth in \cite{FPS}. In the following, we give a
  short derivation of the delocalization threshold, assuming with
  \cite{FPS} that $N$ is large.  The maximum value $E_0$ of $E$ in a
  sequence space with $2^N$ possible sequences is given by the
  condition 
  $$\int_{E_{0}}^\infty P(E)d E = 2^{-N},$$ and is $E_{0} = N
\sqrt{\ln 2}$ to leading order in $N$.  The population is localized
around this maximum if there exists a nonzero solution for the
stationary frequency of the maximum:
\begin{equation}
n_0^* = (1-\mu)^N n_0^* e^{kE_0} / \bar W(n_0^*).\label{randomenergy}
\end{equation}
The mean fitness in the localized state must be larger than in
the delocalized state, and approaches the latter as $n_0^*$ decreases
to zero.  In the delocalized state, the growth rate of the optimal
sequence if back mutations are neglected, $(1-\mu)^N e^{kE_0} /
\bar W$, is smaller than 1. The second term in the quasispecies
equation, which describes the generation of the optimal sequence due
to mutations, becomes now important. A typical sequence has a value of
$E$ of the order $\sqrt{N}$ (while $E_0 = N\sqrt{\ln 2}$), and its
growth rate in the absence of back mutations, $(1-\mu)^N e^{kE} /\bar
W$ is vanishingly small. The number of sequences with a given fitness
is therefore determined by the number of mutations to this fitness,
which is proportional to the number of configurations with this a
fitness. This leads to the following expression for the mean fitness
of the delocalized state,
$$\bar W(0) = \int_{-\infty}^\infty e^{kE-E^2/N}dE / \sqrt{\pi N} =
e^{k^2 N/4}. $$ The mean fitness in the localized state cannot be
smaller than this, and the condition that Eq.~(\ref{randomenergy}) has
a solution with nonzero $n_0^*$ breaks down at the error threshold
$$k=2(\sqrt{\ln 2} - \sqrt{\ln 2 + \ln(1-\mu)}).$$

\subsection{Spin-glass models}

This result for the delocalization transition in the uncorrelated
rugged fitness landscape mirrors the phase transition from a phase
with finite occupation of the ground state to a phase with zero
occupation in the random-energy model \cite{der81}. In general, it can
be expected that if the fitness landscape of a quasispecies model is
equivalent to the free energy landscape of a spin model, the phase
transitions in the two models are similar. Of course, there is no
one-to-one mapping between the two. Rather, if one maps a quasispecies
model onto an equilibrium spin model, one ends up with an anisotropic
Hamiltonian, with the time index in the quasispecies model
corresponding to the row index in the spin model \cite{leut86}, and
with the stationary state of the quasispecies model corresponding to
the surface configuration of the spin model \cite{tara92}. Using the
transfer matrix method, various analytical and numerical results can
be obtained. Thus, Tarazona \cite{tara92} studied among other models the quasispecies version  of the Hopfield model,
$$W(\sigma) = \exp\left(\frac{K}{2N^2} \sum_{i\neq j} \sum_{\nu=1}^p
\xi_i^\nu \xi_j^\nu \sigma_i \sigma_j\right),$$ where the number $p$
of randomly chosen ``master sequences'' $(\xi_1^\nu,...\xi_N^\nu)$ is
a small fraction of $N$ (of the order $0.02N$). From the properties of
the Hopfield model we know that the $p$ highest fitness maxima
coincide (almost) with the $p$ master sequences and their
complementary sequences (which have opposite spin signs). There are
secondary maxima and high-lying saddle points between them, which are
produced by the overlap of different sequences, forming a complex
network percolating through configuration space. For small mutation
rates, the quasispecies is centered around one of the master
sequences. As the mutation rate is increased, two phase transitions
occur, just as they exist in the Hopfield model.  The first one leads
the population from the localized state to a ``spin-glass'' state that
percolates through sequence space along the secondary peaks and ridges
generated by the superposition of master sequences. Only at a higher
value of $\mu N$ does the system make a second transition to a fully
disordered mixture of sequences. ``Spin-glass'' states for the
quasispecies model are also expected when the fitness landscape is
equivalent to the free-energy landscape of other spin-glass models.

\subsection{Discussion}

Finding that many fitness landscapes show an error threshold beyond
which the population is no longer in the neighborhood of a fitness
maximum, raises the question whether such error thresholds also exist
in real biological systems, and whether they play a role in
determining mutation rates. Cells possess sophisticated proof-reading
and error-correcting mechanisms in order to reduce the frequency of
mutations due to copying errors and chemical or radiation damage.
Since maintaining these mechanisms costs energy, it can be expected
that these mutations are kept at a level which is just low enough that
it does not lead to the loss of adaptation. In general, mutation rates
seem to never be much higher than 1 mutation per coding parts of the
genome per generation \cite{drake} and often far less, so that a
considerable fraction of the offspring always has unchanged genes.
Experiments with artificially increased mutation rates show that a
threefold increase is lethal for certain RNA viruses, while heavily
mutagenized Drosophila populations can survive in the laboratory,
although they are weak competitors with nonmutagenized strains. The
literature on these and other experiments, and a general discussion of
spontaneous mutation rates can be found in \cite{drake}.

How the error threshold might have been avoided in
prebiotic evolution (which the inventors of the quasispecies model had
in mind) is an object of discussion. Up to now, the process that led
to the formation of the first cells is unknown.

Finally, let us mention that there exist different definitions of the
error threshold in the literature. In this section, we have adopted
the understanding of a phase transition that occurs strictly speaking
only in the infinite genome limit $N\to \infty$. However, one can also
find the understanding that the error threshold is the critical
mutation rate above which the genotype with highest fitness is lost,
assuming that $N$ is finite. For the sharp-peak landscape, the two
concepts are equivalent. However, for the multiplicative
Fujiyama landscape there is no delocalization phase transition,
although there exists a mutation rate beyond which the tip of the peak
is occupied only due to rare back mutations. This loss of the fittest
genotype will be discussed more thoroughly in the next section, where
finite populations are considered. The finite population size brings
new aspects to the problem, since accidental fluctuations can
contribute to the loss of fit genotypes.

\section{Finite populations}
\label{finite}

Some of the results of the previous section will be fundamentally
modified when the finite population size is taken into account, due to
random fluctuations, which cause {\it genetic drift} in the
absence of selection, and  {\it stochastic escape} from fitness
peaks. For instance, if the fitness advantage $(W_0-W_1)/W_1$ in the
sharp-peak landscape or in the uncorrelated rugged landscape is only
of the order of the inverse population size, the population can escape
from it easily through random fluctuations. In fact, a fitness
advantage that is too small will not affect the population at all,
because selection effects will be drowned out by random fluctuations. Our
results for the random landscape with fitness values evenly
distributed over a finite interval will also change for finite
populations: The result that all sufficiently high peaks are populated
was based on the assumption that enough individuals exist to
occupy each one of them with a sufficiently large number so that
fluctuations can be neglected. It was furthermore based on the
assumption that all high peaks can be discovered because eventually a
mutation will lead individuals to each one of them. For a finite
population, however, only one or a few peaks can be occupied
simultaneously, and the population will move in a random fashion
between peaks. In a similar way, in a ``spin-glass'' landscape the
population will not be spread in a stationary distribution all over
the ridges of the landscape, but it will wander as a cloud of finite
extension along these ridges from one peak to the next. In a
completely flat landscape, the population will wander as a cloud of
individuals through the entire configuration space.

In the following, we will study the effects of finite population sizes
in a quantitative manner.  In order to make quantitative statements
about the effects of fluctuations, one has to build a model specifying
how the genetic composition of a population results from that of the
previous generation. A widely used model is the Wright-Fisher sampling
method: The next generation of $M$ new individuals is determined by
drawing an individual $M$ times from the parent generation, with a
probability proportional to its fitness. After each drawing, the
parent individual is put back to the pool of parents such that it can
be drawn again, and a copy of it is added to the new generation;
this copy may also be mutated with a probability determined by the
mutation rate. For sexual, diploid organisms, $M$ is the number of
gametes, which is twice the population size. The next generation of
gametes can be obtained by a drawing procedure similar to the one just
described, provided that mating is random, that males and females are
equal in number, and that the fitness of a diploid genotype is the
product of the fitness values of the two haplotypes it contains.  In more
complicated situations, many of the results obtained under the ideal
assumptions of fixed population size, random mating, Wright-Fisher
sampling, etc, can be generalized if one introduces effective
population sizes.

\subsection{Drift, fixation, and the diffusion equation method} 

One consequence of finite population sizes and fluctuations in the
composition of a population is that genes get lost from the gene pool.
If there is no new genetic input through mutation or migration, the
genetic variability within a population decreases with time. After
sufficiently many generations, all individuals will carry the same
allele of a given gene. This allele is said to have become fixed. In
the absence of selection, the probability that a given allele will
become fixed is proportional to the number of copies in the initial
population. Thus, if a new mutant arises that has no selective
advantage or disadvantage, this mutant will spread through the entire
population with a probability $1/M$, $M$ being the population size.
If the individuals of the population are diploid, each carries two
sets of genes, and $M$ must be taken as the number of sets of genes,
i.e., as twice the population size.  Further above, we have seen that
the probability that a mutant that conveys a small fitness increase by a
factor $1+s$ has as probability of the order $s$ to spread
through a population. In populations of sizes much smaller than $1/s$,
this selective advantage is not felt, because mutations that carry no
advantage become fixed at a similar rate.  In the same manner, a
mutation that decreases the fitness of its carrier by a factor $1-s$,
is not felt in a population much smaller than $1/s$.
An interesting consequence of these results is that the rate of
neutral (or effectively neutral) substitutions is independent of the
population size. The reason is that the probability that a new mutant
is generated in the population is proportional to $M$, while its
probability of becoming fixed is $1/M$.

Due to drift and fixation, two individuals of a population where
selective forces can be neglected have a common ancestor $M$
generations back on an average (because both choose the same parent
with probability $1/M$ in Wright-Fisher sampling). The mean number of
mutations separating two individuals is therefore $2M\mu$, where $\mu$
is the mutation rate. If this number is small, the ``cloud'' of
individuals has a small extension in genotype space, while it becomes
more spread for larger $\mu$. Calculations of various other properties
of a drifting cloud, like overlaps and correlations, can be found in
\cite{der91,higgs92}. A recent calculation of properties of ``clouds''
 of sexual individuals is given in \cite{der99}.

General mathematical expressions for fixation probabilities, fixation
times, and other quantities related to changes in gene frequencies can
be obtained using diffusion approximations, which are good
approximations if the changes in gene frequencies from one generation
to the next are small. (See, e.g., \cite{crow}).  Let us consider a
genetic locus with two alleles, $A_1$ and $A_2$, and let $\phi(p,x;t)$
be the probability density that the frequency of $A_1$ is $x$ at time
$t$, given that it was $p$ at time zero. If $B(x)$ denotes the mean
change in allele frequency $x$ per generation (which is our time
unit), and $V(x)$ its variance, $\phi(p,x;t)$ changes according to the
Fokker-Planck equation
\begin{equation}
\frac{\partial \phi(p,x;t)}{\partial t} = \frac{1}{2}\frac{\partial^2}{\partial x^2}\left[V(x)\phi\right]-\frac{\partial}{\partial x}\left[B(x)\phi\right]\, .
\label{FP}
\end{equation}
For Wright-Fisher sampling and for a fitness ratio $(1+s)$ between
alleles $A_1$ and $A_2$, we have $B(x)=x(1-x)s$ and $V(x)=x(1-x)/M$,
with $M$ being the population size.

Crow and Kimura \cite{crow} gave a quick derivation of Eq.~(\ref{FP}):
Introducing a transition probability $g(x-\xi,\xi;t,\delta t)$ from allele frequency $(x-\xi)$ to $x$ during a short time interval $(t,t+\delta t)$, one finds
\begin{eqnarray*}
\phi(p,x;t+\delta t) &=& \int\phi(p,x-\xi;t)g(x-\xi,\xi;t,\delta t)d\xi\\
&\simeq&\phi(p,x;t)\int g(x,\xi;t,\delta t)d\xi-\frac{\partial}{\partial x}\phi(p,x;t)
\int\xi g(x,\xi;t,\delta t) d\xi\\
&+& \frac{1}{2} \frac{\partial^2}{\partial x^2}\phi(p,x;t)
\int\xi^2 g(x,\xi;t,\delta t) d\xi.
\end{eqnarray*}
Using $\int g d\xi=1$, $\int \xi g d\xi= B(x) \delta t$ and $\int
\xi^2 g d\xi= V(x) \delta t$, and expanding $\phi(p,x;t+\delta t)$ to
first order in $\delta t$ one obtains Eq.~(\ref{FP}).

In order to calculate the probability of gene fixation, a diffusion
equation where $p$ is variable and $x$ is fixed is useful:
\begin{equation}
\frac{\partial \phi(p,x;t)}{\partial t} =
\frac{V(p)}{2}\frac{\partial^2 \phi}{\partial p^2}+B(p)\frac{\partial
\phi}{\partial p}\, .\label{KBE}
\end{equation}
This equation can be derived under the condition that the transition
probabilities $g$ do not depend on the time $t$, leading to 
\begin{eqnarray*}
\phi(p,x;t+\delta t) &=& \int\phi(p+\xi,x;t)g(p,\xi;\delta t)d\xi\\
&\simeq&\int g(p,\xi;\delta t)\left[\phi(p,x;t)+\xi\frac{\partial \phi}{\partial p} + \frac{\xi^2}{2}\frac{\partial^2\phi}{\partial p^2}\right]d\xi
\end{eqnarray*}
Introducing again $B$ and $V$ and expanding in $\delta t$, one obtains
Eq.~(\ref{KBE}).

The probability of gene fixation is $u(p,t)=\phi(p,1;t)$, and can be
obtained using the boundary conditions $u(0,t)=0$ and $u(1,t)=1$
(assuming that there are no mutations). In a situation without
selection, and with random drift due to Wright-Fisher sampling, i.e.,
$B(p)=0$ and $V(p)=p(1-p)/M$, the solution is
$$u(p,t)=p+\sum_{i=1}^\infty (2i+1)p(1-p)(-1)^i
F(1-i,i+2,2,p)e^{-i(i+1)t/2M}$$ with $F$ being the hypergeometric
function,
$$F(1-i,i+2,2,p)=\sum_{n=0}^{i-1}p^n(-1)^n\frac{(i+n+1)!}{(i-n-1)!i(i+1)n!(n+1)!}.$$
The ultimate probability of fixation is obtained by letting $t\to
\infty$ in Eq.~(\ref{KBE}), resulting in $\partial u/\partial t = 0$
and
$$u(p,\infty)=\frac{\int_0^pe^{-2\int_0^qB(x)/V(x)dx}dq}{\int_0^1e^{-2\int_0^qB(x)/V(x)dx}dq},$$
which becomes for $V(x)=x(1-x)/M$ and $B(x)=x(1-x)s$
$$u(p,\infty)=\frac{1-e^{-2Msp}}{1-e^{-2Ms}}.$$ This reduces to $2Msp$
if $Ms$ is large. If there is initially one individual with the
advantageous mutation, this mutation becomes fixed with a probability
$2s$. The result equation (\ref{fix}) obtained earlier is different
from this because it was obtained for a situation where $V(x)$ and
$B(x)$ are different, namely $V(x)=2x(1-x)/M^2$ and $B(x)=sx(1-x)/M$.

The average number of generations until fixation is 
\begin{eqnarray*}
\bar t(p)&=&\frac{\int_0^\infty t (\partial u(p,t)/\partial t)dt}{\int_0^\infty  (\partial u(p,t)/\partial t)}\\ 
& =& \int_0^\infty \left[1-u(p,t)/u(p,\infty) \right]dt\\
&=& \int_p^1\frac{2u(x,\infty)(1-u(x,\infty))}{V(x)\partial u(x,\infty)/\partial x}dx + \frac{1-u(p,\infty)}{u(p,\infty)}\int_0^p \frac{2u^2(x,\infty)}{V(x)\partial u(x,\infty)/\partial x}dx\, .
\end{eqnarray*}
In order to go from the second line to the third line, one has to
multiply both sides with $u(p,\infty)$ and then to apply the operator
$V(p)\partial^2/\partial p^2 + B(p) \partial/\partial p$ on both
sides. Using Eq.~(\ref{KBE}) one can then eliminate the time variable
from the right-hand side. Solving the resulting differential equation
for $\bar t(p)$ leads to the third line.

 For random drift with no selection, we have $u(p,\infty)\simeq p$ and obtain
 $$\bar t(p)=2M(1-1/p)\ln(1-p) \simeq 2M \hbox{ for } p=1/M\, .$$
 For
 weak selection and sufficiently large population size $M$, we have
 $u(p,\infty) \simeq (1-e^{-2Msp})$, and the main contribution to
 $\bar t(1/M)$ comes from the upper limit of the first integral,
 $$\bar t(p=1/M) \simeq (1/s)\ln M.$$
 (The integral does not diverge
 if one remembers that fixation is reached when $p$ differs from 1 by
 less than $1/M$.)

Recently, the diffusion equation method was applied to a model where
the fitness is a Gaussian function of a vector of character values,
and where mutations occur with a probability which is Gaussian in the
distance in character space spanned by the mutation. The diffusion
equation could be mapped on the mean-field theory of Bose
condensation, leading through an analytical calculation to a
stationary state where a non-vanishing fraction of the population has
an identical genome
\cite{cop99}.

\subsection{Muller's ratchet}

Muller \cite{muller} first pointed out that small asexual
populations might be at a high risk of accumulating deleterious
mutations. Assuming that back mutations are rare, the sequence with
highest fitness can get lost through random drift, and cannot be
restored.  The population thus glides down the fitness peak on which
it was initially placed in a ratchet-like manner.  This process of
iterated loss of the fittest sequence is known as ``Muller's
Ratchet''.

The model usually employed to illustrate this process is a
multiplicative Fujiyama landscape, where all genotypes with $n$
mutations away from the peak have the same fitness $W_n=(1-s)^n$
\cite{fel74}, with $s$ being small. As we have seen above, such a
fitness landscape has a stationary population distribution in the
infinite population size limit, where each site has the same
probability of having a mutation. If we also assume that the genome
length $N$ is infinite, and that the mutation rate $U=\mu N$ per
genome is fixed, we obtain from Eq.~(\ref{fujiyama}) the probability
$\mu/s$ for having a mutation at a given site, leading to the
following distribution of individuals with $k$ mutations in the
stationary state:
$$P(k) = \frac{(U/s)^k}{k!}e^{-U/s}.$$
This result can be found at
various places in the literature. The sequence with the highest
occupation is the sequence with zero mutations, which sits at the top
of the peak. This situation changes drastically when the population
size is finite. If we preserve the infinite genome limit and start
with a population that sits at or near the peak, all mutations that
occur are downhill mutations, which decrease the fitness. Because the
best sequences keep getting lost through fluctuations, the population
moves downhill with a constant speed $R$ per generation, and with a
constant width $\Delta k$ \cite{HW}. The reason why $\Delta k$ does
not become larger with time is that the common ancestor of all
individuals is only of the order of $M$ generations back, and
consequently all individuals have the same mutations that this
ancestor had, plus a limited number of additional new ones. In spite
of heavy efforts, no simple analytical expression could be derived
that gives a good approximation to $R$ as function of $s$, $U$, and
$M$. A careful study that also contains some analytical work and
important references to previous literature was done by Higgs and
Woodcock \cite{HW}. One of their findings is that the rate of decrease
in fitness, $R \ln (1-s)$, is largest for intermediate selection
strength.  For strong selection, mutations are far less likely to
become fixed, while for weak selection a fixation leads only to a
small fitness loss.

The ratchet can be halted or prevented by a variety of mechanisms not
taken into account in the simple model that we just described.  First,
its rate might be so slow that its effect does not manifest itself
during the lifetime of a population. With increased population size
and selection strength, and decreased mutation rate, the ratchet
becomes slower. Second, if the genome length $N$ is finite, the
ratchet can be brought to a halt even on the Fujiyama landscape,
because the rate of back mutations increases with increasing $k$. Very
generally, one can show on the Fujiyama landscape that if there is a
small fraction $p$ of favourable mutations, they will accumulate if
$p$ is larger than a threshold value $p^*$, while disadvantageous
mutations will accumulate for $p<p^*$ \cite{woodcock}.  Third,
realistic fitness landscapes do not consist of a single infinitely
high peak. For this reason, not all mutations are deleterious, but a
certain fraction is neutral or even advantageous. One can expect that,
even in situations where a population initially moves downhill, it
will very soon explore more of the landscape and find other peaks or
high ridges. We will discuss the case of neutral mutations further
below, which suggests that realistic fitness landscapes have many
neutral ridges with high fitness.  Fourth, the phenotypic effects of
many mutations can be compensated by other mutations that reverse this
effect. This holds in particular for quantitative traits that are
determined by the effects of many genes.  A quantitative genetic model
introduced by Wagner and Gabriel \cite{WG} is an example where
compensatory mutations can set in at a short distance from the peak.
A fifth way of escaping the ratchet is via recombination. This process
combines parts of genomes of different individuals and occurs not only
in sexual species, but also in bacteria. The process of recombination
can combine two parts of genomes that have no deleterious mutations,
thus restoring genomes of higher fitness.

In conclusion, there are many reasons why reasonably sized populations
should not accumulate deleterious mutations. The question remains
whether Muller's ratchet becomes relevant in certain situations.  It
is often assumed that the ratchet might be involved in the degradation
of the Y chromosome (which cannot undergo recombination but may have
been located relatively high in the fitness landscape immediately
after its creation), and in the fate of small asexual populations.
(For a discussion, see \cite{gordo}.)  If a small population
accumulates deleterious mutations, the ratchet is even accelerated by
a process called ``mutational meltdown'': A population that continues
to decrease in fitness, will eventually reach a stage where its
population size is no longer regulated by the carrying capacity of the
environment, but by the low viability of its offspring, leading to a
rapid decrease in population size.  Reference to the literature on
this topic can be found in the review by Baake and Gabriel
\cite{baake}.

Let us finally mention that there are other possible explanations for
the extinction of small populations, besides Muller's ratchet.  For
sexual populations, recessive deleterious alleles are expressed more
often and lead to a higher mortality because of inbreeding in small
populations. For any small population, the small genetic variety may
hinder adaptation to new challenges. And last, the survival of an
individual might depend on a sufficiently high density of other
individuals (e.g.  for finding suitable partners, for hunting
together, etc.). Indeed, All\'ee already emphasized in the 1930s that
under-crowding can have negative effects on animals \cite{allee}.

\subsection{Stochastic escape}

In the previous paragraphs, we have considered the situation of a
finite population gliding down a Fujiyama peak. For Fujiyama peaks of
finite height, and also for the sharp-peak landscape, one can expect
that a finite population has a non-vanishing probability of escaping
the peak altogether, even for parameter values for which it would
remain centered around the peak in the infinite population size limit.
Computer simulations of a finite population in a sharp-peak landscape
can be found in \cite{now89,wie95}.  Reference \cite{wie95} also
contains an analytical calculation based on a diffusion approximation
for the occupation of the wild type which is valid in the limit
$W_0/W_1\simeq 1$, and which is based on the assumption that all
mutant sequences have the same probability of being occupied.  It is
found that the delocalization transition found for the infinite
population limit is still pretty sharp, but moves to smaller mutation
rates by an amount roughly proportional to $1/\sqrt{M}$ compared to an
infinite population. Near the transition, large fluctuations are seen.
For mutation rates below the threshold, the occupation of the wild
type sequence has the same value as for the infinite population, and
spontaneous escape is rare.  Analytical calculations for this model
can also be found in \cite{AF98}, however, these authors neglect
fluctuations.

The escape from the optimal sequence in a Sherrington-Kirkpatrick
spin-glass landscape was studied via computer simulations in
\cite{bon93}. Just as for the sharp-peak landscape, the main effect of
the finite population size seems to be a shift of the error threshold
by an amount of the order $1/\sqrt{M}$. For mutation rates higher than
the error threshold but not too high, the population wanders like a
cloud through genotype space, thereby staying on ridges of high
fitness.

In the two examples for stochastic escape given so far, the escape is
due to large fluctuations in the number of individuals with the best
genotype, leading to an escape within a reasonably short time. A
qualitatively different situation is considered by Peliti
\cite{pel98}, who calculates the probability that a (because of strong
selection) genetically homogeneous population escapes from a high
fitness peak in one step and ends up with a lower maximum fitness.  He
considers a random fitness landscape where each genotype has an
independent fitness value. The probability that the population escapes
from the state of high fitness is given by the product of the
probability that all offspring receive a mutation, and of the
probability that none of them mutates to a higher fitness (which is
supposed to occur with probability $h$ per mutation), $e^{M(\ln U
-h)}$. Zhang \cite{zhang97} considers the evolution of a finite
population in a random fitness landscape with Gaussian distributed
fitness values. He estimates the time it takes the population to reach
another, higher peak by combining the waiting time for a mutation that
brings an individual to this peak with the time it takes the mutated
individual to grow to a population size comparable to that on the
first peak. However, his estimates are based on the assumption that
the occupation of both peaks grows exponentially in time, without
taking into account saturation due to the fixed total population size.

Nimwegen and Crutchfield
\cite{nim2000} consider a population centered
at a sharp peak of fitness $W_0$ in a surrounding landscape of fitness
1 and perform a detailed analytical and numerical analysis of the
probability that it finds a state of higher fitness which is $w$
mutations away from it. They suggest an approximate formula for the
mean waiting time of the form
$$\langle t \rangle \simeq (\ln W_0 / \mu)^{w-1}/w!M\mu\, $$ implying
that the width $w$ of the barrier affects the waiting time much
stronger than the barrier depth $(W_0-1)$. If the mutation rate is
close to the error threshold, this waiting time has to be replaced by
the time that a population diffusing through genotype space needs to
find the second peak. A rough estimate for this time is $\tau \simeq
2^N/MN\mu$.

\section{Neutral evolution and RNA landscapes}

The fitness landscapes mentioned so far were obtained by assigning a
fitness directly to the genotype, and they tend to have local maxima
corresponding to particularly fit genotypes. In general, however, one
can expect that the same phenotype can be realized by a variety of
different genotypes, suggesting that there might be many mutations
that are neutral or nearly neutral with respect to the fitness. The
fitness optima should in this case not be viewed as peaks, but rather
as plateaus or mountain ridges. In this section, we will therefore
consider models that define the fitness as function of a phenotype
which can be realized by different genotype configurations.

\subsection{RNA landscapes}

The only genes for which a calculation of the phenotype from the
genotype has been possible so far, are those genes that code for RNA,
like tRNA or ribosomal RNA. These RNAs are single-stranded molecules
that partly fold back on themselves, thereby forming ``stems'' of
paired bases and ``loops'' of unpaired bases. This secondary structure
determines much of the binding energy of the molecule, and of its
catalytic function. An often cited example is the clover-leaf
structure of tRNA, which has remained virtually unchanged throughout
evolutionary history, even though there is considerable variation
between the primary sequences of different species. For a given
sequence, the secondary structure with minimum free energy can be
successfully calculated using computer algorithms.  It is found that
some secondary structures are realized by many more
sequences than others. Two sequences are called connected if they
differ by only one or two point mutations. A neutral network is then a
set of sequences with identical structure so that each sequence is
connected at least to one other sequence. It is found from computer
simulations \cite{fon93,sel94} that for frequent structures these
networks percolate through sequence space. Due to the high
dimensionality of sequence space, the networks penetrate each other so
that each frequent structure is almost always realized within a small
distance of any random sequence.

A fitness landscape is obtained by assigning a fitness to the
different secondary structures. This is done in \cite{hsf96,fs98} by
assigning to each structure a fitness which is a function of the
distance in shape space to a predefined target structure. A computer
simulation of the evolution of a population of finite size shows that
even for very small mutation rates the population moves in a
diffusion-like manner along the neutral network that corresponds to
the target structure and splits into several well defined
subpopulations. When the mutation rate exceeds a critical value, the
{\it phenotypic error threshold}, the target structure is lost from
the population. If the simulation is started with a homogeneous
population consisting of a single random sequence, evolution proceeds
in epochs, during each of which the population is dominated by one
secondary structure (or by several structures of equal fitness) and
diffuses on the neutral network belonging to this structure, until a
mutation occurs in an individual that allows the population to conquer
a structure of higher fitness.

A neutral network for model proteins is numerically explored in
\cite{bas2000}. It is found that the network percolates through 
sequence space, suggesting that proteins also may undergo extended
neutral evolution.

\subsection{Holey landscapes}

A particularly simple example of a fitness landscape with many neutral
directions is a ``holey'' landscapes where all fitness values are
either 1 (viable) or 0 (not viable). Gavrilets and Gravner \cite{GG96}
numerically studied diploid sexual populations on such a landscape,
where each genotype is randomly assigned one of the two fitness
values.  As long as the fraction of viable genotypes is above the
percolation threshold, a percolating network of viable genotypes
exists that spans the entire genotype space. The main focus of
\cite{GG96} is the existence of different ``species'' on the viable
network (or, if the system is below the percolation threshold, on
viable clusters) that are reproductively isolated because matings
between them cannot lead to viable offspring.

The relationship between connectivity properties of a neutral network
and the populations that evolve on it is studied analytically and
numerically in \cite{nim99}. The main finding is that a population
does not move over a neutral network in an entirely random fashion but
tends to concentrate at highly connected parts of the network,
resulting in phenotypes that are relatively robust against mutations.

\subsection{A quantitative genetic model with nonlinear gene interaction}

Apart from RNA landscapes, hardly any models try to construct fitness
landscapes by going through an intermediate phenotype stage, although
this is particularly important when dealing with quantitative traits
that are affected by many genes. One such model was recently
introduced by Higgs and Taylor \cite{TH00}.  Genotypes are represented
by binary variables, and each of several traits is the additive effect
of a given number of loci. Each trait has an optimum value, and the
fitness is a product of Gaussian functions of the distances to the
optimum values. The authors show from computer simulations that
qualitatively different regions in parameter space exist. For certain
parameter values, neutral percolation through genotype space is
possible, while for another range of parameter values lost fitness
can be restored through compensatory mutations, allowing a population
to explore a large part of genotype space. For a third parameter
range the configurations of high fitness are clustered together in
genotype space.

\subsection{The neutral theory}

The neutral theory was first suggested by Kimura in 1968 \cite{kim68}
as a possible explanation of the surprisingly large genetic variation
within and between species. This theory holds that the great majority
of evolutionary mutant substitutions are not caused by positive
Darwinian selection, but by random fixation of selectively neutral or
nearly neutral mutants. Much of the intra-specific variability at the
molecular level, such as is manifested in the protein polymorphism, is
believed to be selectively neutral or nearly so, and maintained in the
species by a balance between mutational input and random extinction or
fixation of alleles. This theory is supported by the finding that in
general the molecular changes that are less likely to be subject to
natural selection occur more rapidly in evolution. Thus, nucleotide
changes that cause no amino acid changes (called synonymous or silent
substitutions), and nucleotide changes in the non-coding regions occur
at much higher rates in evolution than those which lead to amino acid
changes. Furthermore, it was found that for each protein, the rate of
evolution in terms of amino acid substitutions is approximately
constant per year and per site for various lines along phylogenic trees,
as long as the function and tertiary structure of the molecule remain
essentially unaltered. This phenomenon is called the molecular clock
and is in strong contrast to the great variability of evolutionary
rates on the phenotypic level, which are believed to be governed by
positive natural selection. A defense of the neutral theory is given
in the book by Kimura \cite{kimura}.

Within the neutral theory, fitness landscapes with many neutral
directions are more realistic. Of course, the neutral theory
also assumes that many mutations are deleterious and are quickly
removed by negative selection, and that a few mutations are clearly
advantageous and positively selected, just as in
the RNA landscape discussed above.

Gillespie \cite{gill} presents some arguments against the
applicability of the neutral theory to amino-acid substitutions in
proteins. He lists various examples of naturally occurring variation
in proteins that has functional significance and allows for
micro-adaptation of enzymes to the body temperature or to the oxygen
concentrations in different environments. He also argues that, while
the mean amino acid substitution rates are constant, their variations
are much larger than predicted by the neutral theory and hint at
episodic bursts of substitutions, with periods of quiescence in
between. (However, the above-mentioned computer simulations by
Bastolla et al \cite{bas2000} have shown that even neutral evolution
can be accompanied by large temporal fluctuations in substitution
rates.)  Furthermore, the fact that the silent substitution rate
depends on the generation time, while the amino acid substitution rate
does not, needs an explanation.  The picture favored by Gillespie is
one of selection in a temporally fluctuating environment. According to
him, the plateaus in the neutral landscape have humps and bumps, which
fluctuate in time.

\section{Extensions}

There are innumerable possibilities to extend and generalize the above
models for evolution in fitness landscapes in order to take into
account features of the biological reality neglected so far. Some of
these generalizations are mentioned in the following.

\subsection{Demographic structure of the population, and large-scale evolutionary trends}

The age structure of a population, and the fertility and death rate as
function of the age of an individual, were not considered in the
studies reviewed so far. They are, however, of interest when a
mutation affects features relevant for the demographic structure, like
the mean age of reproduction, or the age-dependent mortality. In this
case, the relative fitness of different genotypes has to be calculated
by performing statistical averages over the age structure of the
population, and the outcome of the calculation may depend on the state
of the population. A formalism for calculating stationary
demographic structures as well as demographic changes due to mutations
was introduced by Demetrius (for a review see \cite{dem97}).  This
formalism is similar in spirit to the formalism of thermodynamics; its
most interesting achievement is to explain the
large-scale trend to increased body size in evolutionary lineages:
Scaling arguments and energetic considerations suggest that the
generation time increases with the power 1/4 of the body size, while
the total number of offspring of an individual is proportional to its
body size. Starting from these relations, one can show that
individuals with a larger body size will replace those with a smaller
body size under conditions where the population is stationary or
slowly growing. In a rapidly growing population smaller individuals
have an advantage.

\subsection{The effect of recombination}

So far, we have mainly discussed asexual species. One of the most
important features of sexual species is recombination, i.e., the
combination of different parts of different parental genomes into a
new genome. Since sex is so widespread, and since clonal lineages
generally have a shorter life time than sexual ones, it is widely
believed that there is an evolutionary advantage to sex. However,
where this advantage precisely lies, is difficult to pinpoint, the
favoured hypotheses being the capacity to create fitter
combinations in uncertain environments \cite{wil75,bel82}, and the
ability of escaping parasites by creating new combinations
\cite{ham90,pot91} (see section \ref{parasites}). Several recent
computer simulations of models with a changing environment show an
advantage of sexually reproducing
populations\cite{sta96,ber97,pek00,mar00}. Recombination can generate
good gene combinations as well as destroy them. Which of the two
effects is more important, depends on the properties of a
population and its environment.  (For a recent review, see
\cite{bar98}.)

We have already mentioned the positive effect of recombination at
halting Muller's ratchet.  The first computer simulations
demonstrating the advantage of recombination in a finite population
and multiplicative fitness landscape were done by Felsenstein
\cite{fel74}. He demonstrated that recombination makes the
fixation of favorable mutants more likely, and retards random fixation
of unfavorable mutants. Using analytical calculations and computer
simulations, Christiansen et al~\cite{chr98} showed that the time it
takes for the first advantageous double mutant to occur is generally
shorter in the presence of recombination. Kauffman
\cite{kauffman} studied the effect of recombination in NK landscapes
and found that recombination helps to find higher peaks only if $K$
is small enough such that peaks tend to be clustered.

Higgs pointed out a negative effect of recombination on neutral
evolution in RNA landscapes \cite{hig97}: The replacement of a base
pair requires two mutations, with the intermediate form having a
slightly lower fitness.  If recombination occurs within a gene coding
for RNA, it can combine genomes with the new and the old base pair,
generating the intermediate form. This reduces the mean fitness of the
population, and slows down the fixation of the new base pair.

Recombination is an important ingredient of genetic algorithms.
Genetic algorithms seek good solutions for complex problems by
evolving a population of possible solutions on a ``fitness landscape''
which is a measure of the quality of each solution.  The reason for
including recombination is the belief that good solutions are composed
of good building blocks, which can be easily combined by
recombination. (For an introduction to genetic algorithms, see, e.g.,
the books by Mitchell and Goldberg \cite{mit96,gol89}.) The ``royal
road function'' is a fitness function that is constructed explicitly
to consist of building blocks \cite{nim99a}. The fitness is the sum of
the fitness values of blocks of a given number of binary digits. The
fitness of each block is 1 if all digits are 1, and zero
otherwise. Computer simulations of the evolution of an initially
random population showed that recombination is useful only during the
early stages of evolution, until all finished blocks have been
combined that were present in the initial population. From then on,
the speed of evolution is determined by the time necessary to finish
further blocks.

A statistical-physics analysis of evolution in a fitness landscape
that corresponds to the free energy of a one-dimensional spin glass
was performed in \cite{sha94}. The authors showed analytically that
recombination reduces the higher moments in the fitness distribution
of the population, making the distribution broader and the effect of
selection stronger.

\subsection{Mutable mutation rates}

Mutation rates can be increased in a heritable manner through
loss-of-function mutations in polymerase (enzymes that weave DNA or
RNA strands) or in mismatch repair genes. It is estimated that under
normal growth conditions colonies of E. coli bacteria have a fraction
of $10^{-5}$ of such mutator bacteria \cite{Ninio}.  We have seen
above that for well adapted populations an increased mutation rate is
a disadvantage because it increases the distance of the population
from the fitness peak. However, in populations that are under adaptive
pressure individuals with increased mutation rates have an advantage
because they are more likely to produce an advantageous mutation.
Indeed, computer simulations by Taddei et al \cite{tad97} of a
population in an additive fitness landscape show that the frequency of
mutators increases as long as the population is away from the peak,
and decreases when the population is adapted. The mutator can become
fixed in sufficiently small populations. The model allows for mutation
towards the mutator state as well as for back mutations and uses
parameters that are realistic for E. Coli colonies.  Kessler and
Levine \cite{kes98} performed an analytical study of the speed of
adaptation in an additive fitness landscape. In their model, the
fitness is given by the number of 1 bits in a binary genome.
Transitions to the mutator state and back occur with rates
$\sigma_f$ and $\sigma_b$, while the mutation rate of the mutator is
increased by a factor $\lambda$.  The authors find that the speed with
which the mean fitness of the population increases is larger by a
factor $(\lambda \sigma_f +
\sigma_b)/(\sigma_f + \sigma_b)$ compared to a population without
mutators. After some time, the fraction of mutators during
hill-climbing converges to $\sigma_f/\sigma_b$, just as if there were
no fitness degrees of freedom, and it decreases to a value
proportional to $\sigma_f/\sqrt{\lambda\mu}$ when the population is
centered around the peak.

Apart from heritable mutators, bacterial populations may also contain
transient mutators that have an increased mutation rate only for one
or two generations. In particular stressed bacteria are believed to
suffer increased DNA damage,  temporarily leading to a higher mutation rate 
\cite{bri98}.

Variable mutation rates are used extensively in the genetic-algorithms
literature. The above-mentioned textbooks \cite{mit96,gol89} contain
various citations of studies about optimal mutation rates for different
fitness landscapes. In complex fitness landscapes, one has to find a
compromise between the negative effect of increased mutation rates
(destruction of genotypes of high fitness) and their positive effect
(finding of new peaks).

The concept of a modifier gene that does not directly affect fitness
but that instead determines some aspect of the reproduction process
(like the mutation rate or the recombination rate \cite{fel97}) or that influences the
effect of another gene is widely used in population-genetics
literature.

\subsection{Density- and frequency-dependent selection}

Throughout this chapter, we have assumed that the total population
size is limited by the carrying capacity of the environment.  This
assumption breaks down when different peaks in the fitness landscape
correspond to different ecological niches, and not just to different
genotypes realizing the same phenotype. A model where population sizes
in different niches are regulated independently, is
more appropriate in this case.

However, the more relevant situation is that of frequency-dependent
selection, because the fitness of an individual is likely to depend
not only on the number of individuals of the same genotype, but also
on the frequency of this genotype or on what other genotypes are
present in the population. Interactions like division of labour or
competition for food, territory, and partners, affect the fitness of
an individual.  Frequency-dependent selection opens the door to the
huge field of coevolution, the subject of the next chapter.  Let us
conclude this section by pointing out only one important consequence
of frequency-dependent selection: The population generally does not
evolve towards a fitness maximum. Fisher's theorem, which opened the
discussion of theoretical models, is consequently not valid in the
presence of frequency-dependent selection.

In order to understand the phenomenon that the fitness is not
maximized, we consider the example of frequency-dependent selection
given by Kimura \cite{kimura}: Let us assume that there are two
environments available for a population, and that the fitness is
determined by one genetic locus that has two alleles, $A_1$ and $A_2$.
We further assume that the population is diploid and mates randomly.
Let $p$ be the frequency of $A_1$, and $q=1-p$ the frequency of $A_2$.
Let further $W_{11}=1-s(p-c_1)$ be the fitness of the $A_1 A_1$
individuals (which have a frequency $p^2$ and are adapted to
environment 1), and $W_{22}=1-s(q-c_2)$ the fitness of the $A_2 A_2$
individuals (which have a frequency $q^2$ and are adapted to
environment 2), and $W_{12}=W_{21}=1$ the fitness of the heterozygotes
$A_1 A_2$ (which have a frequency $2pq$).  The mean fitness of the
population is given by
\begin{eqnarray*}
\bar W &=& p^2W_{11} + 2pqW_{12}+q^2W_{22}\\
&=& 1-sp^2(p-c_1)-sq^2(q-c_2),
\end{eqnarray*}
and the frequency of $A_1$ in the next generation is given by
$p'=[p^2(1-s(p-c_1))+pq]/\bar W$.  A stationary point is reached when
$p'=p \equiv p^*$, implying $p^*=(1-c_2)/(2-c_1-c_2)$.  This point is
stable if $c_1,c_2<2$.  The mean fitness of the population at this
fixed point is $\bar W^* = 1-sp^*(p^*-c_1)$. For almost all
combinations of $c_1$ and $ c_2$, this fitness is smaller than the
maximum possible fitness, which is reached for
$p=(3-2c_2)/(6-2c_2-2c_1)$. (This last result is obtained by differentiating $\bar W$ with respect to $p$.)

\chapter{Models for coevolution}

Models for coevolution take into account that individuals belonging to
the same or to different species can mutually affect each other's
fitness. Just as for evolution in fitness landscapes, the space in
which coevolution shall take place must be specified as part of the
model. Among the many interactions an individual is participating in,
only one or a few are usually taken into account by a model. The
traits relevant for an interaction and the space of their possible
values must be given. Fitness values must be assigned as function of
frequencies of traits. All this is often done with little or no
knowledge of the genetic foundation of the considered traits, but is
rather an intelligent guess based on a certain understanding of the
biological system under consideration. Therefore, the objective of
coevolutionary models is not so much a precise reproduction of nature,
but rather an illumination and illustration of mechanisms and
principles shaping coevolution. A simple model, although neglecting
important aspects of reality, may nevertheless give valuable insights
into how interaction between or within species can affect their
evolutionary change. A model can also help to establish general
conditions under which one or another type of evolutionary outcome is
expected. 

It is impossible to give an exhaustive overview over the vast field of
coevolution. The examples below were chosen because they cover a wide
range of  concepts and methods and therefore allow the reader
to gain an overview of the different theoretical approaches to
coevolution. These include game theory as well as discrete and
continuous genetic models, and the concepts of kin selection, group
selection, and sexual selection. Among the possible dynamical patterns
we will find single fixed points, lines of fixed points, runaway,
limit cycles, and chaos.

We will begin with evolutionary game theory, which focuses on
conflicts of interests and searches for stable configurations where no
individual can improve its fitness by adopting a different
strategy. Next, we will present different theoretical approaches to
the phenomenon of altruism, where individuals have traits that reduce
their own fitness while increasing that of other individuals. These
approaches are kin selection, group selection, and a certain class of
game-theoretical models. The next two sections are concerned with
models for sexual selection and speciation, both of which involve an
explicit genetic representation, and sexual reproduction. Finally, we
will discuss models for parasite-host coevolution, which is
theoretically interesting because it involves two different levels of
selection, namely within and between hosts.

Models for many other coevolutionary systems are beyond the scope of
this review. However, they usually bear some resemblance to the models
covered in this chapter. Predator-prey coevolution and symbiosis are
conceptually akin to host-parasite coevolution, and models are
reviewed by Roughgarden in Chapter 3 of \cite{FS83}. A recent
study of predator-prey models which uses quantitative genetics
was published by Gavrilets \cite{gav97}, and a recent publication on models
for symbiosis was written by Frank \cite{frank97}. Models for
ecological character displacement and for coevolution of flowers and
their pollinators usually involve models of quantitative genetics
similar to the ones discussed in the context of sexual selection and
sympatric speciation.  Ecological character displacement is reviewed
by Roughgarden in Chapters 3 and 17 of \cite{FS83}, and references to
the newer literature are found for instance in \cite{DM99}.  A basic
model for flowers and pollinators is due to Kiester et al
\cite{kie84}.

\section{Game theory and evolutionary stable strategies}
\label{game}

Many problems of coevolution which involve some kind of conflict of
interests are conveniently formulated in the language of evolutionary
game theory. Game theory was first formalized by von Neumann and
Morgenstern \cite{neu44} in order to model human economic behaviour.
According to them, each ``player'' in a ``game'' adopts one of several
possible ``strategies'', and the ``payoff'' obtained by a player
depends on the strategies chosen by the other players. If we equate
``player'' with ``individual in a biological population'',
``strategy'' with ``genotype'' and ``payoff'' with ``fitness'', such a
game becomes an ``evolutionary game'', which is essentially a model
for frequency-dependent selection. Evolutionary game theory was first
introduced under this name by Maynard Smith and Price in the context
of animal conflict \cite{MP73}, although ideas related to evolutionary
game theory can already be found in the older literature. The main
objective of evolutionary game theory -- apart from finding good
models -- is to find so-called ``evolutionary stable strategies''
(ESS's) \cite{MP73}. An ESS is a strategy such that if most of the
members of a population adopt it, there is no other strategy or
linear combination of strategies that would give higher reproductive fitness.

Evolutionary games can be classified according to whether individuals
play against one partner at a time, or ``against the field''
\cite{MS82}, which consists of all other individuals of the
population. An example for a pairwise game is a fight between two male
stags for a female. If one assumes that the pairwise encounters do not
depend on the genotype of the partners and that the population size is
large, the mean payoff of each strategy is a linear function of the
frequencies of all strategies, with the coefficients being
proportional to the expected payoffs of the pairwise encounters.  An
example for a game against the field is a game where mothers determine
the sex ratio of their offspring. Producing children of the rarer sex
is an advantage, however, the fitness function is not linear in the
frequencies of the different strategies (see section
\ref{sexualselection} below).

In the following, some general properties and important examples of
evolutionary games and their application to biological situations are
described. Several additional models will be presented in later
sections. For an introduction to evolutionary game theory, the reader may
wish to consult the book by Maynard Smith \cite{MS82} and the review
by Riechert and Hammerstein \cite{rie83}. A recent formal discussion
can be found in \cite{wei95}. An overview which includes many useful
hints to the literature can be found in the chapters 6,8,9,11 of
\cite{kre97}.

\subsection{Evolutionary games and genetics}

Usually, evolutionary game theory makes no assumptions about the
genetic makeup underlying a strategy, apart from the assumption that
the genotype determines the strategy (i.e., the phenotypic and
behavioural traits relevant to the problem under study). Rather, the
spectrum of possible strategies is given as part of the model, and it
is assumed that the strategies with higher payoff increase in number
relative to those with lower payoff. For asexual populations, the
genotype of a child and therefore its strategy is identical to that
of the parent, except in the rare case of a mutation, where
the child adopts a different strategy. 

For sexual populations, the inheritance of strategies becomes more
complicated and depends on the genetic implementation. The phenotypic
ESS cannot be realized in all genetic models. If it can be realized,
not all initial conditions may lead to it.  This is illustrated in
\cite{MS81}, using a one-locus two-allele model for the Hawk-Dove
game.  In \cite{gar98}, it is proved that models with one locus and
many alleles will evolve to the phenotypic ESS if the genetic system
is able to uniquely realize it. An analytical treatment of models with
many additive loci \cite{hin97} shows that the ESS can be achieved for
at least part of the initial conditions.  Computer simulations of a
general non additive two-locus two-allele model for the Hawk-Dove game
\cite{hin98} showed that the ESS is relevant in the majority of cases.

\subsection{Dynamics of evolutionary games}

The above definition of an ESS does not make any assumptions about the
dynamics of the population, and for many considerations no explicit
dynamical model is needed. Where it is needed, usually simple
replicator dynamics is used, where the growth rate of a strategy $i$
is identical to its payoff $g_i$, leading to a change in the
frequency $p_i$ of strategy $i$ according to
\begin{equation}
\frac{d p_i}{dt} = p_i (g_i - \bar g).
\label{replicator}
\end{equation}
(Note that with this
definition the payoff is different from the fitness, as we defined it
in the previous chapter.)  Taylor and Jonker \cite{tay78} have shown
that an ESS is a stable equilibrium under these dynamics. They have
also shown that not every stable equilibrium is an ESS. The reason is
that the notion of ESS requires that no linear combination of
strategies has a higher payoff than the ESS, even in cases where a
linear combination with higher payoff could not increase in frequency
under the dynamics.  If the dynamics is discrete in time, the
frequencies of the strategies change discontinuously from one
generation to the next, and the ESS is not always an attractor of the
dynamics because of ``overshooting''. 

\subsection{Evolutionary games and learning}

Since evolutionary games make no explicit reference to genetics,
they can as well be interpreted as a learning process, during which
individuals assign an increasing weight to those strategies for which
they had a larger gain in the past \cite{har81}. Such learning
processes can quickly lead to a stable fixed point, as
demonstrated by various feeding experiments. For instance, six
sticklebacks were put in a tank where the rate of water flea input was
twice as large at one end than at the other \cite{mil79}. After a
short time, four sticklebacks could be found at the end with the
larger flea supply, while the remaining two fish were at the other
end. This was a statistical distribution, with the fish switching from
time to time from one side to the other, allowing for a quick
adaptation to a changed feeding situation.

\subsection{Prisoner's dilemma, and formal definition of ESS}

The prisoner's dilemma is a two-strategy game for two partners. Each
partner must choose to cooperate or to defect, without knowing the
choice of the other. If both cooperate, each of them obtains the
``reward'' $R$. If one cooperates and the other one defects, the
defector obtains the ``temptation'' $T$, and the exploited cooperator
the ``sucker's payoff'' $S$. If both defect, each of them obtains the
``punishment'' $P$. The payoffs are such that 
$$T>R>P>S\, .$$

Clearly, a population where everyone cooperates has the highest
possible mean payoff $R$ per individual, if the additional condition
$T+S<2R$ is satisfied. However, this configuration is unstable,
because an individual that defects has a higher payoff than the
others and increases therefore in number, while the mean payoff of
the population decreases. The only stable fixed point is the one where everyone defects, and where the mean payoff of the population is $P$. 

Generally, $I$ is a stable strategy if a population where all
individuals adopt it cannot be invaded by any other strategy $J$. If
$E(I,J)$ is the expected payoff for strategy $I$ playing against
strategy $J$, the condition that $I$ is a stable strategy means that for all $J$ either
$$E(I,I) > E(J,I)$$
or, if $E(I,I) = E(J,I)$,
$$E(I,J) > E(J,J)\, .$$

From the point of view of this simple model, it always pays to cheat
and to exploit others. For the explanation of cooperation and
unselfishness, other models are needed (see section \ref{altruism}
below). A simple way to obtain cooperation would be to change the
ranking of the four payoffs such that $R$ is the largest, in which
case it does not pay to cheat.  This ranking is probably appropriate
for many situations. For instance, two lionesses that hunt together
obtain a large payoff. If one lioness refuses to hunt, the other one
is likely to be unsuccessful, and the payoff is low for both.

Recently, an experiment with a RNA virus revealed a situation similar
to prisoner's dilemma \cite{tur99}. If a bacterium is infected by
several individuals of the bacteriophage $\phi_6$, a mutant of the
virus occurs which manufactures fewer of the intra-cellular products
needed for replication of the phages. It therefore replicates faster
within the bacterium than the original virus, as long as enough copies
of the original virus are present. But when the fraction of the
original virus becomes small, all viruses replicate slower than a
population of $\phi_6$. Turner and Chao measured a density-dependent
fitness of the $\phi_6$ virus and a mutant strain $\phi H 2$ by
infecting bacterial cells with a given mixture of the two types, and
by determining the ratio of the two types after virus multiplication
within the bacteria. Although this is not a two-partner game but
rather a game against the field, the fitness values are not far from
being linear in the two viral densities. If the $\phi_6$ is equated
with the cooperator in the prisoner's dilemma, and $\phi H 2$ with the
defector, the approximate fitness values are found to be $R=1$ (by
normalization), $T=1.99$, $P=0.83$, and $S=0.65$. The ranking of the
fitness values is that of the prisoner's dilemma, and the due outcome
of virus multiplication within a bacterial cell is a population
dominated by the defector type $\phi H 2$.

However, competition between two variants of a virus is not all this
experiment is about. The mutation to the defector always occurs in
situations where bacteria are infected by several viruses
simultaneously. Thus, switching to the defective type means switching
to slower growth in a situation of high viral density. This can well
be a programmed strategy of the virus, and perhaps it facilitates its
long-term survival.  The evolutionary role of the ability to switch to
the defective type still needs to be clarified.

\subsection{The hawk-dove game, animal conflicts, and mixed strategies}

The hawk-dove game is formally similar to the prisoner's dilemma, but
with the ranking of the payoffs being given by
$$T>R>S>P\, .$$ This is the same ranking as for the ``chicken game'',
which is familiar in conventional (i.e., not evolutionary) game
theory. The Hawk-Dove game was introduced by Maynard Smith and Price
in order to model animal conflict \cite{MP73}. Such conflicts
usually involve a display of the strength of the animals, but rarely
lead to an escalation and serious injury to one or both
partners. Thus, snakes wrestle with each other instead of biting, and
mule deer crash or put antlers against antlers, but do not hurt the
unprotected side of the partner.  In the hawk-dove game the hawk
strategy means to escalate a conflict, while the dove strategy means
to display first and to retreat if the partner escalates. If two doves
meat, each of them gets the payoff $V$ with probability 1/2. The
payoff is an increase in fitness by obtaining the female or the
territory for which the fight was performed. If a hawk meets a dove,
the hawk gets the payoff $V$. If two hawks meet, each of them gets the
payoff $V$ with probability 1/2, and incurs a large cost $-C$ with
probability 1/2. If $C>V$, we obtain the above ranking
$S>P$. Obviously, a population of hawks can be invaded by a dove, and
a population of doves can be invaded by a hawk.  The ESS is therefore
a mixture of hawks and doves, with the frequency of hawks being
$V/C$. If the cost of injury is very high, the population is dominated
by doves. This is a suggested explanation for the rare occurrence of
escalation in animal conflicts.

Generally, if $I$ is an ESS consisting of a mixture of several strategies
$A$, $B$, $C$, etc., then each of these strategies must earn the same
payoff on an average,
$$E(A,I)=E(B,I)=E(C,I)=...=E(I,I),$$ because otherwise their
frequencies would not remain constant (see \cite{BC78} for a proof).
Stability against fluctuations in the frequencies of these strategies
requires the additional condition
$$E(I,J) > E(J,J)$$
where $J$ stands for all pure strategies contained
in $I$, and for any mixture thereof. In addition, stability against
invasion of any strategy not contained in the ESS must be shown.

At the ESS, the fitness of hawks and doves is equal. There are two
conceptually different ways to realize such a mixed equilibrium,
depending on whether individuals use ``pure'' or ``mixed'' strategies.
For pure strategies, the population consists of a mixture of
individuals that are genetically hawks and behave like hawks at each
encounter, and genetic doves that always behave like doves. For mixed
strategies, the population consists of genetically identical
individuals, each one behaving like a hawk with probability $V/C$ and
like a dove with probability $1-V/C$. In this case, the probability of
behaving like a hawk is the parameter under natural selection.

The original paper by Maynard Smith and Price \cite{MP73} also allows
for more complicated strategies. In order to define such a strategy,
the game is decomposed into small steps. At each step, one of the
partners makes a move which can be either conventional (not harmful)
or dangerous (aggressive). The opponents may adopt strategies that
depend on the partner's last move, like those with the telling names
``bully'' or ``retaliator''. A pattern emerges similar to the
findings of the simpler hawk-dove game: aggressive strategies fare
less well than ``limited war'' strategies if the cost of injury is
high.

Of course, even these more sophisticated models fall short of the far
more complicated reality which involves many more aspects,  like
experiences in early life (important for the formation of dominance
relationships), effects of kinship,  learning, recognizing
individuals, and many more. (See Ch.~11 of \cite{kre97}.)

As for the prisoner's dilemma, there is also a viral analogue of
the hawk-dove game, which occurs for instance with the influenza virus
\cite{hua70}. While they are a minority, the mutant defective viruses 
replicate much faster than the normal viruses (i.e., $T>R$). But since
they lack a large part of RNA they cannot replicate at all in the
absence of the normal viruses ($P=0$).  Thus, a mixture of the two
types of viruses is reached which replicates slowly, leading to a slow
progression of the disease.  The fact that the host cells influence
the generation of defective viruses, suggests that the ``true'' game
is not the one between the normal and the defective virus, but the one
between the virus and its host.

Recently, a further analogue of the hawk-dove and the
prisoner's-dilemma game was found among myxobacteria
\cite{vel00}. When close to starvation, these bacteria develop
fruiting bodies and form spores, whereby only a minority of the
original bacterial cells turn into spores, while the others die. Some
defective mutant bacteria form more spores than the
wild type in the presence of the wild type, but few or no spores in
the absence of the wild type.

A realization of the hawk-dove game among insects is found in some
species of fig wasps. Each fig-wasp species has either winged males,
wingless males, or both types of males. Winged males leave their natal
fig fruits and mate elsewhere with females that have already dispersed
from their own natal fig fruits. Wingless males mate with females
within the closed receptacle of their natal fig fruit, which they
never leave. Species with large broods tend to have wingless males,
while species with small broods tend to have winged males. In species
with intermediate brood sizes, both male forms sometimes occur
\cite{ham79,cook97}. The wingless males are often fighting males
(corresponding to the ``hawk'' type), while the winged males do not
fight for females (they correspond to ``doves''). Hamilton's data
\cite{ham79} indicate that the fraction of migrating female fig wasps
is equal to the fraction of winged males, implying that both types of
males have the same chance to mate, i.e., their frequencies have
achieved an ESS. Examples for male dimorphism in other species are
given in \cite{cook97}. However, in the majority of cases the
dimorphism is not the result of frequency-dependent selection leading
to an evolutionary equilibrium, but due to qualitative differences
between males, with high-quality males adopting one mating tactic and
lower-quality males another \cite{cook97}. 

\subsection{Rock-paper-scissor games, and cycling dynamics} 

The rock-paper-scissor game is a game with three strategies, none of
which is an ESS. Instead, a population of ``rock'' strategies can be invaded by ``paper'', ``paper'' by ``scissors'', and ``scissors'' by ``rock''. An example for a payoff table for this type of game is
\begin{tabular}{r|ccc}
&R & S & P\\
\hline
R & 1 & 2 & 0\\
S & 0 & 1 & 2\\
P & 2 & 0 & 1
\end{tabular}\, .
The equilibrium frequency distribution is 1/3 for each strategy. This
is, however, no stable fixed point, because each pure strategy plays
against itself as well as the mixed strategy. A more general version
of this game has some payoff $\alpha>1$ instead of 2 in the above
matrix. If we use the simple replicator dynamics,
Eq.~(\ref{replicator}), we find that the fixed point is stable for
$\alpha>2$ and unstable for $\alpha<2$. For $\alpha=2$, the system
cycles on a periodic orbit; for $\alpha>2$, it spirals into the fixed
point; for $\alpha<2$, it spirals away from the fixed point, and ends
up with only one strategy surviving. 

A variant of the rock-paper-scissor game is played by the males of a
lizard species \cite{sin96}. This species has three types of
males. Those with a blue throat defend a territory containing of the
order of one female. Males with an orange throat are very aggressive
and defend large territories that contain several females. If the
population consists mainly of blue males, the orange males achieve
more matings that the blue males and therefore increase in number. The
third class of males have yellow stripes on their throat and resemble
receptive females. They can sneak into the territory of orange males
and mate with their females. In a population dominated by orange
males, the ``sneakers'' can increase in number. To close the cycle,
blue males have an advantage in a population dominated by sneakers,
because they guard their females well. The frequencies of the three
types of males oscillate with a period of about 6 years. The
oscillation appears to be slightly damped. The authors also made an
attempt to obtain a payoff matrix from field data, however, the
agreement of the model dynamics with the empirical data is only
qualitative.

\subsection{The war of attrition, and continuous sets of strategies}

The ``war-of-attrition'' model was introduced by Maynard Smith
\cite{May74} in order to analyse animal contests without escalation.
That is, the contestants display their strength, and the payoff goes
to the one who persists longer. Clearly, there must be a cost
associated with displaying, which increases with time. The choice open
to an individual is to select a length of time for which he is
prepared to continue, and an associated cost, $m$, he is prepared to
pay. If player $A$ chooses a cost $m_A$ larger than that of player
$B$, $m_B$, the payoff for player $A$ is $V-m_B$, while that for
player $B$ is $-m_B$. It is easy to see that a stable equilibrium
cannot be a pure strategy, but that it must either involve the
willingness of individuals to pay a variety of different costs with
appropriate probabilities, or a mixture of individuals that differ in
the cost they are willing to pay.  Such a mixed strategy is given by
the probability density function $p(x)$ for accepting a cost $x$. At
the ESS, the expected payoff must be the same for each pure strategy
present in the mixed set, implying
$$ \int_0^m (V-x)p(x)dx-m\int_m^\infty p(x)dx =C$$
independently of $m$. The solution is 
$$p(x) = e^{-x/V}/V,$$
and $C=0$. In order to show that this fixed
point is stable, one has to show that if one of the pure strategies
increases in weight, its payoff is smaller than that of the other
ones. The main consequence of one pure strategy $m$ increasing in weight is
that everyone has to play against this strategy more often. If the
expected payoff of the mixed strategy against this strategy,
$$ \int_0^m -x p(x)dx + \int_m^\infty(V-m)p(x)dx$$
 is
larger than that of this strategy against itself, 
$$(V/2) - m\, ,$$
its weight will be reduced again, and the fixed
point is stable. It is easy to check that this condition is satisfied.
For a more general proof of stability also against simultaneous fluctuation in the frequency of several pure strategies, see \cite{BC78}.

\subsection{Breaking the symmetry between contestants}

If there is an asymmetry between two contestants which affects their
probability of winning the fight (e.g., due to greater body size) or
their perception of the value of the payoff to be obtained, this will
modify the outcome of a contest. If this asymmetry can be perceived by
the two contestants, it can be used to settle the conflict. Maynard
Smith and Parker have shown that asymmetries can be used in this way
even if they do not affect the payoffs or the winning chances
\cite{MSP76} (see also \cite{ham81}). For this purpose, they
introduced a third strategy into the hawk-dove game, called
``Bourgeois'', $B$. This strategy plays hawk if owner of the territory
for which the opponents fight, and dove if intruder. It is not
difficult to show that $B$ is an ESS, and the only one. The opposite
of $B$ would also be an ESS, but it seems unlikely to be realized in
nature (but see \cite{bur76} for a candidate for such a strategy among
certain spiders).

\subsection{Beyond mean-field theory}

The models and calculations discussed so far, are mean-field models
where spatial structure and fluctuations are neglected. There are
also studies where an evolutionary game is put on a lattice, with each
site being occupied with a given (pure) strategy. The score of each
individual against all of its neighbours is calculated, and the
individual is then replaced by its most successful neighbour. Computer
simulations of the hawk-dove game were done by Killingback and
Doebeli \cite{kil96}, and of the prisoner's dilemma by Nowak and May
\cite{now92a}. For the prisoner's dilemma, it was found that the
cooperators do not die out, and for the hawk-dove game the doves
turned out to be more numerous than in the mean-field model. This
means that spatial structure furthers cooperative strategies. Both
models are essentially two-dimensional cellular automata displaying a
variety of different spatio-temporal patterns and even chaos.

Cressmann and Vickers \cite{cre97} mapped spatial evolutionary games
with migration on reaction-diffusion equations, which are continuous
in space and time. They discussed stable equilibria as well as
travelling waves.

\section{Altruism}
\label{altruism}

In agreement with the population-biology literature, we refer to a
biological trait as  altruistic if it is beneficial to other
individuals, but not to the individual carrier. Examples cited in the
literature are warning calls of birds and primates, the social
behaviour of bees and ants and other insects, collective feeding of
the young by all flock members of the Mexican Jay, or blood sharing of
vampire bats (see \cite{dug97} for these and many more examples).  If
such traits are inherited, the question arises how they can be
maintained in a population, because one might expect that a selfish
individual that does not spend its resources or risk its live
for the sake of others, 
but that nevertheless benefits from the altruism of others has
a higher fitness and therefore takes over.

There are essentially three explanations that answer this question:
kin selection, social compensation, and group selection.  These are
not mutually exclusive but may apply simultaneously to a given
population, and they shall be described in the following.  The fourth
possibility, namely that there is no cost but rather a benefit to an
apparently altruistic act, shall not be discussed here, but see, for
instance, \cite{dug97}. Also, the reader should keep in mind that not
all behavioural traits are under genetic control, and that the
explanatory power of the models presented in the following is therefore
limited.

\subsection{Kin selection} 

The basic observation of the theory of kin selection is that an
individual that helps to increase the fitness of a relative propagates
copies of its own genes present in the relative, including the genes
for altruism. The precise formulation of the theory of kin selection
is due to Hamilton \cite{ham64} who built on ideas introduced by
Haldane. His famous inequality can be derived by assuming that the
altruism is determined by a single gene $A$, and that helping another
individual reduces the fitness of the carrier of this gene by $c$, but
increases the fitness of the recipient by $b$ \cite{MSS}. The
relatedness $r$ between the two individuals is defined to be the
fraction of genes that the two have inherited from the same recent
ancestor. Thus, there are different copies of gene $A$ in the
population, each one stemming from an ancestor a few generations ago.
The probability that the copy of gene $A$ present in the altruist is
also present in the recipient, is $r$. The frequency of this copy in
the population will therefore increase only if
$$b/c > 1/r,$$
which is Hamilton's inequality. The same holds, of
course, for all other copies of gene $A$, so that gene $A$ will take
over the entire population, and ``selfish'' alleles will be
eliminated. 


\subsection{Social compensation}

There are a variety of social mechanisms by which individuals
without the altruistic traits can be barred from taking advantage of
altruists. Altruistic behaviour may be enforced by individuals of
higher rank, defectors might be discriminated against, or altruism may
be based on reciprocity. The evolution of such complex behaviour
cannot plausibly be due to natural selection acting on just one or a
few genes, in particular since social behaviour emerges at a high
organizational level (which can be expected to be affected by many
genes), and since learning and cognitive faculties are involved in
many cases.

In spite of these facts, a model for evolution of cooperation based on
natural selection acting on the (genetically determined) behaviour of
individuals has found wide-spread interest, not just in biology, but
also in economics and social science. This model is the repeated
prisoner's dilemma introduced by Axelrod and Hamilton
\cite{axe81,axe84}, where individuals play the prisoner's-dilemma game
repeatedly against each other. If the prospects of having to play
against the same player again are sufficiently high, cooperation can
evolve in this model. Each individual has the capability to remember
whether the opponent cooperated or defected at the last encounter(s)
and adopts a strategy that is a function of the opponent's past
behaviour. A computer simulation that let various strategies play
against each other and kept track of the average payoff of each
strategy (or, in a later version, that evolved the community of
players by reproduction and natural selection), showed that ``tit for
tat'' (TFT), which cooperated at the first encounter and then copied the
opponent's last behaviour, was the winner. Its strength is that it
does not allow for exploitation, and that it enjoys an unbroken series
of cooperation with itself and other ``friendly'' strategies.
Nevertheless, a population of TFT is not a true ESS,
because any friendly strategy does equally well in a TFT
population, and can therefore increase in number by random drift. Also,
TFT can be invaded by a combination of strategies
\cite{boy87,ben95}.

Nowak and Sigmund \cite{now92} point out that if individuals ``err''
sometimes by not behaving according to their strategy, TFT has the
disadvantage of leading to a long round of retaliations. Therefore, a
strategy that is more forgiving than TFT and cooperates with nonzero
probability $q$ if the opponent defected at the last encounter can
invade a TFT population in which errors occur.  Nowak and Sigmund
\cite{now92} suggest that the optimum value for $q$ is
$$q=\hbox{min}(1-(T-R)/(R-S),(R-P)/(T-P)),$$
because it gains the
highest payoff among forgiving strategies that are immune to invasion
by exploitative strategies. (The expression for $q$ can be obtained by
requiring stability against the two most dangerous exploiters of
friendly strategies, namely a strategy that always defects, and a
strategy that does the opposite of the opponent's last step.)
However, in an initial mixed population with many non-cooperative
individuals it cannot thrive. Only after the TFT strategy has
eliminated the non-cooperative strategies, the forgiving TFT can take
over.

Nowak and Sigmund also found \cite{now93} that if players take into
account their own last moves and not only those of their opponents,
stronger strategies fundamentally different from TFT exist.  A model
where individuals carry ``scores'' that contain information about
their past behaviour leads to periodic cycling of the entire
population between cooperation and defection \cite{NS98}.  However,
such a cycling does not occur if there are always some individuals
that are incapable of cooperation \cite{LFS99}.

Several behavioural patterns found in different species have been
claimed to be examples of the ``tit for tat'' strategy.  Axelrod and
Hamilton \cite{axe81} quote the example of sea bass, a hermaphrodite
where partners stay together a long time and change sex roles on a
regular basis. It is found \cite{fis80} that pairs tend to break up if
sex roles are not divided evenly.  Breakup of a joint enterprise upon
onset of exploitation is also found in sticklebacks who approach their
predator, most likely in order to gain some information, in pairs of
two.  If one partner deserts and stays behind, the other one, who is
now exposed to a greater risk, withdraws also to a safer distance
\cite{mil87}.  These findings do not imply that the individual that
defected is genetically programmed with a strategy that tends to
defect, while the other individual that withdrew upon exploitation is
programmed with ``tit for tat''. But they indicate that if for
whatever reason cooperation is given up by one individual, the other
individual does not tolerate exploitation. (For a recent publication
on predator inspection, and a theory of it, see \cite{fish99}.)

Breeding tree swallows are usually quite tolerant of non-breeding
individuals getting close to them. Only if a non-breeder harms them
or their brood, do they respond with aggression \cite{lom85}. This
observation, which was also claimed to be an example of ``tit for
tat'', demonstrates the same pattern: if one side gives up
cooperation, the other one does not accept exploitation.

Finally, the vampire bat shares his bloody meal with roost-mates that
are close to starvation. They do not share with individuals they do not
know, and they are more ready to share if they have been saved from
starvation before \cite{wil84}. Again, we see a behaviour that, while
not really being ``tit for tat'', benefits cooperators and protects a
cooperative group from possible exploitation.

\subsection{Group selection}

Using the words of Wade \cite{wad78}, ``Group selection is defined as
that process of genetic change which is caused by the differential
extinction or proliferation of groups of organisms.''  Because groups
that contain more altruists grow faster and/or survive better,
altruists can survive in the population and even become fixed under
certain conditions, even though their proportion decreases on an
average within a given group. Models for the maintenance of altruism
under group selection are reviewed in \cite{wad78,wil83}, and examples
of very recent studies published in physics journals are given by
\cite{don,silfon}. An application to groups of replicating molecules
in prebiotic evolution is given in \cite{sza87}. All models have the
following ingredients: Individuals live in small groups, called {\it
demes}. Parasites on a host may comprise such a group, mice in a
haystack, birds in a breeding colony, bats using the same roost, etc.
The fitness of individuals carrying the altruistic genotype (either a
single gene, or, in diploid individuals if altruism is recessive, a
pair of altruistic genes) is lower than that of the other group
members by a fixed amount; the fitness of all individuals in the deme
increases with increasing number of altruists.

A central
requirement for the models to function is that groups are sufficiently
small. This has two effects: (i) Randomly assembled groups differ
significantly in the proportion of altruists, leading to significant
fitness differences among newly formed groups; (ii) Genetic drift
allows the proportion of altruists to increase within some groups. 

The dynamics of group-selection models has two components: The growth
within a group, and the formation and extinction of groups. In many
models, migration of individuals between groups is also allowed, and
some models also contain mutation between the two genotypes. The
dynamics within demes is usually determined by Wright-Fisher sampling.
In many models, the absolute fitness values play a role and the deme
size changes (in some models with the overall carrying capacity being
fixed); in other models, the deme size is fixed, and only the
fractions of altruists change. Deme formation is implemented by
splitting large demes, and/or by repopulating empty patches (where a
deme has become extinct) with a copy of another deme, whereby both
processes depend on the total fitness of the deme. Deme
extinction occurs either by its size decreasing to zero, or by a death
probability that depends on the deme's fitness. In many models, groups
do not split or become extinct but are rather dissolved after a
certain growth time, and then reassembled at random.

Eshel \cite{esh72} gave a simple proof that altruism can become fixed
in a certain class of models: Assume that there are infinitely many
demes of finite size, and that groups evolve according to
Wright-Fisher sampling and are killed with a probability that is
larger if the proportion of altruists is smaller. Clearly, there is a
non-vanishing probability (that might be small) that a deme becomes
fixed for the altruistic genotype. If this happens, the deme remains
completely altruistic. Together with the fact that altruistic demes are
killed at a smaller rate than other demes, this leads to a
steady increase of the fraction of altruistic demes, resulting in
fixation. Eshel also showed that a sufficiently small amount of
migration does not change this result. For large migration, the
selfish genotype becomes fixed: An infinite migration rate means that
demes become randomly reassembled at each time step. Clearly, the
mean fitness of altruists is lower than that of selfish individuals in
this case, leading to a fixation of the selfish genotype. For
intermediate migration rates, coexistence of the two genotypes is
possible.

The main reason why many biologists do not assign an important
evolutionary role to group selection is that it works only for a
limited parameter range, which appears to be too restrictive. However,
Wilson \cite{wil83} points out that altruism might reduce the fitness
of an individual only by a very small amount, making group selection
more probable. He also gives evidence for group selection in nature:
Sex ratios biased towards females, as seen in small arthropods in
subdivided habitats, have a natural explanation if group selection
occurs: groups with more females can produce more offspring and
therefore grow faster, while selection within a single group tends to
drive the sex ratio towards 1:1 (see section \ref{sexualselection}).
Another example of group selection mentioned by Wilson is the decrease
of virulence of the myxoma virus which was introduced into Australia
to control the European rabbit. Within each rabbit, more virulent
viruses multiply faster, but rabbits with more virulent viruses die
faster, leading to a net advantage of less virulent viruses (see
section \ref{parasites}).

Wade argues \cite{wad78} that the variation between demes is probably
larger than predicted by these simple one-locus models because
different demes have different genotype composition and are exposed to
somewhat different biotic environments, leading to different selection
effects on the same allele in different demes.  Group selection
experiments with {\it Tribolium} (a beetle) performed by Wade confirm
his view that the effects of group selection are complex and different
in different groups. One consequence is that it is not always obvious
which traits are altruistic and decrease the fitness of an individual
within a group, while benefitting the group.

Also, kinship association or assortative association with respect to a
trait (e.g., if altruists associate with other altruists) leads to
increased variation between groups and to stronger selection. A model
by Wilson and Dugatkin \cite{WD97} shows that if altruism is a
quantitative trait and if altruists like to form groups then altruism
can evolve in an initially rather selfish world.

\section{Sex ratios and sexual selection}
\label{sexualselection}

The term sexual selection was coined by Darwin \cite{darwin} in order
to explain the frequent occurrence of conspicuous traits in males that
are unlikely to be due to ordinary natural selection for improved
survival. These traits are sexual differences such as greater strength
and weapons in males, bright colouring and other ornaments. He
reasoned that these traits improve the mating success of their
carriers because they can win fights for females, or because females
prefer to mate with them.

Competition of males for females occurs because each male has enough
sperm to fertilize the eggs of many females, while the sex ratio
of most species is 1:1. Evolutionary theory has focussed on several
aspects of selection related to the difference between the two
sexes. The first is the origin of anisogamy itself, i.e., of the
difference between eggs and sperms, eggs being large and less
numerous, and sperm being small and abundant. The model by Parker,
Baker, and Smith \cite{PBS} showed that coexistence of two gamete
sizes is favoured over one gamete size if there is a sufficiently high
advantage in producing large, well provisioned zygotes. (For an
overview, see Chapter 6 of \cite{kre97}.) The second is the ratio
between the two sexes. This will be discussed in the next
subsection. The third is the competition between males for fertilizing
females, including direct fights between males, competition between
sperms of males that fertilize the same female, and female choice of
their mating partner(s). Fights between males were the motivation for
the introduction of the hawk-dove game discussed above. Sperm
competition will not be discussed here, but the interested reader can
consult Chapter 6 of \cite{kre97}. Models for evolution due to female
preferences will be discussed in subsection \ref{preferences}.  Sexual
conflict resulting from the different mating interests of males and
females is also believed to drive evolutionary change, and has inspired
several recent models like the ones by Parker and Partridge
\cite{PP98} and by Gavrilets \cite{gav00}. They will not be discussed
further in this review.

\subsection{Sex ratios}

The sex ratio of many species is 1:1. An evolutionary explanation for
this finding was first given by Fisher \cite{fisher}, who showed that
individuals producing more children of the rarer sex have a higher
fitness because they will have more grandchildren. If the sex ratio in
the population is $q : p$ (female : male), a son contributes on an
average $1/p$ sperms to the generation of grandchildren, while a
daughter contributes on an average $1/q$ eggs (assuming $p+q=1$ and
fixed population size).  Thus, if $q<p$ ($q>p$), individuals that have
more daughters (sons) will have more grandchildren, and the sex ratio
will tend towards 1:1.  A beautiful confirmation of Fisher's mechanism
was obtained from a six-year experiment with an experimental {\it
Drosophila} population that initially had a bias towards producing
daughters and that evolved by natural selection in the direction of a
1:1 sex ratio \cite{car98}.

There are examples of sex ratios different from 1:1, in which cases
assumptions underlying the above argument must be violated. One
assumption is that the number of grandchildren is on an average
proportional to the number of copies of the allele that causes the
modified sex ratio. This is correct only if the allele is transferred
to children irrespective of their sex. Another assumption is that
no circumstances exist that affect the viability or fertility of
sons and daughters differently, and that the investment for producing
a daughter equals that for producing a son.  A third assumption is
that mating is random throughout the entire population.  This
assumption is violated whenever mating occurs within subgroups of the
population, leading to a female-based sex ratio. (For an overview over
factors affecting sex ratios, see \cite{har97}.)

An often-cited model for the latter situation was suggested by Hamilton
\cite{ham67}. He assumed that mated females form groups of size $n$
and have $k$ offspring each. The offspring mate randomly within the
group. Mated females disperse throughout the population and form new,
randomly assembled groups that restart the cycle. If $r$ is the
proportion of sons in the population, an allele causing the production
of a proportion $r'$ of sons will be passed to $k^2(1-r')$ eggs and
$k^2 r'(n(1-r)+r-r')/(nr-r+r')$ sperms contributing to the generation
of grandchildren. (In how many grandchildren the allele will be
present depends on the proportion of matings between eggs and sperms
that are both carriers of the allele.) Maximizing the sum of eggs and
sperms contributing to the generation of grandchildren, Hamilton found
his ``unbeatable'' sex ratio $r=(n-1)/2n$. He admitted, however, that
this result is not exact, because the calculation does not calculate a
stable stationary state. Maximizing the number of gametes contributing
to the generation of grandchildren does not necessarily maximize the
number of progeny after several generations. An exact calculation 
apparently does not exist.

These two models by Fisher and Hamilton are precursors of evolutionary
game theory, because they search for strategies such that they are
unbeatable if they are adopted by the entire population. A very
influential paper on sex-ratio theory for social insects is
\cite{tri76}, which explicitly takes into account the close relatedness 
of insects within a colony.

\subsection{Evolution due to female preferences}
\label{preferences}

There is now much evidence that females often choose their mates, and
that their choice favours conspicuous male traits. The exact ways in
which female choice selects for such traits are still debated, and so
are the ways in which female preferences evolve. Males with a strong
or frequent signal might attract more mates simply because they are
noticed more quickly or farther away, or because females are sensitive
to the signal because it also occurs in other contexts (like a
pheromone with a smell similar to that of food). Or, a female
preference for a male trait is under selection because her mate choice
affects her survival or fecundity (e.g., if males provide resources
for the female, or if some males are more efficient at fertilizing her
eggs). Or, a female preference is under indirect selection because her
preferences are genetically correlated with a male trait which is
under direct selection (e.g., if males with a certain trait tend to
sire healthier offspring). There are examples in nature for all these
scenarios, as well as theoretical models. (See, e.g., Chapter 8 of
\cite{kre97} and \cite{and94}.)

Females of several species choose traits that negatively affect the
viability of the males. For instance, bright colouring increases the
risk of predation. A much discussed hypothesis to explain these risky
traits is due to Fisher \cite{fisher} who suggested a ``runaway''
process due to mutual reinforcement of female choice and male trait:
if sufficiently many females prefer males with a certain trait, males
with this trait will have more offspring, and the offspring will carry
not only more of the male trait genes, but also more of the female
preference genes, etc. This reinforcement can work even in situations
where the male trait confers a survival disadvantage. Several
theoretical models for this process exist, two of which shall be
presented in the following. They were chosen because they are
instructive examples of theoretical models for coevolution in general,
in situations where genetic modelling is required. One of these models
involves a few genetic loci and alleles, the other model is a
quantitative genetic model. An overview of a wider range of models
can be found in the book by Andersson
\cite{and94}, which is an excellent and broad review of the huge
field of sexual selection.

\subsubsection{The discrete model by Kirkpatrick}

The model by Kirpatrick \cite{kir82} is a haploid sexual model, i.e.
each individual has one set of chromosomes. Each genome has a
``trait'' locus and a ``preference'' locus, with two alleles $T_1$ and
$T_2$, and $P_1$ and $P_2$. Males that carry the $T_2$ allele have
some conspicuous trait which reduces their viability by a factor $1-s$
compared to males that carry the $T_1$ allele and do not have this
trait. Females that carry the $P_1$ allele show no mate preference,
while females that carry the $P_2$ allele are by a factor $a_2$ more
likely to mate with a given $T_2$ male than with a given $T_1$ male.
The $T$ gene has no effect in females and the $P$ gene has no effect
in males, and there is no correlation between the sex of an offspring
and the alleles transmitted to it.

\begin{figure}
  \centerline{\epsfysize=0.6\columnwidth{{\epsfbox{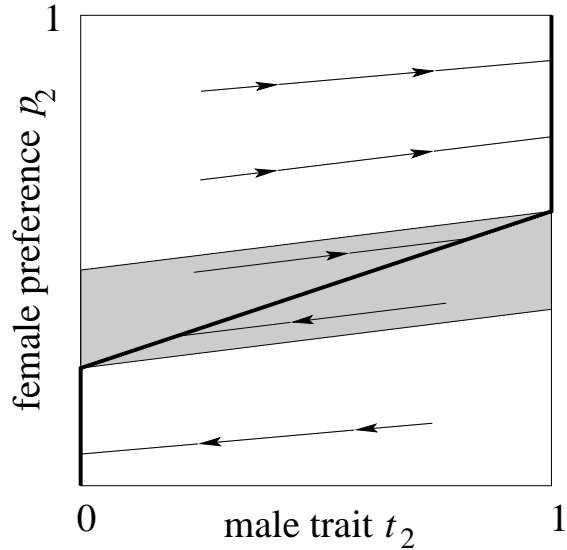}}}}
  {\caption{Schematic representation of the line of fixed
  points in Kirkpatrick's model, and of trajectories for allele
  frequencies. The shaded area contains all those initial conditions
  that lead to a coexistence of the two male trait alleles. All other
  initial conditions lead to a fixation of one male trait
  allele. \label{kirk} }}
\end{figure}

It is not difficult but tedious to write down the recursion relations
of the four frequencies $x_1, x_2, x_3, x_4$ of $T_1P_1, T_1P_2,
T_2P_1, T_2P_2$ individuals from one generation to the next, assuming
random mating and a recombination probability $r$ between the parent
genotypes. Kirpatrick expressed these recursion relations in terms of
$p_2=x_2+x_4$ (the frequency of $P_2$) and $t_2=x_3+x_4$ (the
frequency of $T_2$), and of the linkage disequilibrium
$D=x_1x_4-x_2x_3$ which is a measure of the correlation between $P$
and $T$ alleles within the same individual. He found analytically that a
line of fixed points $(t_2^*,p_2^*)$ exists,
$$t_2^* = \frac{1}{s}+\frac{1}{a_2(1-s)-1}p_2^*
-\frac{1}{a_2(1-s)-1}$$
(see Fig.~\ref{kirk}) which is independent of
the recombination probability $r$. Stability analysis revealed that
this line of fixed points is stable. The system, if perturbed from the
line, will return to it, but there is no selection against changes of
gene frequencies along the line.

Thus, if the initial female preference is large enough, the system
evolves to even larger preferences, and to a high frequency of the
male trait, even though there is no selection for the female
preference, and negative natural selection against the male trait.  At
a fixed point, the relative mating advantage of preferred males is
balanced by the disadvantage under natural selection. 

This haploid model was chosen for its analytical tractability,
although it is not very realistic. Analytical treatment of the diploid
version is extremely complicated, but computer simulations indicated
that its behaviour is qualitatively similar to that of the haploid
model
\cite{kir82}.

\subsubsection{The continuous model by Lande}

Lande suggested a model based on the assumption that both the female
preference $y$ and the male trait $z$ have normal distributions
$p(z)$ and $q(y)$ with means $\bar z$ and $\bar y$ and variances
$\sigma^2$ and $\tau^2$ \cite{lan81}. (Quantitative traits are indeed
often Gaussian distributed if scaled properly, usually by taking the
logarithm.) The change in the mean male trait $\bar z$ from one
generation to the next is the result of selection (natural selection
as well as female preference) and inheritance. If reproduction was
clonal, and if the trait was 100 \% heritable (i.e., not affected by
the influence of the environment), the change in $\bar z$ from one
generation to the next would be given by
\begin{equation}
S=\frac{\int (z-\bar z)p(z)W(z)dz}{\bar W}\, ,
\end{equation}
where $W(z)$ is the fitness of an individual with trait $z$.  If the
effects of all genes contributing to $z$ were additive, and if
reproduction was sexual, the change in $\bar z$ from one generation to
the next would be given by $S/2$, because a son inherits only half of
his genes from the father. (We have again assumed that $z$ is
determined only by the genes, not by the environment.) Now, since the
effects of genes are generally not additive, and since the environment
also has an effect on the trait, the complete expression for the
change in $\bar z$ from one generation to the next is
\begin{equation}
\Delta \bar z = \frac{GS}{2\sigma^2}\, , \label{barz}
\end{equation}
where $G$ is the so-called {\it additive genetic variance} of the male
trait. (The additive genetic variance of the trait is defined in a
similar way to the additive genetic variance in fitness, see section
\ref{fundamentaltheorem}. It is the contribution to the trait variance that
indicates the potential for change in the trait value. The other
contributions to the trait variance are due to the environment and to
dominance effects. For an introduction into the theory of quantitative
genetics, see the book by Bulmer \cite{bul80}.)  $G$ and $\sigma^2$
are assumed by Lande to be constant.  As there is assumed to be no
selection on female preferences, they evolve as a correlated response to
male preferences,
\begin{equation}
\Delta \bar y = \frac{BS}{2\sigma^2}\, , \label{bary}
\end{equation}
with $B$ being the {\it additive genetic covariance} between $z$ and $y$. 

From equations (\ref{barz}) and (\ref{bary}) follows that trajectories in the $\bar z-\bar y$ plane are straight lines of the slope $\bar y/\bar z = B/G$. The condition $S=0$ defines a line of fixed points. If the initial point lies above (below) this line, $\bar z$ and $\bar y$ increase (decrease) with time.

Now, the remaining task is to define an appropriate expression
for the fitness $W(z)$, and to search for fixed points of the
dynamics. Lande chose
\begin{equation}
W(z) = W_{nat}(z)\int dy q(y) \frac{\psi(z|y)}{\int
  dz \psi(z|y) p_w(z)},
\label{lande}
\end{equation}
where $$W_{nat}(z)=e^{-(z-\theta)^2/2\omega^2}$$ is the effect of
natural selection, which is assumed to drive the male trait to an
optimum $\theta$, and $p_w(z)$ is the trait distribution after natural
selection, $p_w(z) = e^{-(z-\theta)^2/2\omega^2} p(z) / \int dz
e^{-(z-\theta)^2/2\omega^2} p(z)$, and has a mean $\bar z_s=(\bar z
\omega^2 + \theta\sigma^2)/(\omega^2+\sigma^2)$.  $\psi(z|y)$ is the
relative preference of $y$ females for $z$ males. Using three
different analytical forms of the female preference, $\psi(z|y)
\propto e^{yz}$, $\psi(z|y) \propto e^{(z-y)^2/2\nu^2}$, or $\psi(z|y)
\propto e^{(z-(y+\bar z_s))^2/2\nu^2}$, Lande performed an expansion
for weak selection, $\sigma^2,\tau^2 \ll \omega^2, \nu^2$, for which
\begin{eqnarray}
S/\sigma^2&\simeq& \left(\frac{\partial\ln W}{\partial z}\right)_{z=\bar z}\nonumber\\
&\simeq&\left(\frac{\partial \ln(W_{nat}(z) \psi(z|y))}{\partial z}\right)_{\bar z,\bar y} .\label{weak}
\end{eqnarray}
The stationarity condition $S=0$ defines a straight line of fixed points,
$$\bar y = (\alpha+\epsilon) \bar z - \alpha \theta\, ,$$
with $\alpha
\simeq \nu^2/\omega^2$ for the second and third model of female
choice, and $\alpha = 1/\omega^2$ for the first model, and
$\epsilon=1$ for the second model and zero otherwise.  

For $B<G(\alpha+\epsilon)$, these results imply a scenario equivalent
to that in Figure \ref{kirk}. The line of fixed points becomes
unstable if this inequality is not satisfied, leading to a runaway
process where $\bar y$ and $\bar z$ both increase without limits. Of
course, the runaway process will eventually be stopped, either because
there is no additive genetic variance left, or because negative
natural selection becomes very strong, both of which effects are not
included in the model's assumptions. A runaway process can also result
in Kirkpatrick's discrete model if females choose mates from a small
group of males (for instance among those males that live in the
territory covered by the female), and if $P_2$ females mate with $T_2$
males wherever possible \cite{seg85}.

The point of both models is that they show that a male trait that is
disadvantageous under natural selection can nevertheless be maintained
in a population due to female preference, even if this preference is
not favoured by selection. One might ask how this preference came to
be in the first place. An intriguing explanation is that there is an
initial advantage to a preference because it facilitates recognition
of males belonging to the same species. The topic of speciation (and
the possible role of sexual selection) is discussed in the next
section.

Not all models of the two types just described lead to lines of fixed
points. A basic requirement for a line of fixed points is that there
is no direct selection acting on females. If female mate choice is
costly, $P_2$ females have a lower fitness than $P_1$ females, and the
preference for a male trait cannot be maintained in the population.
However, if natural selection favours a certain degree of female
preference, a stable fixed point can occur.

\begin{figure}
  \centerline{\epsfysize=0.6\columnwidth{{\epsfbox{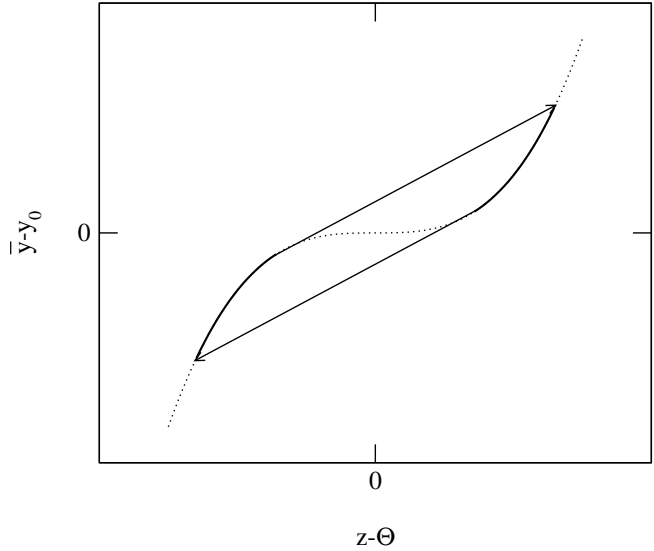}}}}
  {\caption{Schematic representation of the trajectories in the model
  by Iwasa and Pomiankowski. The dotted line is the line of fixed
  points for $b=0$. The solid line marks the attractor of the dynamics
  for a small positive $b$.
\label{cycle} }}
\end{figure}

Under certain conditions male trait and female preference can cycle
indefinitely along a stable trajectory. In \cite{iwa95}, a model
similar to the one by Lande, but with $\psi(z|y)\propto e^{a ( y-y_0)
(z-\bar z)}$ and $W_{nat}=e^{-c(z-\theta)^4}$ is introduced. This
model additionally contains a female fitness function
$W_f=e^{-b(y-y_0)^2}$ representing a weak Gaussian cost to mate
choice. ($b$ is assumed to be small.)  With $b=0$, and assuming weak
selection, Eq.~(\ref{weak}) leads to a line of fixed points
$$-4c(\bar z-\theta)^3+a(\bar y - y_0)=0.$$
A fast runaway process drives the population towards this line. 
In the central region around $y_0$, this line is unstable because the
slope is very small (see Fig.~\ref{cycle}). 
For $b\neq 0$, the change in male trait and female preference is given by
\begin{equation}
{{\Delta \bar y}\choose {\Delta \bar z} } = \frac{1}{2}
\left(\begin{array}{cc} B & G_y \\ G_z & B \end{array}
\right)
{(\partial \ln(W_{f}(y)/\partial y)_{\bar y}\choose (\partial \ln(W_{nat}(z) \psi(z|y))/\partial z)_{\bar z,\bar y}}+{0\choose u} .
\end{equation}
Here, a mutation rate $u$ that decreases the male trait is also
included.  For small $b$ and $u=0$, the runaway process is essentially
the same as for $b=0$. Once the line of fixed points is reached, the
equations slowly drive the  population along this line towards $y_0$, until
the unstable region is reached and a new runaway process starts.  The
system thus has alternating episodes of fast runaway towards large
deviations from the optimum under natural selection, and slow decay in
direction of the optimum in male trait and female preference. If the
mutation rate $u$ is sufficiently strong, the cycle is replaced by a
stable fixed point.

An indirect benefit
results for the female if the male trait reveals information about the
viability of its possessor, for instance if an ornament that is in
good shape indicates resistance to parasites. For an overview of the
literature on this and other variants of female choice models, see
chapters 2 and 3 of \cite{and94}.

We thus have seen that female preferences can drive coevolution
between male traits and these preferences. The dynamical scenario
depends on details of the model, and it has yet to be seen which of
the models describe reality best.

\section{Speciation}

Speciation is the process by which a species splits into two. For
sexually reproducing taxa, species are often defined as genetically
separated units. This means that matings can occur and lead to viable
and fertile offspring between individuals that belong to the same
species whenever they live in the same geographic area, but not
between individuals of different species. With this definition of
species, a theory of speciation has to explain how two reproductively
isolated groups can be formed out of one. Based on the fundamental
work by Mayr \cite{mayr}, it was accepted for a long time that
speciation usually occurs in allopatry, i.e., when two populations
become separated by geographic barriers. They then evolve
independently from each other, and after enough time become
sufficiently different that they do not mate or cannot produce fertile
offspring when they come into secondary contact. A variant of this
model is that of parapatric speciation, where a small population
living at the boundaries of the geographic area of a species evolves
rapidly and becomes reproductively isolated from the main species.
Only recently has it become widely accepted that sympatric speciation
(i.e., speciation without geographic isolation) may have generated
many species of insects, fishes, and birds.  Finally, in plants
polyploidity constitutes an efficient isolating mechanism that
generates new species. Sometimes gametes with more sets of chromosomes
than in the parents are formed, leading to plants that cannot mate
with those belonging to the parent species.

For species that reproduce asexually or mainly through inbreeding, the
above species concept is inadequate, and Templeton suggests to use the
criterion of cohesion instead \cite{tem}: individuals belong to the
same species if they are part of the same evolutionary lineage, in
which new genetic variants arise, spread and replace old variants
through micro-evolutionary forces such as gene flow, genetic drift, and
selection. This definition essentially implies that individuals of a
species occupy the same ecological niche, or, in the language of the
previous chapter, sit near the same peak of the fitness landscape.
Speciation occurs whenever a local population moves from one
fitness peak to another one. This topic has been discussed in the
previous chapter. In this chapter, we will focus on models for
speciation in sexual species.

Not all species are ``good'' species in the sense of the above
definitions. There are many instances where reproductive isolation is
incomplete or where species boundaries fluctuate. This is certainly
the case wherever a population is in the process of speciation.

We will first discuss Wright's shifting balance theory for how a
sexual population can move from one fitness peak to another if mixed
genotypes have low fitness. Then, we will consider speciation
resulting from random mutations and drift in models with flat fitness
landscapes. Next, we will present deterministic models where sexual
selection causes rapid divergence between different populations.
Finally, we will give an overview of models for sympatric
speciation, all of which involve assortative mating and the
availability of several ecological niches.

\subsection{Wright's shifting balance theory}

Due to nonlinear interactions between genes, mating between
individuals of two different well-adapted genotypes may result in
offspring with a mixed genotype that has low fitness. In such a
situation one can expect that natural selection favours the
establishment of reproductive isolation between the two genotypes,
leading to two different species. Wright's shifting balance theory is
concerned with the question of how a second well-adapted genotype can
become established in the first place, given that individuals of the
new genotype are initially rare and mate mainly with individuals of
the old genotype.  A simple realization of this situation is given by
a two-locus model, with one locus having the alleles $A$ and $a$ (with
$A$ being dominant), and the other locus the alleles $B$ and $b$ (with
$B$ being dominant). Let the genotypes $aabb$ and $A*$$B*$ have high
fitness, and the mixed genotypes $A*$$bb$ and $aaB*$ lower fitness,
and let the $a$ and $b$ alleles be rare initially. Clearly, selection
then prevents them from becoming numerous in a large population with
random mating and recombination. Wright emphasizes that in order to
shift the balance from $A*$$B*$ to $aabb$ in the population, a species
must be divided into local populations that are sufficiently small to
allow for large stochastic fluctuations in their genetic
composition. These fluctuations may then move one of the small
populations to the new genotype.  Wright further reasons that this
genotype may become established in a broader geographic area if it has
a fitness superior to that of the old genotype. (For a review by
Wright himself on his theory, see \cite{wright82}.)

\subsection{Speciation in neutral fitness landscapes}

As we have discussed in the previous chapter, the establishment of a
new genotype may occur without crossing deep valleys in the fitness
landscape, because realistic fitness landscapes probably have
high-lying neutral plateaus or ridges. For this reason, several
authors have suggested and studied models of speciation in neutral
fitness landscapes. These models assume that the genotype can be
represented by a string of $+1$ and $-1$ alleles of length
$N$, and that all genotypes have the same fitness. However, matings
between two individuals lead to viable and fertile offspring only if
the parent genotypes differ in at most $k$ digits.

Let us first consider the case $k=N$ with random mating, which was
discussed by Serva and Peliti \cite{ser91}. The authors assume that
each individual of the new generation has two parents randomly chosen
from the previous generation, and that crossover occurs at every place
in the genome with probability 1/2. This means that each allele is
inherited from either parent independently with probability 1/2.
Furthermore, the authors add a small mutation rate $\mu$ per site and
generation. It is easy to show that in the stationary state the mean
overlap between two randomly chosen individuals of a population of
size $M$ is
\begin{equation}
\bar q = \frac{1}{1+M(e^{4\mu}-1)}
\label{barq}
\end{equation}

Next, let us assume that $k<N$, which is the situation studied by
Higgs and Derrida \cite{higgs92}. The first parent of each individual
of the new generation is chosen at random, and the second parent is
chosen randomly among those individuals that have an overlap $\le k$
with the first parent. If $k> (1-\bar q )/2 $ (with $\bar q$ given by
Eq.~(\ref{barq})), there is hardly any difference to the random mating
case. However, if $k< (1-\bar q )/2 $, computer simulations show that
the population eventually splits into two populations that drift
further apart and become reproductively isolated. Due to random
fluctuations, each of these species may become extinct or increase in
size and split further, leading to a continuous chain of speciations
and extinctions. It should be emphasized that this model does not contain
geographic isolation and therefore describes sympatric speciation. 

The allopatric version of this model was studied by Manzo and Peliti
\cite{man94}. These authors assume that two populations exist on two
different islands, with a small migration probability $\epsilon$ per
individual and generation. If $k$ is sufficiently large, the mean
overlap between two individuals of the same island in the stationary
state is
$$\bar q_w = \frac{2\mu + \epsilon}{(4\mu M+1/2)\epsilon+4\mu}\, ,$$
and the overlap between individuals of different islands is
$$\bar q_b = \frac{\epsilon}{(4\mu M+1/2)\epsilon+4\mu}\, .$$
If $ (1-\bar q_b )/2<
k< (1-\bar q_w )/2 $, matings between individuals of the same island
are essentially random, while matings between individuals of different
islands are rarely possible. This leads to a divergence between the
two populations in spite of migration, and to two different species.
The authors emphasize that this type of allopatric speciation occurs
much faster and for less restrictive parameter ranges than the sympatric
speciation in the previous model.

Gavrilets \cite{gav99} studied a variant of these models where matings
occur according to a different rule: both parents are chosen at
random, and only if they differ at no more than $k$ sites, an
offspring is generated. As a consequence, boundary genotypes with few
matching partners generate less offspring than genotypes that have
many potential partners. For this reason, sympatric speciation as
found in the model by Higgs and Derrida cannot occur in this model.
The focus of \cite{gav99} is on the allopatric case, with $m$
subpopulations, each of the same size $M$. An analytical calculation
leads to an implicit expression for the time evolution of the overlap
within and between populations. It is based on the assumptions that
mutation and migration rates are small and that within each
subpopulation each locus is close to fixation for one allele. The main
result is that depending on the initial conditions and the parameter
values the overlap between populations can remain finite or decrease
towards zero.

\subsection{Genetic divergence driven by sexual selection}

We have already seen in section \ref{sexualselection} that sexual
selection can cause a rapid change in a male trait. If populations in
different geographic areas have different dynamics and reach different
equilibria under sexual selection, they may become reproductively
isolated. This plausibility argument that sexual selection can drive
speciation is supported by a recent investigation of speciation and
feather ornaments in birds \cite{mol98}. It was found that genera
containing more species have a higher proportion of ornamented species
than genera with fewer species, suggesting that ornamentation
facilitates speciation.

In the following, we present two models for speciation driven by
sexual selection.  Lande made his model for sexual selection
\cite{lan81} space-dependent by assuming that the optimum male
phenotype is a space-dependent function $\theta(x)=\bar \theta +
\tilde \theta(x)$, and that migration of individuals can be captured
by diffusion terms. Equations (\ref{barz}) and (\ref{bary}) then
become
\begin{eqnarray*}
\frac{\partial \bar z}{\partial t} &=& \frac {GS(x)}{2\sigma^2}+\frac{l^2}{2} \frac{\partial^2 \bar z}{\partial x^2}\\
\frac{\partial \bar y}{\partial t} &=& \frac {BS(x)}{2\sigma^2}+\frac{l^2}{2} \frac{\partial^2 \bar y}{\partial x^2},
\end{eqnarray*}
where $l$ is a characteristic scale for migration \cite{lan82}.
Assuming weak selection and splitting from $\bar z$ and $\bar y$ the
solution resulting for spatially uniform $\theta(x)=\bar \theta$,
\begin{eqnarray*}
 \bar z(t) &=& v(t) + f(x,t)\\ 
\bar y(t) &=& u(t) + \frac{B}{G}f(x,t),
\end{eqnarray*}
one obtains (using the female preference functions and notation introduced in section \ref{sexualselection})
$$\frac{\partial f}{\partial t} = \frac{G}{2\omega^2}\left[\tilde
  \theta(x)-\frac{f}{A}\right] + \frac{l^2}{2} \frac{\partial^2
  f}{\partial x^2}$$
with $A=\alpha/(\alpha+\epsilon-B/G) > 0$.  For
$B/G<\alpha+\epsilon$, the time evolution converges to
$$f(x,\infty)=\frac{A}{2L} \int_{-\infty}^\infty \tilde \theta(\xi)
e^{-|x-\xi|/L} d\xi$$
where $L=l\sqrt{A\omega^2/G}$ is the length
scale over which the male phenotype is correlated. In the less realistic case
$B/G>\alpha+\epsilon$ geographic variation increases at an
accelerating rate. 

\begin{figure}
  \centerline{\epsfysize=0.6\columnwidth{{\epsfbox{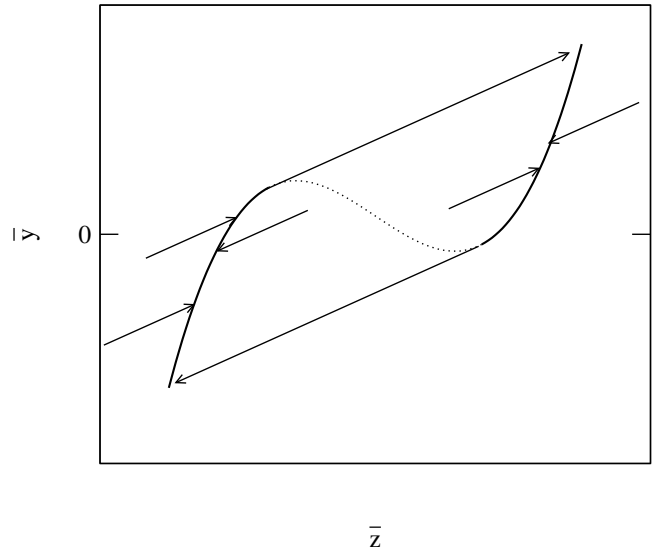}}}}
  {\caption{Schematic presentation of the curve of fixed points in
  Lande and Kirkpatrick's model (the dotted segment is the unstable
  region), and of the linear trajectories of the allele
  frequencies. The fitness maxima and minima coincide with the zeros
  of the fixed point curve. \label{jump} }}
\end{figure}

Sexual selection cannot only create geographic variation in a trait,
but it can also cause the transition from one niche to another even if
the niches are separated by a fitness valley. This elegant alternative
solution to Wright's shifting balance theory was suggested by Lande
and Kirkpatrick \cite{lan88}. The authors again use Lande's model
\cite{lan81} for sexual selection, however with a double-peaked
fitness function for the male trait instead of the Gaussian function
in \cite{lan81}. Using the stationarity condition Eq.~(\ref{weak}), one
obtains a curve of fixed points that has an unstable middle section
(see Fig.~\ref{jump}).
A population which has initially evolved to the left branch of the
equilibrium line may move by random drift to the stability boundary,
from where sexual selection rapidly drives it to the other branch,
which corresponds to the other niche. The authors also study a version
of the model that has in addition to male trait and female preference
a female equivalent of the male trait which is also under natural
selection. Again, they find that sexual selection can drive the
transition from one niche to another.

\subsection{Sympatric speciation}

Sympatric speciation is defined as the establishment of two
reproductively isolated daughter populations within the dispersal
range of the parent population. Such a process can only occur if
mating becomes assortative. In the sexual selection models discussed
so far, an entire population evolves together as a result of female
preferences.  In contrast, assortative mating means that several
subgroups exist that have different traits and preferences. This can
for instance happen if males come in different colours, and if
different females have a heritable preference for males of different
colours. There is good evidence that colour polymorphism has driven
rapid sympatric speciation in certain chichlid lineages occurring in
the Great Lakes of East Africa. Seehausen at al \cite{see99} found
that the colour patterns of most of these species can be classified
into three male and three female patterns, suggesting that the
potential to express these three patterns is ancient and inherited,
and that whenever more than one pattern become expressed in a species,
speciation can occur.

A model showing that sexual selection for two colours can cause
reproductive isolation was introduced by Higashi et al \cite{hig99}.
These authors performed a computer simulation of a model where male
trait $x$ and female preference $y$ can both range from $-2m$ to $+2m$
as the result of additive effects of $m$ loci with the alleles
-1,0,+1. (The male trait could for instance be a colour ranging from
black ($-2m$) to white ($+2m$), with many grey shades in between.) With
a mating preference function of the form $e^{\alpha x y}$, a model
population that initially has only weak preferences and color
variation experiences two simultaneous runaway processes leading to
the establishment of a pronounced bimodal distribution in preferences
and traits, without mating between the two groups. 

A speciation event generally leads also to ecological
diversification. If both species occupied the same niche, drift or
competition would lead to the extinction of one of them. The
establishment of reproductive isolation as described in the previous
paragraph must then be followed by ecological diversification.  Most
models of sympatric speciation, however, consider the possibly more
relevant case where ecological diversification drives reproductive
isolation. In principle two different scenarios are possible. Either
assortative mating is directly coupled to the choice of the ecological
niche, or it is based on a trait that is unrelated to niche choice but
becomes correlated with it during the speciation process. An example
for the first scenario would be a situation where mating occurs on
host plants or host animals or in other types of habitats which
constitute an ecological niche. In the second scenario, alleles for
the expression and preference of traits allowing for assortative
mating (like several colours or pheromones) receive a selective
advantage because they can become correlated with the ecological niche
and thus reduce matings between individuals adapted to different
niches, which would lead to offspring with low fitness.

There are a variety of different models for sympatric speciation
which contain both ecological diversification and assortative mating.
Some models use natural selection based on one locus with two alleles,
others base it on a quantitative character, using a continuous
formalism or many additive loci each with two alleles. According to
the two mentioned possible scenarios, some models base assortative
mating directly on the trait under natural selection, while others
have an additional mating trait, which can again be described using a
single locus or quantitative genetics. 

The first model of this type was introduced by Maynard Smith \cite{MS66}.
The first part of his paper shows that a population can
simultaneously maintain two genotypes adapted to two different niches
if certain conditions are satisfied. One example which he calculates
explicitly is the case of one locus with two alleles, $A$ and $a$,
where $A$ is dominant. The fitness of the $A*$ genotype in niche 1 is
$1+K$ and in niche 2 is 1. The fitness of the $aa$ genotype in niche 1
is 1 and in niche 2 is $1+k$. The population size is regulated
independently in both niches; mating is random, and females lay their
eggs with probability $(1+H)/2$ in the niche where they were born.
There is a stable equilibrium value different from 0 and 1 for the
frequency of the $A$ allele even in the case $H=0$ if $k(1-K)<K$
and $k(1+K)>K$. For $H=1$, this condition becomes $k(1-3K)<K$ and
$k(1+1.5K)>K$. For small selective advantages $k,K$, these conditions
are pretty restrictive, but for stronger selection they are more
easily satisfied. Maynard Smith then argues that once this
polymorphism is established, reproductive isolation can evolve either
through mating in the habitat, or through pleiotropic effects causing
$A$ bearers to mate with $A$ and $aa$ with $aa$, or through the rise
of a modifier allele at another locus which causes assortative mating
with respect to locus A, or through preexisting assortative mating
with respect to a locus B becoming correlated with the A locus. For
this latter case, he performed a computer simulation in order to show
that evolution can indeed converge towards an $AABB$ and $aabb$
population (or an $AAbb$ and $aaBB$ population).

A model using a continuous formalism somewhat similar to the one for
sexual selection presented above, and inspired by models on ecological
character displacement, was suggested by Drossel and McKane
\cite{dro00}. The underlying picture is that of a species invading a
new island or lake with a broad spectrum of food sources. Individuals
with a different character size $z$ can exploit different parts of the
food spectrum. If $K(z)$ is the carrying capacity of the environment
(i.e., a measure of how many individual of size $z$ the environment
could support if no other individuals were present), and
$\alpha(z-z')$ a measure for the strength of competition between two
individuals of size $z$ and $z'$ (usually chosen to be a Gaussian),
then the population distribution after selection is related to the
population distribution before selection via $p_w(z,t) =
W(z,t)p(z,t)/\bar W$ with
$$W(z,t)=1+r-(r/ K(z))\int_{-\infty}^\infty dz'\,
N(t)p(z',t)\alpha(z-z')\, .$$ This linear formula is useful for small
growth rates $r$, or close to a stationary state where fitness values
are close to 1. Far away from a stationary state, the last term may be
large, leading to unrealistic negative fitness values. Therefore, the
better behaved expression
\begin{equation}
W(z,t)=\frac{1+r}{1+(r/ K(z))\int_{-\infty}^\infty dz'\,
N(t)p(z',t)\alpha(z-z')}\, 
\label{fitness2}
\end{equation}
was used in most places in \cite{dro00}. (Other authors use similar
expressions.) The next generation is produced through assortative
mating after selection,
\begin{equation}
p(z,t+1) = C \int
dz_1 \int dz_2\frac{p_w(z_1,t)p_w(z_2,t)\exp[-(z_1-z_2)^2/2V_m]\, 
g(z; z_{1}, z_{2})}
{\int dz_3 p_w(z_3,t)\exp[-(z_1-z_3)^2/2V_m]}\, , 
\end{equation}
where the trait of the children has a Gaussian distribution around the mid-parent value, 
$$
g(z; z_{1}, z_{2}) \propto e^{\left[-\frac{1}{V_{g}} \left( z - 
\frac{(z_{1}+z_{2})}{2} \right)^2\right]}.
$$
Computer simulations and analytical considerations show that $p(z)$
soon develops two or more peaks (depending on the parameters), because
individuals with extreme $z$ values have a selective advantage due to
less competition. In the stationary state, these peaks may become very
pronounced with $p(z)$ being almost zero in the valleys for certain
parameter values, even if the carrying capacity function does not have
multiple peaks, but is sufficiently flat in the center and falls
steeply off at the edges. A conceptual difference to the model by
Maynard Smith is that competition between individuals with similar trait
values drives the expansion of the population into empty niches, a
phenomenon called ``competitive speciation'' by M. Rosenzweig
\cite{ros1,ros2}. In contrast, in the model by Maynard Smith natural
selection is only density-dependent, but not frequency-dependent.

Finally, let us briefly describe two models that appeared recently in
the same issue of {\it Nature} \cite{kk,dd}. Both models have natural
selection based on a trait which is due to the additive effect of
several genes, and assortative mating based on a different trait, or a
male trait and a female preference, which is also the additive result
of several genes. In the model by Kondrashov and Kondrashov \cite{kk},
the fitness function is not frequency dependent and has maxima at
extreme trait values, while the model by Dieckmann and Doebeli
includes the effect of competition between similar phenotypes, and
works with a unimodal carrying capacity. Computer simulations for both
models show that assortative mating can become correlated with
trait values, leading to a separation into reproductively more or less
isolated subpopulations occupying different ecological niches.

Many more models and publications on sympatric speciation exist, and a
good selection of references can be found in the three papers just
discussed.

\section{Hosts and parasites}
\label{parasites}

Many ecologists define a parasite as an organism that lives in or on a
host, from which it derives food and other biological necessities, and
that reduces the fitness of the host by causing morbidity or death, or
by decreasing its reproductive success (see, e.g.,
\cite{cla97}). Parasites include viruses, bacteria, protozoa
(single-cellular eukaryotes like amoebas), fungi, helminths (for
instance worms and leeches), and arthropods (ticks, mites, lice,
fleas, etc). The interactions between hosts and parasites are very
complex. Parasites often have sophisticated multistage life cycles
that may include several intermediate hosts. Others have inert stages
(like spores) that ensure their long-term survival during times
without a host. On the other hand, hosts have a broad repertory of
defense mechanisms, ranging from immune response to behavioural
defenses (like grooming, the avoidance of parasitized sexual partners
and of infested nesting sites, and using the leaves of
parasite-repelling plants in building nests).

Models on host-parasite coevolution have to make many simplifying
assumptions, and they focus on a few fundamental questions like (a)
Which degree of virulence (i.e., damage done to the host) allows the
parasite to multiply best? (b) How many resources should a host
allocate to the defense against parasites? (c) What are the long-time
patterns of host-parasite coevolution within simple genetic models?

In the following, we will discuss these three topics. 

\subsection{Evolution of parasite virulence}

If the parasite draws too heavily on the host's resources, the host
may die quickly, and with it the survival chance of the parasite. This
argument has lead to the expectation that parasites should evolve to a
moderate level of virulence. The best documented example of the
evolution of virulence is that of the Myxoma virus, a very virulent
strain of which was introduced in 1950 into Australia in order to
check the European rabbit, which in the absence of natural enemies had
become an agricultural pest since its introduction in 1859. During the
first summer, the virus spread dramatically, killing more than 99\% of
the infected rabbits within less than two weeks of infection. However,
after the second summer, the virus had changed, and the most common
strain had a reduced virulence, with 70-95\% fatality, and a typical
sickness duration of three weeks or longer. Also, the rabbits became
more resistant to the virus. Laboratory tests investigating the innate
resistance of rabbits found that over seven years the case-fatality
rate for the moderate virus strain decreased from 90\% to 30\%. In
response, the more virulent virus strain increased in number in
regions where rabbits had the highest innate resistance. (For a
relatively recent review, see the article by Fenner and Kerr
\cite{fen94}).

Models for virulence must take into account the qualitative difference
between the dynamics within a host and between hosts. A common class
of models use equations for the numbers $x$ and $y_i$ of uninfected
and infected hosts, with $i$ being the index of the parasite strain:
\begin{eqnarray}
\dot x &=& k -ux-x\sum_i \beta_i y_i\nonumber \\
\dot y_i &=& y_i(\beta_i x -u -v_i). \label{par_dyn}
\end{eqnarray}
$k$ is the growth rate due to birth, recovery, and migration. Some
authors use more complicated expressions, without a change in the main
conclusions. $u$ is the death rate in absence of the parasite, $v_i$
is the increase in death rate due to infection by strain $i$ (i.e.,
the virulence), and $\beta_i$ is proportional to the probability of
transmission of the parasite during a contact between a healthy and an
infected individual. The model assumes that a host cannot be infected
with more than one strain at the same time. If there is only one
strain of parasites, a stable fixed point exists with $x^* =
(u+v_1)/\beta_1$ and $y^* = [k\beta-u(u+v_1)]/[\beta_1(u+v_1)]$. A
second strain can invade if
$$\beta_2/(u+v_2) > \beta_1/(u+v_1)$$ and in this case replaces the
first strain.  Thus, the evolutionary stable strategy is the one with
the highest value of $\beta_i/(u+v_i)$ or, equivalently, of
$R_0=k\beta_i/u(u+v_i)$, which is the number of individuals which a
single infected host can infect when invading a susceptible
population. One can expect that there is a relation between virulence
and transmission: For very small virulence, the parasite density in
the host is low, and so is the transmission rate. With increasing
virulence, the transmission rate increases and can be expected to
increase only slowly or to saturate when virulence becomes high.
Analytical expressions found in the literature and chosen for their
simplicity are $v=c_1 +c_2\beta + c_3 \beta^2$, or
$\beta=c_1v/(c_2+v)$.  Both expressions lead to an optimum of $R_0$ at
some intermediate level of virulence, where transmission rates are
high and host lifetimes long. As pointed out by Lensky and May
\cite{LM94}, the ESS strategy, which has the highest $R_0$ and
lowest $x^*$, is not the one which initially spreads fastest in a
susceptible population and is characterized by the highest $\dot y/y$.
For a review of virulence models of this type, see Chapter 9 of
\cite{FS83} by May and Anderson, and \cite{bull94,GA94}.

If a host can become infected by several strains of the parasite which
have different virulence, it can be expected that the more virulent
strains win over the less virulent strains during growth within the
host, leading to an increase in the mean virulence of the parasite. A
model for ``superinfection'', where the more virulent strains always
replace the less virulent strains within a host, was introduced by
Nowak and May \cite{MN94}. The equation for $\dot y$ in
Eqs~(\ref{par_dyn}) is replaced in their model with
$$\dot y_i = y_i(\beta_ix-u-v_i+s\beta_i\sum_{j=1}^{i-1}y_j-s\sum_{j=i+1}^n
\beta_jy_j),$$ where the degrees of virulence are ranked according to
$v_1<v_2<...<v_n$, and $s$ is the factor by which the transmission
rate is reduced if the new host is already infected by another strain.
Numerical simulation and analytical solution of special cases show
that for $s>0$ several strains can coexist in the population, and that
all of them have a virulence larger than the ESS value found above for
the case of single infection. The same authors also studied a
different model \cite{MN95}, where the more virulent strains do
coexist with the less virulent strains within the same host without
replacing them, but where the virulence perceived by the host is its
most virulent parasite. In this case, the long-term winners are those
parasite strains that have a virulence close to the maximum possible
value for which $R_0$ is still larger than 1.

A different formal approach which combines within-host and
between-host selection was taken by Frank \cite{frank96}. He suggests
a simple expression for the fitness $w_i$ of a parasite as function of
its transmission rate $z_i$ (which is assumed to increase with
increasing virulence):
$$ w_i=\frac{z_i}{\bar z}(1-\alpha \bar z).$$ The first factor
represents the within-host selection and gives an advantage to those
parasites with higher than average virulence. The second factor is due
to between-host competition and confers higher fitness to lower
virulence. $c\bar z$ is the average virulence of parasites within the
considered host. The fitness maximum results from $\partial
w_i/\partial z_i = 0$, leading to
$$\alpha \bar z = 1-\frac{z_i}{\bar z}\frac{d\bar z}{dz_i}.$$ The
winning strategy $z^*$ is the one satisfying $\alpha z^*=1-r$, with
$r=d\bar z/dz_i$. How the mean $\bar z$ changes with the $z$ of one
parasite depends on the degree of relatedness $r$ which is the
probability that two parasites are identical by descent. If all
parasites belong to the same strain (i.e., have
the same genome), a change in the virulence of one parasite implies an
identical change in all parasites, i.e., $r=1$, and the optimal
strategy is $z=0$, or low virulence. On the other hand, if a host is
usually inhabited by many types of unrelated parasites (i.e., the
virulence values of which are independent from each other), $r$ is
small, and $\bar z$ evolves to a large value. These conclusions are
essentially the same as those of the models mentioned before. An
important condition for the validity of this result is that
transmission of the parasite is horizontal, i.e., between hosts
belonging to the same generation. For vertical transmission from
parents to their offspring, one can expect that a low degree of
virulence evolves. This is confirmed by experiments \cite{bul91}.

Let us conclude this subsection by mentioning that within-host
competition does not necessarily lead to increased virulence. In
section \ref{game} we have discussed the example of defective viral particles
which can invade and replace the normal virus strain, but which have a
lower growth rate than the original wild type population once the
original strain is reduced to low frequency within the host.

\subsection{Optimal defense strategies}

Capturing host defense against parasites within a simple model is
impossible, since it involves the immune system (which is very complex
and the object of a whole field of research) as well as defensive and
parasite-avoiding behaviour. Nevertheless, a model may be useful for
gaining some first insights into the question of how resources are
best allocated if higher investment in defense decreases the
investment in reproduction. A simple model which contains a birth rate
$b$ and a recovery rate $\gamma$ was introduced recently by van Baalen
\cite{baa98} and is given by the equations
\begin{eqnarray}
\dot x &=& b(x+y) -ux-x\beta y + \gamma y\nonumber \\
\dot y &=& y(\beta x -u -v -\gamma). \label{baalen}
\end{eqnarray}
It is assumed that $b$ is a decreasing function of $\gamma$. This
model has a stable fixed point $x^*=(u+v+\gamma)/\beta)$,
$y^*=(b-u)(u+v+\gamma)/\beta(u+v-b)$, if $u<b<u+v$. Whether the
population is stable against the invasion of a host $x_2$ with
different $\gamma$ and $b$ can be tested by a linear stability
analysis of the extended model,
\begin{eqnarray}
\dot x_1 &=& b_1(x_1+y_1) -ux_1-x_1\beta (y_1+y_2) + \gamma_1 y_1\nonumber \\
\dot y_1 &=& \beta x_1(y_1+y_2)-y_1( u +v +\gamma_1). \nonumber \\
\dot x_2 &=& b_2(x_2+y_2) -ux_2-x_2\beta (y_1+y_2) + \gamma_2 y_2\nonumber \\
\dot y_2 &=& \beta x_2(y_1+y_2)-y_2( u +v +\gamma_2), \nonumber \\
\end{eqnarray}
around the fixed point $(x_1^*, y_1^*,0,0)$. The result is that this
fixed point is stable if $y_1^* > y_2^*$. Thus, the host that can
maintain the highest frequency of infected individuals wins.

This condition for an ESS of the host can now be combined with the
findings of the previous subsection for the evolutionary stable
parasite strategy. Depending on the analytical form of $\beta(v)$ and
$b(\gamma)$, one or even two fixed points of the complete system may
exist. A specific model investigated by van Baalen has two stable
fixed points for the host-parasite coevolution. At one fixed point,
the hosts invest little in defense and parasites are common but
avirulent. At the other fixed point the hosts are heavily defended
against rare but virulent parasites.

\subsection{Continued host-parasite coevolution}

So far, we have considered situations where a stable fixed point
exists with only one host strategy and one parasite strategy present.
This fixed point was found by maximizing the basic reproductive rate
$R_0$ (for the parasite) or the number $y^*$ of infected individuals
(for the host).

Continued evolution like limit cycles or chaotic trajectories becomes
possible if we take into account that a given parasite genotype may be
best adapted to one host genotype, and less adapted to other host
genotypes. In this case, one can expect a coevolutionary chase where
the parasite evolves to become adapted to the host, and the host
evolves in order to escape the parasite. Models of this type usually
contain one or a few loci with several alleles for the host, and a
matching number of genotypes for the parasite. Furthermore, it is
assumed that each host genotype is matched by a parasite genotype so
that the host fitness is minimal if the frequency of its matching
parasite is 1, and the parasite fitness is maximal if the frequency of
its matching host genotype is 1. Finally, a small mutation rate is
included which randomly changes the genotype of an individual host or
parasite.

A particularly transparent discussion of such a model is given by
Seger \cite{seg88}. In his model, all $n$ genotypes are haploid. Host
genotype $i$ has the frequency $H_i$ (with $\sum_iH_i=1$) and the
fitness $W_i=1-sP_i$, and parasite genotype $j$ has the frequency
$P_j$ (with $\sum_jP_j=1$) and fitness $V_j=1-t(1-H_j)$.  The
population size is assumed to be infinitely large, resulting in
deterministic recurrence equations for the allele frequencies. The
parasite has a small mutation rate $m$. If there is no recombination
(i.e., if children have a genotype identical to their parent, apart
from possible mutation), the fixed point where all genotypes have the
same frequency is unstable when the mutation rate $m$ is smaller than
a threshold value $m^*$ which scales roughly as $1/n^2$ for large
$n$. For $n=2$, trajectories are limit cycles, while they appear
chaotic for $n \ge 3$. If the host genotype is characterized by two
two-allelic loci (i.e., $n=4$), with a recombination rate $r$ between
them, computer simulations reveal a very rich scenario of different
trajectories, depending on the parameter values.

An earlier discussion of host-parasite coevolution which also reports
limit cycles and chaos can be found in \cite{MA83b}. A recent field
study of clonal freshwater snails and their trematode parasites
revealed oscillations similar to the ones predicted by these models
\cite{dyb98}. 

Some scientists believe that the main use of sexual reproduction is to
evade parasites. Recombination can quickly create new genotypes which
might be better at resisting the parasite.  This advantage must be
large enough to offset the ``cost of males'', which is due to the fact
that males often contribute little more but their genetic information
to the next generation, while in a population of parthenogentic
females each individual produces offspring. Computer simulations of a
model with a host population consisting of sexual and asexual
individuals are reviewed by Hamilton et al \cite{ham90}. In this
model, sexual individuals produce only half as many offspring as
asexual individuals; mutation causes the switching between the two
modes of reproduction. Each host is inhabited by one individual of
each of $n$ parasite species. The parasite genome is characterized by
$k$ binary loci, and the host genome by $nk$ binary loci, with $k$
loci for each parasite, and by an additional locus determining whether
reproduction is sexual or asexual. The fitness of each parasite is
given by the number of matching alleles with the host-defense sector,
while the host fitness is reduced by the same amount. A computer
simulation using a finite population size, truncation selection (i.e.,
the least fit host and parasite individuals are eliminated), and small
mutation rates (which are ten times faster in the parasite than in the
host), shows that for a sufficiently large number $n$ of parasite
species the sexually reproducing genotype wins.  Discussions of the
experimental evidence for this hypothesis can be found in
\cite{col99,pen99}.

Let us conclude this section with a model of parasite evolution
in a spatially extended host-parasite system introduced by Savill et
al \cite{sav97}. The main point of this model is to discuss the effect
of spatial patterns on evolution. Each site of a two-dimensional
lattice is assigned a host density and a density of each of the 20
parasite types. Part of the hosts disperse randomly to neighbouring
lattice sites, and parasites disperse preferentially to neighbouring
lattice sites with higher host density. The parasite types differ in
the degree to which they prefer sites of high host density. Host
fitness at a site decreases with increasing total parasite density,
while parasite fitness increases with host density and decreases with
parasite density. Parasites can mutate between the different
types. This model shows spiral patterns and turbulent patterns even in
the absence of mutation, which are due to the nonlinear population
dynamics. These patterns are the equivalent of periodic oscillations
and chaos in population dynamics models of the Lotka-Volterra type,
without spatial structure (which are not discussed here since they do
not involve evolution). In the presence of mutations, the most
successful parasite genotype is determined by the local spatial
pattern and is different for spirals and for turbulence.  Thus,
parasite evolution is enslaved to the large-scale spatio-temporal
structure of the system. This example illustrates that the outcome of
evolution on smaller scales can be determined by properties of the
system on larger scales. This is often ignored in a world of
evolutionary thought where the ``selfish gene'' is the most important
unit.

\chapter{Modelling extinction patterns}

This chapter focuses on models of large-scale evolution that operate
on the time scale of extinction and origination of species.  They are
motivated by findings in the fossil record that species extinction
occurs in avalanche-like events, where often many families of species
become extinct simultaneously.  It has therefore been suggested that
nature itself is in a so-called self-organized critical state, at the
boundary between chaos and stability, where even small events may
sometimes trigger a large avalanche, such that avalanches on all
scales occur. The models presented in this chapter are all very simple
and contain only a few ingredients. All of them share the property of
showing extinction avalanches on all scales.

The outline of this chapter is as follows: First, we will give an
overview of our knowledge of extinction events from the fossil data.
Then, we will define the concept of self-organized criticality and
present various models that can be found in the literature. Finally,
in the last subsection, we will discuss what these models have in
common, and to what extent they are relevant to reality.

\section{The fossil data}

The data for extinction analysis are the distribution in space and
time of fossil species from the Cambrian to the present. This time
period covers 544 million years and is called the Phanerozoic. The
data show that extinctions are clustered in time. Not only are there
several profound mass extinctions like the one at the end of the
Cretaceous, but relatively sudden and rapid turnovers occur at lesser
scales as well.  Both large and small extinctions were used by
19th-century geologists to define boundaries in the time scale
\cite{raup86}.  Thus, the ``Big Five'' extinctions occurred at the ends
of the Ordivician, Devonian, Permian, Triassic, and Cretaceous epochs,
and it is estimated that each of them killed at least 70\% of all
existing species \cite{jab94}. The largest extinction event at the end
of the Permian epoch probably killed around 95\% of all
species. However, these five extinctions account for less than 10\% of
all species that ever became extinct.  Mass extinctions are often
associated with global causes like changes in sea level or in climate,
or meteorite impacts.  Thus, there is a lot of evidence that a large
meteor impact occurred at the end of the Cretaceous epoch
\cite{alvarez}.  Raup and Sepkoski
\cite{raup84} analysed the temporal distribution of 12 major
extinctions during the past 250 million years and found a statistically
significant periodicity of 26 million years. Such a long period can
best be explained by assuming a cosmic cause for these extinctions.
However, a considerable number of mass extinctions do not follow this
pattern, and other authors doubt the existence of any periodicity
\cite{benton95}. During large extinctions, species loss is not random,
and some families are more affected than others.  Also, species-poor
families are not necessarily at greater risk than species-rich ones
\cite{jab94}. For instance, the extinction intensity of sea urchins
during the end-Cretaceous event is not correlated with the number of
species in a genus, and not even with geographical range \cite{mar98}.
However, species with a larger geographic range generally have a
larger chance of survival. This holds even more during quiet times,
where ``background extinctions'' usually affect only few
species simultaneously.  In the following, we will present statistical
data on the size distribution and temporal correlations of extinction
events.

\subsection{The size distribution of extinction events}

\begin{figure} \begin{center}
  \includegraphics[width=0.8\columnwidth]{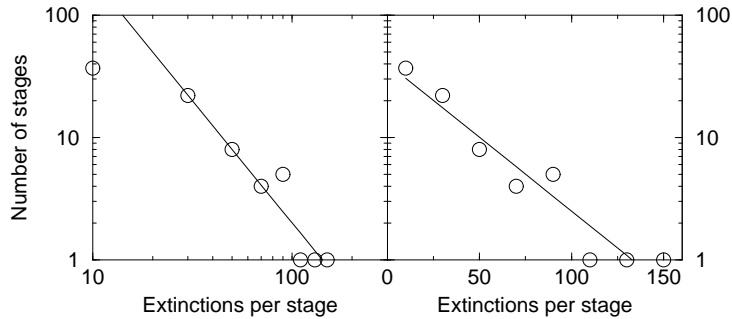}
     \end{center} \caption{The
      frequency distribution of extinction sizes during 79 geological
      stages. The left plot is double logarithmic, and a power law
      with exponent -2 is also shown (solid line). The right plot is
      linear-logarithmic, and the line is an exponential fit with a
      decay constant 0.028. }
\label{figraup} 
\end{figure}

Usually, it is very difficult to date the origin and death of a
species, because not enough of its individuals become fossilised.  It
is easier to date the extinction of an entire group of species, and
for this reason usually data for genera or families are listed. Raup
\cite{raup86} gives a histogram of the frequency of extinction sizes
during the 79 generally recognized geologic stages of Phanerozoic
time, based on recorded times of extinction of 2316 marine animal
families.  His data are shown in Fig.~\ref{figraup}.
These data show that the size distribution of extinction events is
broad, and that extinction events of all sizes have occurred. Because
data points are few, and statistics are not very good, one could fit
the data with a power-law size distribution as well as with an
exponential decay \cite{sol96,new96}, or with some intermediate function.

\begin{figure} \begin{center}
\includegraphics[width=0.5\columnwidth]{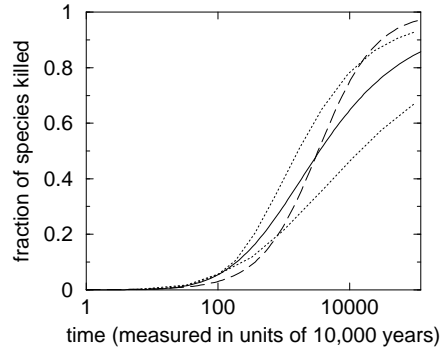}
\end{center} \caption{The kill curve used by Raup (solid line), and the corresponding uncertainties (dotted lines). The dashed line is an alternative kill curve suggested by Newman, and equally compatible with the data.
}
\label{killcurve} 
\end{figure}

\begin{figure} \begin{center}
\includegraphics[width=0.5\columnwidth]{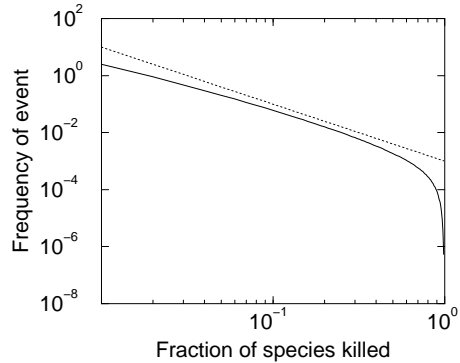}
\end{center} \caption{The distribution of species  extinction sizes 
according to Raup's analysis. (Data from Mark Newman.) The dotted line is a power law with the exponent -2. 
}
\label{ns_newman} 
\end{figure}

A more sophisticated analysis based on more data was done by Raup
several years later \cite{raup91}. He evaluated the stratigraphic
ranges of 17,621 phanerozoic marine genera. These genera can be grouped 
into 68 ``cohorts'', one for each geological stage. He found that the mean
fraction $G$ of genera surviving for at least a time $t$ (measured in
millions of years) can be well fitted by a curve of the form 
$$G=1-[q(e^{(p-q)t}-1)]/[pe^{(p-q)t}-q]$$
with $p=0.249$ and
$q=0.250$.  Such a function can be obtained if one
assumes that each genus initially consists of one species, and that
within a genus each species gives rise to a speciation event with a
rate $p$, and becomes extinct with a rate $q$. The result $p=0.249$ means that the
mean lifetime of a species is around 4 million years. Next, Raup considered
the survival times of genera separately for each ``cohort'', and found
that the data for the different cohorts scatter broadly, much more
than a model with constant speciation and extinction rates would
predict. Only if one assumes that species extinctions do not occur
randomly, but are clustered,  can one reproduce the same degree of scatter.
In particular, Raup tried a ``kill curve'' of the form
$$kill/10,000 yr = (\ln t)^a/\left[e^b+(\ln t)^a\right]$$ for the mean
time interval between extinction events that kill at least the
fraction $kill$ of all species (see Fig.~\ref{killcurve}). The unit
time 10,000 yr was chosen because it appears small enough that no
interval contains more than one extinction event of the size range
relevant to this discussion. In general, it is a very difficult
problem to choose time intervals for defining extinction probabilities
properly
\cite{foote94}.
Raup found that performing a computer simulation
based on a kill curve with $a=5$ and $b=10.5$ reproduces the fossil data
pretty well. As pointed out by Newman \cite{new96}, such a ``kill
curve'' is related to the size distribution of species-extinction
events $n(s)$, via the equation
$$\int_{kill}^1 n(s/S) d(s/S) = 1/t(kill).$$
Here, $s$ is the number
of species killed during an event, and $S$ is the total number of
species.  This size distribution of species-extinction events that
corresponds to Raup's kill curve is shown in Fig.~\ref{ns_newman};
it is not far from a power law with the exponent -2.
Newman \cite{new96} also showed that a power-law size distribution of
extinction events with an exponent $-2$ would be equally suitable for
reproducing the fossil extinction data (see dashed line in
Fig.~\ref{killcurve}). In contrast, an exponential size distribution
of extinction events which has the same mean species lifetime of 4
million years would correspond to a kill curve so flat that it would
lie in the lowest part of Fig.~\ref{killcurve} and could not reproduce
the large extinctions.  To summarize so far, there seems to be good
evidence that the size distribution of extinction events is not far
from a power law with the exponent -2.  However, this result has to be
taken with some caution, since the derivation of the kill curve is
based on a couple of simplifying assumptions, the effect of which has
not been fully explored: First, it was assumed that species
extinctions probabilities are the same for all genera, and that they
are uncorrelated in time.  These assumptions are not generally
correct. Second, speciation rates are assumed to be constant in time
and identical for different genera. This assumption is in contrast to
the known facts that speciations occur at accelerated rate after
mass extinctions, and that speciation rates may vary by one order of
magnitude between different genera \cite{sep98}.

It is therefore worthwhile to note that a simple consideration leads
us also to the conclusion that the size distribution of extinction
events $n(s)$ should be close to a power law with exponent $-2$. The
only assumptions we must make are that the extinction distribution is a
broad function, and that both small and large extinction events kill a
considerable fraction of species. These assumptions seem well
justified by the findings in the fossil record. The condition that a
non-vanishing fraction of all species die during mass extinctions that
kill some finite fraction $1/M$ of all species, means that
$$\left[\int_{S/M}^S s n(s) ds\right] / \left[\int_{1}^S s n(s) ds\right] $$
is not small
even though $S$ is very large, and $n(s)$ can therefore not decay
faster for large $s$ than  a power law with the exponent $-2$. On
the other hand, the condition that a non-vanishing fraction of all
species die during small, geographically confined extinctions that
kill less than $S/M$ species, even when $M$ is chosen rather large,
means that the above expression is not close to 1, and $n(s)$ can
therefore not decay slower for small $s/S$ than a power law with the
exponent $-2$. Together, these conditions give a function similar to
$n(s) \sim s^{-2}$.

\subsection{Lifetime distributions of species and temporal correlations among extinctions}

The lifetime distribution of species and temporal correlations in the
extinction events are related to each other. Imagine that on an
average one extinction event occurs during each time period $\Delta
T$, and that it kills on an average the fraction $x$ of all species,
and that there are no correlations between the events. Clearly, the
probability for a given species to become extinct during a time
interval $\Delta T$ is then given by $x$, and the lifetime
distribution of species is an exponential function $\propto
\exp[-xt/\Delta T]$. Thus, an exponential lifetime distribution
implies that there are no temporal correlations between extinction
events, and that extinction is blind with respect to the age of a
species. Indeed, this seems to be the case if one considers the
lifetime distribution of species within a given genus. Figure
2 of \cite{sol96}
shows the survivorship of a Paleozoic Ammonoidea genus, which is
well described by an exponential function. 
(The paleozoic period covers the time span 570 --225 million years
ago.)
Similar curves, albeit with different decay constants are obtained for
other genera, and often also for the lifetime distribution of genera
within a given family. Because the data in these graphs scatter only a
little, the majority of extinction events contributing to them must be
small ``background extinctions''. Van Valen \cite{valen} suggested an
explanation now known as the Red Queen hypothesis. It says that
species in ecosystems are in constantly evolving interaction. Thus,
the degree of adaptation of older species is not larger than that of
younger ones, because the environment of a species changes all the
time, forcing the species to change all the time. At any time and for
any species, the stress on a species might become larger than it can
bear, and the species then disappears, no matter how long it has
already existed.

\begin{figure} \begin{center}
\includegraphics[width=0.5\columnwidth]{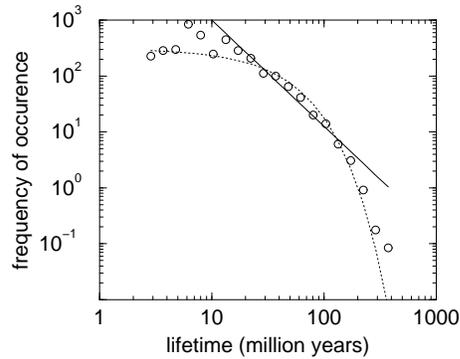}
\end{center} \caption{Frequency distribution of marine genus lifetimes in the fossil record (after \protect{\cite{new99a}}). The dotted line is an exponential function, and the solid line is a power law with the exponent $-1.9$.}
\label{lifetime} 
\end{figure}

As we have mentioned in the context of Raup's ``kill curve'' in the previous
subsection, the contribution of larger extinction events leads to a
much broader scatter of the survival probabilities, although the mean
lifetime distribution of species, averaged over all genera, is still
an exponential function in his model. 
However, several authors have
suggested that on larger time scales a deviation from an exponential
law becomes visible. Fig.~\ref{lifetime} shows the frequency
distribution of marine genus lifetimes in the fossil record. 
On time scales beyond 20 million years, a power-law fit seems to be a
good approximation to the data, if one neglects the last few data
points, which are too low because the lifetimes of genera that still
live today are not yet known and therefore not included. If this power
law is indeed present, it implies long-time correlations in the sizes,
in the selectivity, or in the occurrence of different extinction
events. Such long-time correlations were found by Sol\'e et al
\cite{sole97,sole98}, who considered the Fourier-transform of the correlation
function for extinction sizes, the so-called power spectrum, and found
it to be close to a $1/f$ behaviour. This result could not be
reproduced in \cite{kirch98}. These authors, however, considered the
combined curve for extinction and origination as function of time, and
not just the extinctions. Because large origination events tend to
follow large extinction events, and because they contribute with the
opposite sign, they destroy any long-time correlations that might be
present in the extinctions. A more recent analysis by Newman and Eble
\cite{new99} confirms that the power spectrum can be fitted with a
$1/f$ law for the smaller frequencies, i.e. for larger times. However,
they point out that an exponential fit would work equally well.
As an aside,  none of the
three mentioned studies of the power spectrum found a clear peak at
the frequency 1/(26 million years) or any other frequency.  The
above-mentioned periodicity in the large extinction events found by
Raup and Sepkoski \cite{raup84} is therefore not confirmed by other
authors studying power spectra.

To summarize this subsection, the lifetime distribution of species
within a genus follows an exponential and is mainly due to smaller
extinction events. On larger time scales, where large extinction events
become more important, there is no agreement on whether the lifetime
distribution is still exponential, or a power law with an exponent
close to 2. 

\subsection{The fractal nature of taxonomy}

Although not directly related to the statistics of extinction events,
the size distribution of taxa is a quantity of interest to the topic
of this chapter, because it is shaped by speciation and extinction
processes. In 1922, Willis \cite{willis} noticed that if he counts the
number of species within each genus, and then plots the number of
genera that contain a given number of species as function of the
number of species, he obtains a power law. A more recent and
comprehensive study by Burlando \cite{bur90,bur93} that extends also
to higher taxa and to extinct taxa, confirms this finding. The
exponent of the power law varies between 1.5 and 2.3.

\subsection{The concept of punctuated equilibrium}

The concept of punctuated equilibrium was introduced in 1971 by Gould and
Eldredge \cite{gould77,gould93} in order to interpret the observation
that many species are found to be stable for millions of years.  They
suggested that species originate by rapid branching from an existing
species, and that they tend to remain stable thereafter. Up to then,
it was generally assumed that species change continually, and the
missing fossil evidence for this was ascribed to the incompleteness of
the fossil record. According to Gould and Eldredge, phyletic change
occurs in the majority of cases not by a continuous
transformation of species, but by a sequence of speciation and
extinction events. Their hypothesis has been confirmed by the finding
in the fossil record of many instances of punctuated branching, with
the survival of the ancestral species.  If a species was simply
transformed into a new form, there could be no coexistence of the old
and new form.

In 1993, the concept of punctuated equilibrium for single species was
generalized by Kauffman \cite{kauffman} to evolution as a whole, where
periods of relative quietness alternate with active periods, in which
large extinction events and subsequent speciation occur.

\subsection{Large trends}

Let us conclude this section on the statistical evaluation of the
fossil record by mentioning that several large-scale trends have been
identified for the past 600 million years. Thus, Benton
\cite{benton95} pointed out that the average species diversity has
increased (see also
\cite{jab99}), and Raup and Sepkoski \cite{raup82} found that the mean
extinction rate for genera and families declined during the
Phanerozoic. Similarly, Sepkoski \cite{sep98} showed that rates of
origination have generally declined throughout the Phanerozoic, with
the exception of accelerated speciations during rebounds from mass
extinctions.

\section{Self-organized critical models}

\subsection{The concept of self-organized criticality}

In 1987, Bak, Tang, and Wiesenfeld introduced the idea of
self-organized criticality to explain the frequent occurrence of
power laws in nature \cite{bak87}. Their prototype model is that of a
sandpile. If one slowly drops sand on an initially flat surface, a
pile is built. Initially, the slope of the pile is small, and
occasionally a newly added grain triggers a small avalanche.
Sometimes, a grain reaches the edge of the surface and drops to the
ground. As the pile becomes steeper, the mean size of avalanches
grows. After some time, the pile reaches a stationary state with
avalanches of all sizes. The mean slope of the pile is such that it
allows for some avalanches to extend through the entire system. The
mean number of grains dropped onto the pile per unit time then equals
the mean number of grains leaving the system per unit time. It is
clear that this stationary state with avalanches of all sizes is an
attractor of the dynamics: As long as the slope is smaller than the
critical one, the mean number of grains entering the system is larger
than that leaving the system, and the pile keeps becoming steeper. On
the other hand, if we initially build a very steep pile, a lot of sand
will soon  leave the system in a big avalanche, thus bringing the
pile back to the critical shape.

While real  sandpiles may deviate  from the behaviour sketched  here by
oscillating between  two different slopes \cite{jae89}  or by
having  more large  avalanches due  to their  inertia,  many different
models simulated on the computer show a power-law size distribution of
avalanches. It  is therefore suggested  by Bak and  collaborators that
many slowly driven systems in nature are in such a critical state with
dissipation  events  on  all  scales.  In a  recent  review,  Turcotte
\cite{turcotte} gives many examples  for natural systems that might be
self-organized  critical.  Among  them  are  landslides,  earthquakes,
forest fires, and turpidities. 

It has been suggested that coevolution in extended ecosystems could
also give rise to self-organized critical behaviour
\cite{kauffman,bak93}. If this is true, then extinction events of all
sizes can occur due to the internal dynamics of ecosystems alone,
without the need for any external trigger; large and small
extinctions are caused essentially by the same mechanism.  Models for
networks of interacting species are not automatically self-organized
critical. For example, the model which inspired the Bak--Sneppen
model, is a modification of the NK model introduced in subsection
\ref{NK}; it has a critical point separating a ``frozen'' from a
``chaotic'' phase \cite{KJ,kauffman}. In this model, the contribution
to the fitness of a trait of species $i$ depends not only on other
traits of the same species, but on a certain number, $C$, of randomly
chosen traits of $S_i$ other species to which species $i$ is
connected. For large $K$ (larger than approximately $CS_i$), an
adaptive walk simulation quickly reaches a state where each species
sits at a local optimum. For small $K$, maxima are rare, and the
system keeps evolving if the total number of species is large enough,
because it is impossible for all species to be at a local maximum
simultaneously. Near the boundary between the two phases, a small
change in the system parameters can trigger large coevolutionary
avalanches, starting from a frozen state; their size distribution
seems to be a power law. A modified model was solved analytically and
also shows a frozen and a chaotic phase, separated by a critical point
\cite{fly92}. Self-organized critical systems must possess a property
not contained in this model, which drives the system to the critical
point.

A variety of different models for such self-organized critical
ecosystems were introduced, and a representative selection is reviewed
in the following.  In the last subsection we will then discuss to what
extent these simple models indeed capture the behaviour of nature.

\subsection{The Bak--Sneppen model}

The simplest and oldest self-organized critical toy model for
coevolution was introduced in 1993 by Bak and Sneppen \cite{bak93},
and is discussed extensively in \cite{PMB}. In this model, each point
on a $d$-dimensional lattice with $L^d$ sites represents a species.
The bonds of the lattice define who is interacting with whom. Each
species $i$ is characterized by a single number, $f_i$, which lies in
the interval $[0,1)$ and stands for the fitness of that species.
Initially, the fitness values are attributed at random, and they are
distributed uniformly in the interval $[0,1)$. Now, it is assumed that
species with lower fitness values have a larger probability to become
modified or replaced by a different species than those with higher
fitness values. For this reason, the species with the smallest fitness
is chosen and assigned a new random value.  This step represents
either mutation to a different species, or extinction of a species
followed by replacement with another species in the same ecological
niche. Since a change in one species affects other species that
interact with it, the random numbers on the $2d$ nearest neighbour
sites are also replaced with new random numbers between 0 and 1. Then,
the next species with the smallest fitness value is chosen, and the
same procedure is iterated as long as the computer simulation runs.
Time steps in the computer simulation are not meant to correspond to
time intervals of equal size. Rather, the waiting time until a
mutation or replacement occurs in the species with lowest fitness is
dependent on the value of the fitness. If we equate, for instance, the
fitness with a ``barrier to change'', and if in analogy to
thermodynamics we introduce some kind of (small) temperature $T$, then
the probability for a species to change per unit time can be given as
$\propto \exp(-f_i/T)$, and the species with lowest fitness usually
changes first.

Paczuski, Maslov, and Bak define an $f$ avalanche as comprising all
the updates that occur from the moment where $f$ is the smallest
fitness value in the system, up to the moment where the smallest value
is larger than $f$ for the first time after that.  The size $s$ of an
avalanche is defined to be the number of updates during the avalanche.
Clearly, a site can become part of an avalanche only if one of its
neighbours has been part of the same avalanche before, because only
then the fitness value of the site may have dropped below $f$.  The
larger $f$, the larger the mean avalanche size $\langle s\rangle_f$;
it diverges at a critical value $f_c$, which is $f_c \simeq 0.667$ in
$d=1$.  This critical value plays a special role in the simulations,
because after some transient time, fitness values below $f_c$ occur
with zero weight in the thermodynamic limit. This can be proved by
considering the ``gap'' $G(t)$, which is the largest fitness value
that has been chosen up to time $t$. In the thermodynamic limit $L \to
\infty$, the gap $G(t)$ increases with time according to
\cite{gapequation}
\begin{equation}
\frac{d G(t)}{dt} = \frac{1-G(t)}{L^d \langle s\rangle_{G(t)}}. 
\label{gap}
\end{equation}
The right-hand side of this equation is the ratio between the mean
distance $(1-G(t))/L^d$ between the fitness value $G(t)$ and the next
highest value, and the mean duration $\langle
s\rangle_{G(t)}$ of a $G(t)$ avalanche. From Eq.~\ref{gap} follows
that the size of the gap $G(t)$ increases with time, until it reaches
the critical value $f_c$, where $\langle s\rangle_{G(t)}$ diverges. The system
thus self-organizes itself to a stationary state where the fitness
values above $f_c$ are uniformly distributed, and where fitness values
below $f_c$ occur with zero weight in the thermodynamic limit.

Computer simulations show that the  divergence of the mean avalanche size close to $f_c$  has
a power-law form $\langle s \rangle_f \sim (f_c-f)^{-\gamma}$, with
$\gamma \simeq 2.7$ in $d=1$. Close to $f_c$, one can derive an
expression involving $\gamma$ in the following way: We have
$$\langle s \rangle_{f+df} - \langle s \rangle_{f} =\langle r^d
\rangle \frac{df}{1-f} \langle s \rangle_{f+df}.$$
The first two factors on the right-hand side are the probability that
one of the $\langle r^d \rangle $ sites that has been visited by the
$f$ avalanche has a fitness value between $f$ and $f+df$. If such a
site exists, it takes on an average another $\langle s \rangle_{f+df}$
steps until the $(f+df)$ avalanche is also finished. Using the
definition of $\gamma$, one finds
\begin{equation}
\gamma = \frac{\langle r^d\rangle(f_c-f)}{1-f}, \label{gamma}
\end{equation}
and $\langle r^d \rangle \sim (f_c-f)^{-1}$.

\begin{figure} \begin{center}
\includegraphics[width=0.5\columnwidth]{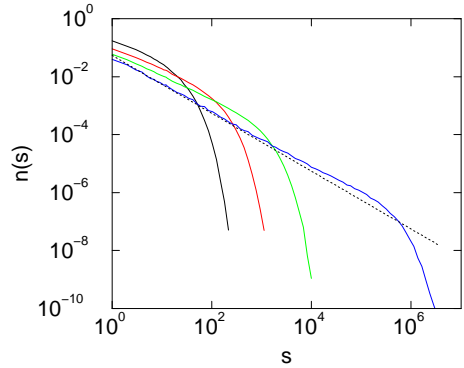}
\end{center} \caption{Size distribution $n(s)$ of $f$ avalanches in the Bak-Sneppen model in $d=1$, for $f=0.446, 0.55, 0.61, 0.66$. The straight line is a power law with slope -1. }
\label{bs_avalanches} 
\end{figure}

Figure \ref{bs_avalanches} shows the size distribution of $f$
avalanches in a 1-dimensional system, for different values of $f$.  At
$f=f_c$, the size distribution of avalanches is a power law $n(s) \sim
s^{-\tau}$, with $\tau \simeq 1.1$ in $d=1$. Because of the slow
convergence towards the asymptotic slope, this type of plot is not the
best way to find the numerical value of $\tau$. Its value can be
obtained more accurately if one measures the number of sites $a(t)$
that are below $f_c$ as a function of the time $t$ since the beginning
of an $f_c$ avalanche.  If one considers only those avalanches that
lasted at least until time $t$, one has $\langle a(t)\rangle \sim
t^{\tau-1}$. The reason is that the average activity at time $t$,
$a(t) \int_t^\infty n(s)ds$, is constant. This can be best understood
if one considers a $(f_c+\epsilon)$ avalanche, with infinitesimally
small $\epsilon$. If such an avalanche has survived for a
sufficiently long time, it will never stop, but approach a state of
constant average activity. The $f_c$ avalanches are just subavalanches
of that infinite avalanche.

A variety of other critical exponents have been introduced that can be
related to $\tau$ and $\gamma$ via scaling relations. Many
modifications of this model have also been studied, and a variety of
analytical calculations have been performed, which cannot be mentioned here.
The interested reader 
can find more references in \cite{PMB,turcotte,newman}.

Let us conclude this subsection by discussing the lessons that can be
learned from this model in the context of biological evolution. The
strongest point of the model is to show that a system consisting of
many interacting units and having some measure of ``fitness'' can
self-organize itself to a critical point, where chains of changes of
all sizes can occur. The more detailed behaviour of the model does not
easily agree with observed data. For instance, there is no easy
way to relate the coevolutionary avalanches of this model with
extinction avalanches in the fossil record. In the Bak-Sneppen model,
mass extinction events with subsequent long periods of recovery do
not occur. Rather, an extinct species is immediately replaced, before
the next one becomes extinct. The avalanche exponent $\tau$ in the
Bak-Sneppen model always lies between $1$ and $1.5$, with the maximum
possible value 1.5 being assumed in a random-neighbor model
\cite{boe94} and in mean-field theory \cite{fly93};  this value is
much smaller than the observed value close to 2 (if there is a power
law at all in the real data). Also, the exponent for the lifetime
distribution of species is 1 in all dimensions
\cite{man98}, which is far from the real value. These problems with
the details notwithstanding, the suggestion that biological nature
might be in a self-organized critical state has inspired the invention
of a variety of other models, several of which will be discussed in
the following.

\subsection{ The Sol\'e--Manrubia model}

A model that allows for true extinctions and for diversification was
introduced by Sol\'e and Manrubia \cite{sol96a}, and studied further
in \cite{sol96b,sol97}. Their model contains $N$ species and a matrix
of couplings $(J_{ij})$ that indicates how each species $j$ affects
each other species $i$. A positive $J_{ij}$ may indicate that species
$i$ can feed on species $j$, or that species $i$ lives in symbiosis
with species $j$. A negative $J_{ij}$ could mean that species $i$ is
eaten by species $j$, or that species $i$ competes with species $j$.
The matrix elements are in the interval $(-1,1)$.
The dynamics of the system consist of the following iterated steps:
(i) For each species $i$, one of its $J_{ij}$ is chosen at random and
replaced with a new value randomly chosen in $(-1,1)$. This mimics
the effect of random changes in the environment, or random drift in
the genetic makeup of a species. (ii) Each species $i$ for which the ``field''
$h_i=\sum_{j} J_{ij}$ is negative, becomes extinct. If $s$ species become
extinct, an extinction avalanche of size $s$ is said to have occurred.
(iii) The extinct species are replaced by slightly modified surviving
species in the following way: A surviving species, $k$, is chosen at
random to become the parent of the new species. For each extinct
species $j$, the couplings $J_{ij}$ and $J_{ji}$ are replaced with
$J_{ik}+\eta_{ij}$ and $J_{ki}+\eta_{ji}$, where the $\eta$ are chosen
randomly from a small interval $(-\epsilon,\epsilon)$.  This mimics
the effect of invasion of an existing species in the now empty niches.

\begin{figure} \begin{center}
\includegraphics[width=0.5\columnwidth]{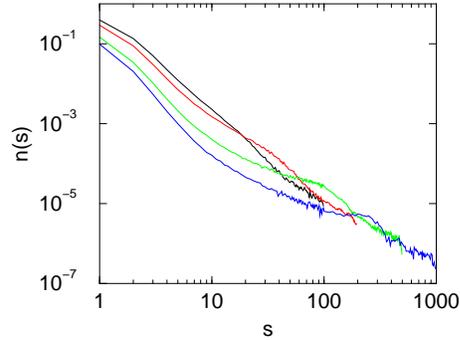}
\end{center} \caption{Size distribution $n(s)$ of avalanches in the Sol\'e--Manrubia model for $N=100,200,500,1000$. }
\label{sm_avalanches} 
\end{figure}

After a transient initial stage, the system evolves to a stationary
state with periods of relative quietness, and periods of large
activity, and with a broad distribution of extinction events. Figure
\ref{sm_avalanches} shows the size distribution of extinction events
for systems with $N=$100, 200, 500, and 1000 species. Although it is
usually claimed in the literature, based on simulations with $N=100$
or 150, that the size distribution is close to a power law with
exponent 2, this is not completely clear from the figure. Rather,
there seems to be a strong dependence on the system size, and it is
not possible to tell from the figure which curve will be approached in
the thermodynamic limit $N \to \infty$.
The lifetime distribution of species appears to follow a power law in
this model \cite{sol96b}. From the branching behaviour of species, a
fractal taxonomy was found which bears resemblance with real
taxonomies \cite{sol97}.

More studies are necessary in order to better understand this
interesting model, and to deduce its avalanche-size distribution in
the large-$N$ limit.  Only when the asymptotic behaviour is known,
can reliable statements be made about the dependence of the results on
the choice of the threshold value of the field for extinction (which
was zero in most simulations), and on other modifications of the model
rules.

\subsection{The Manrubia-Paczuski model}

The model by Manrubia and Paczuski \cite{man98} is to some extent a
simplified version of the Sol\'e-Manrubia model. 
This model contains $N$
species, each of which is characterized by its viability $v_i$, which
is the equivalent of the field in the Sol\'e-Manrubia model, and is an
integer variable. At each time step, the following operations are
performed in parallel for all species: (i) With probability 0.5, $v_i
\to v_i-1$; otherwise $v_i$ is unchanged. This is the equivalent of
the stochastic decrease in the field of a species due to the random
replacement of couplings in the Sol\'e-Manrubia model. (ii) Species
with $v_i$ below a threshold $v_c$ become extinct. For each extinct
species $i$, a surviving species $j$ is chosen at random to become the
parent of a new species that fills the empty niche, and $v_i=v_j$.
This step is similar to the extinction and replacement step in the
Sol\'e-Manrubia model.  (iii) All $N-s$ species that survived
extinction receive a coherent shock $q(s)$ so that $v_j=v_j-q(s)$.
The value of $q(s)$ is chosen randomly in the interval $[-s,s]$ after
each extinction event. The reason for this step is that after
extinction and replacement, the environment and therefore the
viability of the surviving species has changed. In the Sol\'e-Manrubia
model, the fields are automatically changed due to the replacement,
because part of the couplings are changed. 

\begin{figure} \begin{center}
\includegraphics[width=0.5\columnwidth]{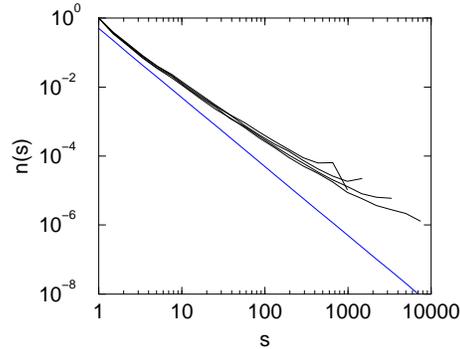}
\end{center} \caption{Size distribution $n(s)$ of avalanches in the Manrubia--Paczuski model for $N=1000,2000,4000,8000$ (from top to bottom). The straight line is a power law with exponent 2.}
\label{mp_avalanches} 
\end{figure}

The dynamics of this model can most easily be understood by focusing
on the species with the largest viability. The viability of this
species decreases slowly due to the random downward drift, and it
experiences upward and downward jumps due to the shocks following
extinction events. On an average, these shocks have size zero and can
therefore not prevent the downward drift. Eventually, the viability of
the top species becomes so small, and the sizes of extinction events
so large, that a shock occurs the size of which is larger than the top
viability, and all species become extinct. This event can be treated
in computer simulations in several ways. One possibility would be to
add the additional condition that the shock size must be smaller than
the maximum viability. Another possibility would be to assign the
viability zero to the $N$ extinct species, who then experience a shock
with $q<0$ that moves them to a high viability value. Both rules
essentially have the same effect, namely that from time to time very
large extinctions occur. Due to shocks that move all species to a
viability value of the order $N$, almost all species subsequently
have the same large viability. A third option, which is the one
chosen by the authors, is to restart the simulation with a random
initial configuration. Figure
\ref{mp_avalanches} shows the size distribution of extinction
avalanches for the second version of the rules. 
One can see that for
small avalanche sizes, the distribution is close to a power law $n(s)
\sim s^{-\tau}$ with the exponent $\tau \simeq 2$, however, the weight
of larger avalanches is higher than this power law. This is different
from the fossil data, where the statistics of large extinction events
do not deviate in the upward direction from a $\tau=2$ power
law. Nevertheless, the curves are closer to the power law for larger
system sizes, and this trend continues to very large system sizes
\cite{MPC}.  Manrubia and Paczuski suggest an analytical treatment of
their model which leads to $\tau=2$, and to an exponent 2 for the
lifetime distribution of species and the size distribution of genera.
Their calculations are based on the assumption that the viability
profile is not significantly affected by extinction events and shocks
in the large $N$ limit, but that it has a stationary shape.  This
assumption is probably good for sufficiently large system sizes, where
system-wide shocks occur rarely, and where the viability profile is
essentially constant over long time intervals.

\subsection{The Newman model}

A particularly simple model that bears some formal resemblance to the
Manrubia-Paczuski model was introduced by Newman \cite{new96,new97}. Like
the Manrubia-Paczuski model, Newman's model does not contain explicit
links between species. Newman characterizes each species $i$ by one
number, $x_i$, which stands for its stress tolerance.  This is
comparable to the viability in the Manrubia-Paczuski model.
Initially, the tolerance values are chosen at random in the interval
$[0,1)$.  The second ingredient of the model is the level of
environmental stress, $\eta$, which is chosen at each time step
independently and at random, from a distribution $p_{stress}(\eta)$
that is a Gaussian or a function with exponential tails. The dynamical
rules consist in the iteration of the following steps: (i) A value for
the stress $\eta$ is chosen, and all species with $x_i<\eta$ become
extinct. (ii) Each extinct species is replaced with a new species with
a random value of $x_i$. (iii) In addition, a small fraction $f$ of
all species obtain a new random value $x_i$. This last step is the
equivalent of the random replacement of part of the bonds in the
Sol\'e-Manrubia model, and the random drift in viability in the
Manrubia-Paczuski model.

After some time, most of the $x$ values are so large that the stress
rarely exceeds them. A requirement for this to occur is that the width
of the stress distribution is sufficiently small, so that the larger
$x$ values are all in the tail of the distribution. The probability
distribution of the tolerance, $\rho(x)$ must satisfy $\rho(x)\sim
1/\int_x^1p_{stress}(\eta)d\eta$ in the stationary state, so that the
mean number of species with value $x$ destroyed is the same for all
$x$, as the mean number of species added at value $x$. We then
obtain the size $s$ of an extinction avalanche triggered by a
stress of strength $\eta$
$$s = \int_\eta^1 \rho(x) dx \sim \rho(\eta) \sim
1/p_{stress}(\eta),$$
and using $n(s) ds = p_{stress}(\eta)d\eta$ we find
$$n(s) \sim s^{-2}.$$  In deriving these results we have twice made use of the
fact that the main contribution to $\int_x^1 p_{stress}(\eta)d\eta$
comes from the lower boundary of the integral. If the distribution was
not Gaussian or exponential, the exponent $\tau$ would be different
from 2.  These analytical results are confirmed by computer
simulations.

The lifetime distribution of species can also be calculated
analytically. Let $p(t>\tau|x)$ be the probability that a newly
created species has at least the lifetime $\tau$, given that it has the
stress tolerance $x$, and let $p_{\tau}(\tau)$ be the lifetime distribution of species. Since a newly created species has each value of $x$ with the same probability, we have
\begin{eqnarray*}
p_{\tau}(\tau) &\sim& -\frac{d}{d\tau} \int_0^1 dx p(t>\tau|x)\\
&\sim& -\frac{d}{d\tau} \int_0^1 dx [1-\int_x^\infty p_{stress}(\eta)d\eta]^{\tau}\\
&\sim& -\frac{d}{d\tau} \int_0^1 dx \exp[-\tau p_{stress}(x)]\\
&\sim& \int_0^1dx  p_{stress}(x)\exp[-\tau p_{stress}(x)]\\
&\sim& \frac{1}{t} \left[ \exp[-\tau p_{stress}(x) \right]_0^1 \sim 1/t.
\end{eqnarray*}
This result is in agreement with the computer simulations by Newman
\cite{newman}, but it does not match the fossil data very well.

As Newman points out, his model shows that a power-law size
distribution of extinction events can occur even if changes in one
species do not affect other species, but under the sole influence of
environmental stress, like climatic changes etc. He also studied
numerically modifications of his model that take into account
interaction between species, correlations between new $x_i$ values and
existing ones, different levels of tolerance for different types of
stress, and taxonomy. It appears that these modifications do not
affect the main results. In the version with taxonomy, the size
distribution of genera has an exponent close to 1.5. Another
modification of the model, where extinct species are not replaced
immediately, but with a rate that depends on the number of present
species, does not affect the size distribution of extinction events
either \cite{wilke97}.

\subsection{The Amaral-Meyer model}

\begin{figure} \begin{center}
\includegraphics[width=0.5\columnwidth]{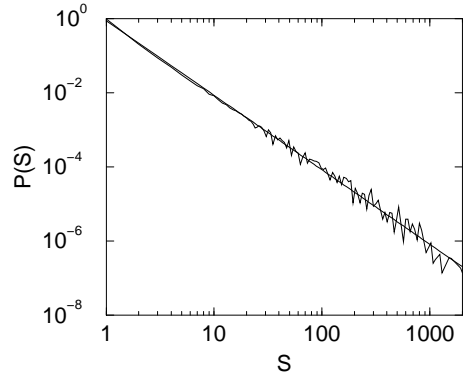}
\end{center} \caption{The size distribution of extinction events in the Amaral-Meyer model for $k=3$, 
$L=6$, $N=999$, $p=0.05$, and $\mu = 0.01$. The straight line is a
power law fit with an exponent $\tau = 2.009$ (After
\protect{\cite{dro98}}).}
\label{amaral} 
\end{figure}

A model that arranges species into food chains was introduced by
Amaral and Meyer \cite{ama99}.  It is defined as follows: Species can
occupy niches in a model ecosystem with $L$ levels in the food chain,
and $N$ niches in each level.  Species from the first level $l=0$ do
not depend on other species for their food, while species on the
higher levels $l$ each feed on $k$ or less species in the level
$l-1$. Changes in the system occur due to two processes: (i) Creation
of new species with a rate $\mu$ for each existing species. The new
species becomes located at a randomly chosen niche in the same level
or in one of the two neighbouring levels of the parent species. If the
new species arises in a level $l>0$, $k$ species are chosen at random
from the layer below as prey. A species never changes its prey after
this initial choice.  (ii) Extinction: At rate $p$, species in the
first level $l=0$ become extinct. Any species in layer $l=1$ and
subsequently in higher levels, for which all preys have become
extinct, also become extinct immediately. This rule leads to
avalanches of extinction that may extend through several layers and
are found numerically to obey a power-law size distribution with an
exponent $\tau \simeq 2$, as shown in Figure \ref{amaral}.

In \cite{dro98}, it was proven analytically that $\tau=2$. The
calculation is particularly simple in the case $k=1$: 
Each species becomes extinct as soon as the bottom species to which it is connected becomes extinct, which happens with a probability $p$ at each time step. The lifetime distribution of species is therefore 
$$p_{\tau}(\tau) \sim \exp[-p\tau].$$ Stationarity requires
furthermore that the probability distribution $p_s(s)$ of the number
of species $s$ connected to a bottom species does not change with
time. When a bottom species is destroyed, $s$ species vanish
altogether. In order to keep the number of connected sets of $s$
species constant, sets of size $s-1$ must grow to sets of size $s$ at
the same rate as species in those sets vanish, leading to
$$ds/dt = ps.$$
Using the identity $p_s(s) ds = p_{\tau}(\tau) d\tau$, we find then 
$$p(s) \sim 1/s^2.$$ 

For $k>1$, one obtains similar results, with $\tau=2$, and with an
exponential tail in the lifetime distribution of species.  A
modification of the model, in which each species becomes extinct as soon
as the first (of several) preys becomes extinct, is not critical \cite{dro98}. One
could also think of a modification in which not only basal species,
but also other species can become extinct spontaneously. Since this allows
for more small extinction avalanches, one can expect the exponent
$\tau$ to become larger than 2.

\subsection{The Slanina-Kotrla model}

The first four models introduced in this section all have a fixed
number of species or niches. Species that become extinct are
replaced with new ones. This situation is different in the model by
Amaral and Meyer presented in the previous subsection, where the total
number of species fluctuates with time. Nevertheless, the total number
of niches available for population is constant in their model. A model
where neither the number of species nor the number of niches is fixed
in advance, was introduced recently by Slanina and Kotrla
\cite{sla99,sla00}.  Each species is assigned a barrier $b$ between 0 and 1
against mutation, and is connected to a couple of other species. The
dynamical rules of the model are as follows: (i) The species with the
lowest barrier $b$ is chosen; $b$ is replaced with a new random value
$b'$, and so are all the barrier values of the species to which it is
connected. This part of the rules is identical to the Bak-Sneppen
model.  (ii) If the new barrier $b'$ is lower than that of all
neighbours, the species and all its links are removed from the
network. The species that subsequently have no links left are also
removed. If the new barrier $b'$ is higher than that of all
neighbours, a daughter species is created which is assigned a random
barrier value, and which inherits all the connections from the mother
species. If the mother species has only one connection, an additional
connection between mother and daughter is created.

Starting with one species, a network is built as the dynamical rules
are iterated, and large fluctuations in the total number of species
and the number of disconnected groups of species are observed.  From
time to time, the network breaks down completely, with only very few
species remaining. The probability distribution for the number of
extinctions (species removals) per iteration step is numerically found
to be close to a power law with the exponent $\tau \simeq 2.3$. The
authors also looked at the number of extinctions during $\lambda$
avalanches that start when the lowest barrier value drops below
$\lambda$, and that end when it is again larger than $\lambda$. These
avalanches are the equivalent of the $f$ avalanches in the Bak-Sneppen
model. The size distribution of $\lambda$ avalanches is characterized
by two different scaling regions, with an exponent close to 2 for
avalanche sizes smaller than a crossover value proportional to
$\lambda^{-3.5}$, and a smaller exponent around 1.65 for the larger
avalanches. These larger avalanches begin and end with a very small
network, and therefore correspond to ``bursts'', during which the
network is built and collapses again.

\subsection{Discussion}

We have described a variety of models all of which show extinction or
mutation avalanches of all sizes as a consequence of the internal
dynamics of the model, without the need for external triggers like
meteorite impacts for large extinction avalanches. Some models can
even be treated analytically.  While the avalanche-size distribution
in some of the models is similar to the ones found in the fossil
record, none of the models reproduces all the features of the fossil
record mentioned in the first part of this chapter. Many other models
can be found in the literature, among them a model based on
Lotka-Volterra type interactions \cite{fer95}. Most models are studied
only with computer simulations and lack a systematic analysis under
which conditions or over which range of avalanche sizes the data
resemble a power law. It is therefore quite possible that many models
which are claimed to be self-organized critical are in fact not. The
reader who is interested in more models is referred to the review by
Newman \cite{newman}.

There are a few ingredients that are common to all the mentioned
models: All models have some measure of the fitness of a species,
sometimes also called viability or barrier against mutation. In the
Amaral-Meyer model the number of prey can be considered as such a
measure of fitness. Furthermore, in all models species cannot smoothly
adjust to changes and thereby maintain a high fitness value that they
have at some moment in time. Rather, the fitness of species with high
fitness values tends to decrease. In the models by Sol\'e and Manrubia
and Manrubia and Paczuski, this decrease in fitness even occurs during
time periods without extinctions, and happens in many small steps. In
the Newman model, even very fit species may be randomly picked and
assigned a different (typically smaller) fitness value. In the
Amaral-Meyer model, species continue to loose prey until all their
prey have become extinct; the rules of the model do not allow a
species to escape this fate by adopting a new species as prey. In the
Bak-Sneppen and the Slanina-Kotrla models, a species with high fitness
usually looses this high fitness when a neighbour
mutates. Furthermore, all models have a threshold in fitness below
which species either become extinct (all mentioned models apart from
Bak-Sneppen model) or mutate spontaneously to a completely different
fitness value (Bak-Sneppen model and Slanina-Kotrla model; this
mutation to a completely different fitness value can of course also be
interpreted as extinction and subsequent invasion of a new species). A
special feature of the Newman model is that the threshold value varies
from time step to time step. In none of the models do the rules allow a
species to steadily improve itself and thus prevent it from eventually
falling below the threshold. Hence, all species have a
finite lifetime. This situation is fundamentally different from that
in the ``webworld model'' by Caldarelli, Higgs, and McKane
\cite{cal98}, where species can adjust to a change in the set of prey,
and where the system reaches a frozen state where all species are well
adapted none become extinct. A modification of this model
\cite{dro00a} includes realistic equations for the dynamics of
the population sizes; it still allows for adaptation of species to a
changed environment, and has only small extinction avalanches.

A fitness decreasing towards a threshold value is in itself not
sufficient to produce a power-law size distribution of avalanches, or
a self-organized critical state.  What is needed furthermore, is the
right degree of coupling, or correlation between species. This can be
best understood by looking at the extreme cases: no coupling or
correlations between the extinction of two species, and perfect
coupling. Imagine a model where the fitness of each species slowly
decreases, without being affected by any other species, until it drops
below the extinction threshold.  In this case, only one or a few
species will fall below the threshold simultaneously and become extinct
together. This means that there are only small, but frequent,
extinction events. In the opposite case of infinitely strong coupling
between species, one species falling below the threshold will pull all
other species with it, leading to the simultaneous extinction of all
species. So far, it is generally not possible to predict on the
basis of the model rules whether the extinction avalanches are mainly
small, or mainly large, or whether their distribution follows a power
law. Only for part of the models presented in this section, a power law can
be deduced by analytical arguments (for the Bak-Sneppen model, the
Newman model, the Amaral-Meyer model, and the Slanina-Kotrla model),
while the Sol\'e-Manrubia model and the Manrubia-Paczuski model are
not yet fully understood.

A question often discussed in the context of extinctions is whether
they are largely due to external influences, or to internal
interactions. The models presented in this section cannot give a clear
answer to this question. The slow or stochastic decline of the species
fitness towards the threshold can be attributed either to changes in
the external environment, like changes in sea level or global cooling
or warming, or to the continuous change in the genetic makeup of the
species network. It might not always be meaningful to distinguish
between the two, as even the climate can be influenced by changes in
the ecosystem. The models do not agree on whether an avalanche is
necessarily driven by the interaction between species. While this is
clearly the case in the Sol\'e-Manrubia model and the Amaral-Meyer
model, the Newman model provides a counter example showing that even
when species do not affect each other's extinction, large avalanches
can occur.  Clearly, the models presented in this section are only a
starting point in our attempt to understand the patterns of
extinctions in the fossil record.

Let us end this section by pointing out that apart from extinction
patterns many other features of large-scale evolution have not been
addressed at all by these models. Among these features are the
large-scale trends to an increase in diversity and complexity and to
less extinctions mentioned earlier in this chapter. All models in this
chapter evolve to a stationary state, where the time-average of
various statistical properties remains constant after an initial
transient period. A model showing an increase of the average fitness with
time, and a decrease of the extinction probability, was introduced by
Sibani et al \cite{sib95}. A model that focuses on the increase in
complexity with time (but not on extinctions and originations) was
recently introduced by Drossel \cite{dro99}. Its main ingredients are
a gain due to specialization, and a cost of interaction (or
communication). Another model for diversification, which is based on 
chemical reaction networks, is suggested in \cite{fur00}.

Recently, new types of models for evolving ecosystems were introduced
 which do not have a time scale separation but are driven by species
 immigration and focus on species diversity and network
 structure \cite{sol00,bas00}. This is a promising direction of
 research, as it tries to model ecosystems in greater detail than the
 simple self-organised critical models.

\chapter{Conclusions}

In this review, we have discussed a broad spectrum of models for
evolutionary processes. All these models have in common that the space
in which evolution takes place is fixed. In models with fitness
landscapes, the genome has a fixed size and is usually represented by
a binary string, with a fitness value assigned to each
configuration. Drift and fixation can be described with these models,
as well as the finding of fitness peaks and the escape from them. In
coevolutionary models, the space is a set of strategies or traits, and
the fitness of an individual depends on the strategies or traits of
other individuals. Evolutionary stable states and attractor
trajectories in trait space can be found. In models of many
interacting species, the degrees of freedom for each species are
usually a set of possible fitness values. Properties of the stationary
state, like extinction avalanches, can be evaluated.

Despite these achievements, a lot remains to be done. Critics of
Neodarwinism keep pointing out its shortcomings and its inability to
explain the large-scale macroevolutionary process. Let us therefore
conclude this review by giving a list of criticisms, challenges and
open problems. 

First, Spetner has tried to calculate time scales for
evolutionary processes based on the assumption that they are the
result of many small steps, each of them being the occurrence and
fixation of a slightly advantageous random mutation. Choosing
realistic numbers for the probability of random
mutations, of the probability that a random mutation confers an
advantage, of the fixation of a slightly advantageous allele, and
estimating the average number of steps needed to generate a new
species, he arrived at the conclusion that Neodarwinism does not
work in this simple form\cite{spe70,spe97}.

Secondly, the space in which evolution takes place is a vanishingly
small subspace of the space of all possible sequences of nucleic
acids, and it has the property that the occurring sequences map
meaningfully on living individuals. The main criticism of
Sch\"utzenberger is that this space (and the mapping) has not been
identified, and that it has not been shown that it has properties such
that it contains trajectories leading from the first cell to organisms
like human beings\cite{schut67,schut96}. An attempt in the right
direction might be the ansatz of some researchers to describe
structure and function of the genome in terms of linguistics (see, for
instance, the article by Ji in \cite{msibe}, pg.~411). There are
indeed striking similarities in design features between the cell's
language and the human language. Thus, only amino acid sequences which are
allowed by the ``semantic'' rules, can occur.

Focusing also on the subspace in which evolution takes place,
Kauffman \cite{kauffman} points out that many characteristics of
organisms are likely to result from of laws of self organization which
determine the generic properties of systems with certain
ingredients. For this reason, natural selection can only choose among
the forms allowed by those laws. A generally acknowledged
manifestation of constraints are pleiotropic effects between
genes. Changing a gene to optimize one trait may negatively affect
other traits that are already at an optimum for a particular
environment.

The existence of constraining natural laws can also be deduced from
the phenomenon of convergent evolution. Equivalent ecological niches
are typically filled by species with similar phenotypic adaptations,
which in many instances have occurred independently. Less known are the
striking examples of convergent evolution on the genetic level. For
instance, ruminants (for example the cow) and colobine monkeys (for
example the langur) independently developed a fermentative foregut
where cellulose is digested with the help of bacteria. The bacteria
themselves are subsequently digested with the help of a lysozyme,
which is related to conventional lysozyme found for instance in
tears. Sequence analysis revealed that several identical amino-acid
substitutions had been made in the conventional lysozyme producing the
lysozyme found in the stomach of both groups of mammals
\cite{ste87}. This also represents a challenge to the neutral theory.

In addition to constraints within an organism, there are also
external constraints, such as natural selection (or
other forces) operating at the higher organizational levels of
groups, populations, and species. Any replicating unit that shows
sufficient stability between its birth and death, and that interacts
as a whole with its environment can be subject to natural
selection. The survival chances of an individual within such a unit do
not merely depend on the properties of that individual, but on the
collective properties of all individuals within this unit. We cited
examples of this in the subsection on group selection (section
\ref{altruism}), and in the model by Savill et al on host-parasite
evolution (section \ref{parasites}), where the large-scale spatial
pattern determines which genotype is advantageous. Gould and Lloyd
\cite{gould99} point out that the variability of a trait in a species 
and its geographic range impart a level-specific component of fitness
to the species as a whole, affecting its survival chances. They also
argue that these traits are not necessarily the result of an
adaptation. The importance of the geographic distribution of
populations for evolutionary dynamics is generally acknowledged and
included in several models. A recent review of metapopulation
biology can be found in \cite{han98}.

There is a fundamental discrepancy between differences in genotype and
in phenotype. We have mentioned in the section on neutral evolution that
the same phenotype may be realized by a variety of different
genotypes. Furthermore, the same genotype can give rise to
different phenotypes, depending on the environmental conditions during
development. There seems to be a considerable phenotypic plasticity
within organisms that does not require any genotypic changes. (See
\cite{west89} for a review.)  On the other hand, genetic changes 
smaller than those between two very similar frog species seem to have
given rise to the large phenotypic differences among mammals
\cite{wil74}.  The genetic changes responsible for macro-evolutionary
changes are generally thought to occur in regulatory genes involved in
embryonic development, where a small change can have a large
phenotypic effect.

Evolutionary innovations typically involve several concerted
changes. For instance, a light-sensitive spot (as the first stage of
eye evolution) is of no use if it does not have the nerve wiring and
brain capacity to transmit and interpret the signal. Similarly, the
``irreducible complexity'' of many molecular biological processes
cannot have arisen through a sequence of random small changes,
according to M.~Behe \cite{behe}. In addition to the just mentioned
possibility of changes in genes involved in development (which are
only relevant for multicellular organisms), many other sophisticated
changes occur in the genome which might be sources of concerted
changes. In section
\ref{mutations}, we have already mentioned transposable genetic
elements. Another example are gene inversions which allow switching
between two modes of operation. For instance, bacteria of the genus
{\it Salmonella} can produce two different types of flagellae,
depending on which of the two responsible genes is turned
on. Recently, it was found that a protein involved in translation
termination in yeast cells can switch
heritably to a different form (called prion), inducing new phenotypic
states \cite{tru00}.

Incorporating these genetic possibilities in theoretical models,
and showing how they can narrow down the subspace in which
evolution takes place and produce adaptations, is one of the great
open challenges in the field.

\bigskip

{\bf Acknowledgement} A variety of people and events have contributed
during the past 5 years to my learning about biological evolution. I
profited a lot from Yaneer Bar Yam's course on complex systems at MIT
in 1996, from the interdisciplinary seminar on evolution and the
collaboration with Alan McKane and Paul Higgs at the University of
Manchester in the years 1997-1999, from the knowledge on genetics and
bacterial biology in Eshel Ben-Jacob's group at Tel Aviv University,
and from the Conference on Biological Evolution and Statistical
Physics in Dresden in May 2000. I thank Mark Newman for sending me the
data for the figures in Section 4.1, Ricard Sol\'e for useful
discussions, Paulien Hogeweg for reference \cite{nim2000}, Mike
Drossel for improving my English, and Uwe T\"auber, Susanna Manrubia
and Ido Golding for helpful comments on the manuscript.  While writing
this review, I first lived on a fellowship from the Minerva
Foundation, and since May 2000 on a Heisenberg Fellowship from the
German Science Foundation, grant number Dr300/2-1.

\end{document}